\documentclass[%
 reprint,
 superscriptaddress,
nofootinbib,
 amsmath,amssymb,
aps,
prd,
]{revtex4-2}

\usepackage{graphicx}
\usepackage{dcolumn}
\usepackage{bm}
\usepackage[
unicode=true,
colorlinks=true,
linkcolor=blue, 
citecolor=blue, 
urlcolor=black, 
hyperfootnotes=true,
psdextra=true,
]{hyperref}

\usepackage{tabstackengine}
\usepackage{mathtools}
\usepackage{mathrsfs}
\usepackage{cancel}
\usepackage{enumitem}
\usepackage{overpic}
\usepackage{nccmath}
\usepackage{mathtools}
\usepackage{orcidlink}
\usepackage{euscript}
\usepackage[
colorinlistoftodos,
prependcaption,
textsize=tiny,
]{todonotes}

\newcommand\pdfmath[1]{\texorpdfstring{$#1$}{#1}}

\newcommand{\RE}{\mathrm{Re}}

\newcommand{\mybar}[3]{%
    \mathrlap{\hspace{#2}\overline{\scalebox{#1}[1]{\phantom{\ensuremath{#3}}}}}\ensuremath{#3}
}

\begin{document}


\title{
Two-Loop Turbulent Helical Magnetohydrodynamics: \\
Large-Scale Dynamo and Energy Spectrum
}


\author{Michal Hnati\v{c}\orcidlink{0000-0003-0051-3036}}
\email{hnatic@saske.sk}
\affiliation{
Institute of Physics, Faculty of Sciences, Pavol Jozef \v{S}af\'arik University, Park Angelinum 9, 040 01 Ko\v{s}ice, Slovakia 
}
\affiliation{
Institute of Experimental Physics, Slovak Academy of Sciences, Watsonova 47, 040 01
Ko\v{s}ice, Slovakia
}
\affiliation{
Bogoliubov Laboratory of Theoretical Physics, Joint Institute
for Nuclear Research, 141 980 Dubna, Moscow Region, Russian Federation
}
\author{Tom\'{a}\v{s} Lu\v{c}ivjansk\'{y}\orcidlink{0000-0001-8529-4893}}%
\email{tomas.lucivjansky@upjs.sk}
\affiliation{
Institute of Physics, Faculty of Sciences, Pavol Jozef \v{S}af\'arik University, Park Angelinum 9, 040 01 Ko\v{s}ice, Slovakia
}%
\author{Luk\'{a}\v{s} Mi\v{z}i\v{s}in\orcidlink{0000-0002-3758-5837}}
\email{mizisin@theor.jinr.ru}
\affiliation{
Bogoliubov Laboratory of Theoretical Physics, Joint Institute
for Nuclear Research, 141 980 Dubna, Moscow Region, Russian Federation
}%
\author{Yurii Molotkov\orcidlink{0000-0002-9459-2099}}
\email{molotkov@theor.jinr.ru}
\affiliation{
Bogoliubov Laboratory of Theoretical Physics, Joint Institute
for Nuclear Research, 141 980 Dubna, Moscow Region, Russian Federation
}%
\author{Andrei Ovsiannikov\orcidlink{0000-0002-6606-6331}}
\email{andrei.ovsiannikov@student.upjs.sk}
\affiliation{
Institute of Physics, Faculty of Sciences, Pavol Jozef \v{S}af\'arik University, Park Angelinum 9, 040 01 Ko\v{s}ice, Slovakia 
}%


\date{\today}

\begin{abstract}



We present a two-loop field-theoretic analysis of incompressible helical magnetohydrodynamics (MHD) in fully developed stationary turbulence.
A key feature of helical MHD is the appearance of an infrared-unstable ``mass-like'' term in the loop diagrams of the magnetic response function.
Physically, this term corresponds to the relevant perturbation of the Joule damping, proportional to  $\boldsymbol{\nabla} \times \boldsymbol{b}$ ($\boldsymbol{b} =$ magnetic field).
Its presence destabilizes the trivial ground state $\langle \boldsymbol{b} \rangle = 0$ and forces us to look for a mechanism for stabilizing the system.
We show that such stabilization can be achieved in two ways: (i) by introducing into induction equation an external mass-like parameter that precisely cancels these dangerous loop corrections (kinematic regime), or (ii) via spontaneous breaking of the rotational symmetry, leading to a new ground state with nonzero large-scale magnetic field (turbulent dynamo regime).
For the latter case, we study the two-loop correction to the spontaneously generated magnetic field and demonstrate that Goldstone-like corrections to Alfv\'en modes along with some other anisotropic structures arise.
Our results also confirm that the emergent mean magnetic field leads to a steeper slope of the magnetic energy spectrum, $-11/3 + 2\gamma_{b\star}$ (with $\gamma_{b\star} = -0.1039 - 0.4202\rho^2$, for $|\rho| \leqslant 1$ as the degree of helicity), compared to the Kolmogorov velocity spectrum of $-11/3$, thereby breaking equipartition.

\end{abstract}


\maketitle

\section{\label{sec:intro}Introduction}

Magnetohydrodynamics studies the interaction of magnetic fields with electrically conducting fluids -- liquids or gases -- considered continuous media.
The theoretical foundation of classical (non-relativistic) MHD consists of the Maxwell equations without displacement currents and the hydrodynamic equations of motion for a continuous medium (usually the Navier-Stokes equations) \cite{LandauLifshitz_FluidMech, Cowling_MHD, Davidson_InttoMHD}.
The connection between these two groups of equations arises, on the one hand, from the induction current generated when a conducting medium moves in a magnetic field; this current must be considered in the Maxwell equations.
On the other hand, the magnetic field's influence on the medium's currents results in an additional electromagnetic body force (the Lorentz force) that must be included in the hydrodynamic equations.
These mutual interactions between the magnetic and velocity fields are significant when describing realistic flows in strongly nonlinear regimes \cite{Moffatt_MagnFieldGen, Davidson_InttoMHD, White_FluidMech}.
Essentially, MHD can be viewed simply as Newton's second law for a conducting fluid, or it can be systematically derived from kinetic theory in the collisional limit as a simple single-fluid description \cite{LandauLifshitz_PhysicalKinetics, Bellan_FundofPlasma, Goedbloed_MHDPlasmaKin}.
Regarding the situation in relativistic MHD, see e.g., \cite{Hattori2022} and the literature cited therein.

Moving conductive fluid with high electrical conductivity (e.g., plasma or molten metals) is a natural component of most celestial bodies, and it is also used in modern large technical facilities: MHD generators \cite{Alemany2021, Dominguez-Lozoya2021, Alemany2023}, tokamaks and stellarators \cite{Xu2016, Pamela2020, Hoelzl2024}, breeder reactors with liquid metal coolant \cite{Rhodes2018, Smolentsev2021, Mistrangelo2021}, metallurgical devices \cite{Hunt1976, Davidson_InttoMHD}, etc. 
From a historical perspective, the development of magnetohydrodynamics as an independent branch of physics is largely due to astrophysical problems, which include the dynamics of cosmic gas masses \cite{Beskin2010, Cranmer2019}, thermal conductivity in cooling flows in clusters of galaxies \cite{Voigt2004, Balbus2008, Parrish2009}, the propagation of cosmic rays \cite{yan2021, Engelbrecht2022}, interstellar scintillations \cite{Gupta2000, FalcetaGonalves2014}, the polarization of light from distant stars \cite{Popova2023}, and many others.
In turn, modern progress in magnetohydrodynamics is largely due to the development of laboratory experiments with liquid metals \cite{Stefani2008, MistrangeloBuhler2021} and numerical simulations 
(see, e.g., bibliography in \cite{Beresnyak_TurbinMHD}).

The need to apply the theory of the interaction between a conducting medium and an electromagnetic field to astrophysical problems was first clearly formulated by Hans Alfv\'en \cite{Alfven_CosmicED}.
He drew attention to the following well-known circumstances.
First, interstellar gas, stellar atmospheres, and matter inside stars are highly ionized and, therefore, are excellent electrical conductors.
Second, many space objects possess magnetic fields.
These facts, along with the observation that despite their extremely low density, the linear dimensions of phenomena considered in astrophysics usually greatly exceed the mean free path of gas particles and can, as a result, be regarded as a continuous medium, formed the basis of the modern theory of cosmic electromagnetic processes (see e.g. \cite{Fleishman_CosmicED, Wald_AdvancedEM}).

Among all mentioned topics, the complex problem of the origin and maintenance of planetary, stellar, and galactic magnetism has always aroused an enduring interest.
Gradually, this problem evolved into a general theory of magnetic field excitation in a moving conducting medium, with a theoretical basis from MHD equations -- commonly known as magnetic dynamo theory \cite{Vainshtein_TurbulentDynamo, Vainshtein_MFinSpace, Zeldovich_MFinAstro, Moffatt_MagnFieldGen, Krause_MFEandDynamo, Ruediger_MProcinAstro, Shukurov_AstroMF, Moffatt_Self-ExcitingDyn, Beresnyak_TurbinMHD}.
This theory is crucial for understanding many astrophysical and geophysical processes, such as planet \cite{Tobias2021} and solar cycle \cite{Tobias2002}, the inversion of the Earth's magnetic field \cite{Jones2011, Roberts2013}, the formation of stars \cite{Sur2010}, jets \cite{Blandford1982, Meier2001} and protostellar disks \cite{Hartmann_Accretion}, accretion \cite{Pringle1981}, etc.

The term ``dynamo'' is associated with the ability of hydrodynamic motions to act like a dynamo machine without wires and windings.
This possibility was originally noted in the famous work of Larmor \cite{Larmor1919}.
The chronological development of magnetic dynamo theory is, in a sense, dialectical.
Larmor's hypothesis collided with a series of so-called ``anti-dynamo'' theorems \cite{Cowling1934, Bullard1954,  Zeldovich1956, Backus1958} that demonstrated the impossibility of a stable dynamo for many classes of symmetric or two-dimensional flows of a conducting fluid (for details, see also \cite{Moffatt_Self-ExcitingDyn}).
Nevertheless, these works were the first to reveal certain physical effects firmly established in modern theory, such as the diamagnetic effect of pushing out lines of force from a turbulent conducting region \cite{Zeldovich1956}.
For some time, the existence of a dynamo was considered infeasible.

The new stage of magnetic dynamo theory development began after the work of Parker \cite{Parker1955}, Elsasser \cite{Elsasser1958}, Braginsky \cite{Braginsky1964a, Braginsky1964b, Braginsky1964c, Braginsky1964d}, and especially Steenbeck et al. \cite{Steenbeck1966}, who showed that a conducting medium with spiral-shaped small-scale motions is capable of generating large-scale magnetic fields.
Then, clear examples of flows leading to dynamos were constructed \cite{Roberts1972flow, Ponomarenko1973}, which led to a revision of the dynamo concept in favor of the fact that movements capable of supporting and generating magnetic fields must
be asymmetrical (usually turbulent) and three-dimensional \cite{Alemany1979, Selyuto1984}.
It is worth noting that mirror-symmetric isotropic or anisotropic turbulence causes only turbulent diffusion of the magnetic field \cite{Steenbeck1966}.
In general, there is no doubt that a turbulent flow with exact statistical isotropy cannot generate a large-scale field. 
Furthermore, all observed large-scale fields, such as those in disk galaxies, are created when the statistical symmetries of turbulence are broken by the system's large-scale asymmetries, such as stratification, rotation, or shear.
Thus, modern dynamo theory describes the excitation of magnetic fields by three-dimensional asymmetric motions of the conducting medium \cite{Moffatt_Self-ExcitingDyn, Beresnyak_TurbinMHD}.

The absence of the displacement current in the MHD equations allows us to effectively use the pre-Maxwell concept of magnetic field lines when considering media with sufficiently high conductivity.
This concept, along with the results on the preservation of the topology of magnetic fields embedded in a highly conductive liquid, due to the effect of ``frozen'' magnetic flux (described back in Alfv\'en's ``Nobel'' article \cite{Alfven1942}, see also modern results about this topic \cite{Eyink2009}), leads to the conclusion that the magnetic field can increase during the movement of the medium if the ``mirror symmetry'' is violated in the system, the simplest measure of which is the kinetic helicity (the product of vorticity and velocity) \cite{Moffatt1969, Berger1984}. 
This quantity also commonly called chirality or gyrotropy, along with other invariants of the ideal MHD equations, such as magnetic helicity, cross-helicity, total energy, and some more specific Casimir invariants, forms the basis of topological invariants that help to understand the mechanisms of energy transfer between hydrodynamic scales using various types of ``cascade'' mechanisms.

Note that in the case of a predominantly turbulent nature of the medium movements the above picture of the frozen magnetic flux is inconsistent with the idea of freely moving turbulent vortices.
Indeed, in the absence of mechanisms for changing the topology of the magnetic field, one would expect the formation of a very chaotic, felt-like structure of the magnetic field, which does not happen.
The fact is that magnetic fields in turbulent fluids do not obey the Alfv\'en theorem and demonstrate the ability to change their topology and connectivity due to the process of magnetic reconnection \cite{Biskamp_Reconnection, Beresnyak_TurbinMHD}.
The discovery of this mechanism made it possible to explain rapid changes in magnetic fields in conducting media and became key to the development of modern magnetohydrodynamics \cite{Priest_MagnReconn, Yamada2010}.

The classification of different types of dynamos is quite extensive. 
Dynamos are distinguished based on various criteria such as the physical mechanisms responsible for their operation -- which include differential rotation ($\omega$-dynamo), the $\alpha$-effect \cite{Vainshtein_MFinSpace, Zeldovich_MFinAstro, Moffatt_MagnFieldGen}, the stretch-twist-fold process \cite{Childress_FastDynamo, Vainshtein_TurbulentDynamo}, etc. -- their dependence on the magnetic Reynolds number into fast \cite{Childress_FastDynamo} and slow dynamos \cite{Zeldovich_MFinAstro}, and the type of motion of the conducting medium, whether laminar (e.g., the Bullard dynamo \cite{Bullard1955}) or turbulent flows \cite{Vainshtein_TurbulentDynamo, Moffatt_Self-ExcitingDyn}.
They are also categorized by spatial scales of action, dividing dynamos into large-scale (mean-field) \cite{Brandenburg2018} and small-scale (fluctuation) types \cite{Brandenburg2005}, as well as by energy sources such as convective or rotational energy.
Additionally, the magnetic field's symmetry and its generation's temporal dynamics may play a role.

In magnetic dynamo theory, it is customary to distinguish between two formulations of the problem: the kinematic (linear) \cite{Zeldovich_MFinAstro, Moffatt_MagnFieldGen, Krause_MFEandDynamo} and the nonlinear dynamo \cite{Childress_FastDynamo, Brandenburg2005}.
In the kinematic problem statement, the magnetic energy is less than the kinetic energy of disturbances at all scales up to the dissipation scale.
The magnetic field is considered passive and does not influence the motion of the conducting medium; the focus is on whether a given velocity field can amplify a magnetic field.
In contrast, in the nonlinear dynamo, the kinetic and magnetic energies reach equilibrium at a certain scale, and the back influence of the magnetic field on the flow is taken into account, leading to a fully coupled system of equations.
The kinematic dynamo is of limited astrophysical importance.
This is because, over a relatively short time (by astrophysical standards), it generates fields strong enough to influence the flow significantly \cite{Vainshtein_TurbulentDynamo, Moffatt_MagnFieldGen}.
Although theoretical studies of the kinematic dynamo are popular due to their relative simplicity, the non-linear dynamo possesses genuine physical significance \cite{Tobias2021}.
Remarkably, the non-linear dynamo can be described quite well using very general statistical approaches, which is one of the topics of this paper.

For geo- and astrophysics needs, a specifically turbulent non-linear dynamo is of greatest interest \cite{Brandenburg2005, Rincon2016, Tobias2021, Brandenburg2023}.
As is known, hydrodynamic turbulence is undoubtedly one of the most widespread distributed non-equilibrium systems in nature.
Indeed, observations of matter motion in the interstellar medium, galaxy clusters, and solar wind have confirmed that turbulent motion in cosmic environments is the rule rather than the exception \cite{Goldstein1995, Armstrong1995, Chepurnov2010}.
Likewise, direct observations of secular magnetic field variations in the planets of the solar system \cite{Finlay2010, Johnson2012, Moore2019, Connerney2018} also indicate intense turbulent processes in their liquid metallic cores.
The causes of turbulence are believed to be various thermal or compositional mechanisms \cite{Brandenburg2005, Roberts1972, Roberts2013, Dormy_MathDynamo}, as well as the development of powerful instabilities such as magnetorotational instability \cite{Balbus1991a, Balbus1991b, Sisan2004, Kawazura2024}.

One cannot overlook the considerable progress made in the numerical study of MHD turbulence \cite{Davidson_InttoMHD, Beresnyak_TurbinMHD, Leidi2023}.
For instance, direct numerical simulations (DNS), which represent ``fully resolved'' numerical experiments, now enable modeling MHD turbulence at Reynolds numbers $\RE$ of
the order of $10^4 - 10^5$ and magnetic Reynolds numbers $\RE_\mathrm{m}$ attaining values up to $10^3$ \cite{Schekochihin2004, Beresnyak2014}.
However, this remains dramatically smaller than those found in real astrophysical conditions, where the Reynolds number may be of the order of $10^8$ or even exceed it.
As a result, a common compromise between the expensive (but exact) DNS and the more rudimentary phenomenological or semi-empirical approaches is to employ large eddy simulations (LES) in astrophysical research \cite{Miesch2015, Kupka2017, Grete2017}.
The core principle of LES in MHD is to explicitly resolve the large scales of the flow while relying on subgrid-scale stress and electromagnetic models to represent the effects of smaller-scale fluctuations, which cannot be directly computed due to computational constraints.
The exact effective Reynolds and magnetic Reynolds numbers attainable with LES depend on the specific subgrid model.
Nevertheless, it is generally accepted that these effective values can surpass those achievable through DNS, enabling a more realistic representation of astrophysical MHD turbulence under limited computational resources \cite{Brandenburg2005, Miesch2009}.
For tasks pertinent to the subject of this study, the following widely used codes can be 
recommended: Snoopy \cite{Lesur2015Snnopy} and Pencil Code \cite{Brandenburg2021PencilCode}.
In addition to numerical simulations, one cannot fail to note the great progress associated with experimental measurements of the turbulent dynamo (see, e.g., \cite{Nornberg2006, Monchaux2007, Stefani2008, Brito2011, Rahbarnia2012}).
Nowadays, theoretical dynamo models are being tested in several large experimental installations, where liquid sodium is usually used (see still up-to-date review \cite{Gailitis2002}).

The ranges accessible to plasma fluctuations in forced MHD turbulence cover many orders of magnitude.
These fluctuations often have spectral distributions in the form of power laws, as in the case of ordinary turbulence.
However, the structure of the energy spectrum in MHD turns out to be much more complex.
The spectrum of MHD turbulence was first independently considered by Iroshnikov \cite{Iroshnikov1963} and Kraichnan \cite{Kraichnan1965}, who proposed a phenomenological framework for describing the turbulent energy cascade in the presence of a guiding magnetic field.
They suggested that MHD turbulence consists of weakly interacting Alfv\'en waves propagating in opposite directions along the magnetic field lines.
Notably, they were the first to point out that MHD turbulence in a strong mean magnetic field -- where the local mean field $\boldsymbol{B}_0$ is much stronger than the fluctuations -- is fundamentally three-dimensional rather than two-dimensional, as previously thought.
This means that the magnetic field cannot be eliminated by a choice of reference frame; thus, each region of the fluid retains a mean magnetic field significantly stronger than small-scale disturbances.
The Kraichnan-Iroshnikov theory predicted an isotropic energy spectrum of the form $E(k) \sim k^{-3/2}$, but it did not account for the anisotropy introduced by a constant magnetic field.
Their initial studies in the strong-field limit concluded that turbulence in the inertial range should be weak.
However, subsequent analyses -- for example, observations of solar wind turbulence \cite{Dobrowolny1980} -- showed that MHD turbulence tends to become stronger, not weaker, during the cascade.
This led Goldreich and Sridhar \cite{Goldreich1995} to propose the existence of another turbulence regime: strong MHD turbulence associated with the so-called critically balanced anisotropic cascade.
Their concept assumes that the timescales of 
non-linear interaction and Alfv\'en wave propagation are equal, leading to an anisotropic turbulence spectrum.
In this regime, the spectrum in the direction perpendicular to the mean magnetic field is similar to the Kolmogorov spectrum, $E(k_{\perp}) \sim k_{\perp}^{-5/3}$, while in the parallel direction, the spectrum is steeper, typically following $E(k_{\|}) \sim k_{\|}^{-2}$.
Because the spectra with slopes of $-5/3$ and $-3/2$ differ by only about $11 \%$, there is an ongoing debate in the scientific community regarding the exact form of the spectrum in specific systems (see, e.g., \cite{Briard2018, Beresnyak_TurbinMHD}).
Over the past 20 years, several modifications of the Goldreich and Sridhar theory have been proposed (see, e.g., \cite{Boldyrev2006}) that predict spectra with a $-3/2$ slope.
However, recent numerical simulations do not support the $-3/2$ spectra \cite{Beresnyak_TurbinMHD}.
Let us also note that, while MHD spectra can be adjusted by accounting for intermittency, the resulting deviations are relatively small and may not be significant in the astrophysical context \cite{Beresnyak_TurbinMHD}.
Note that in the presence of helicity, the ``zoo'' of various energy cascades in MHD becomes even more diverse (see, e.g., \cite{Chkhetiani2006, Pouquet2010}).
The results of this paper, in particular, also allow us to shed some light on the fundamental structure of spectra in the regime of highly developed MHD turbulence, including in the presence of helicity.

The theoretical study of MHD equations employs a wide array of methods and approaches from various areas of modern theoretical and mathematical physics.
The last decades have been marked by several important mathematical results concerning continuous symmetries and similarity solutions \cite{Fuchs1991}, invariants of ``ideal'' MHD \cite{Faraco1991}, and an analog of the ``frozen-in'' theorem \cite{Eyink2009}.
Moreover, for two-dimensional stochastic MHD equations, existence and uniqueness of solutions have been established \cite{Barbu2007, Chueshov2010}.
With the three-dimensional case, the situation is, in all likelihood, the same as in the case of the stochastic Navier-Stokes equation (the existence of a martingale solution has been proven, but the question of its uniqueness is open), except that it is more complicated.
See, e.g. \cite{DaPrato_StochEq} and references therein.

The study of developed turbulence in MHD stands apart.
The theory of ordinary hydrodynamic turbulence was widely developed after the work of Kolmogorov and Obukhov (K41), who showed that the main properties of turbulent flows can be described based on a few assumptions (Kolmogorov's hypothesis) using a certain set of universal parameters. 
Although a general theory of turbulence, which remains one of the most pressing problems of the XXI century, does not yet exist, many of its aspects have already been clarified.
In the context of MHD, the situation is similar.
Despite significant progress in recent decades, most existing approaches to the study of developed MHD turbulence are still either purely phenomenological -- in the spirit of K41 theory -- or are based on the MHD equations in a way that allows us only to establish the occurrence of instability (i.e., increased magnetic field fluctuations) and to study the initial stages of its development by closing the infinite chain of equations at the level of moments (in the simplest case, these are simply linearized equations) \cite{Dormy_MathDynamo, Moffatt_Self-ExcitingDyn, Beresnyak_TurbinMHD}.
Either way, all mentioned methods describe only the initial stages of magnetic field generation, not allowing for full consideration of nonlinear contributions or the study of steady states.
In contrast, the formalism of statistical field theory used in this work is well-suited to describing the final steady state \cite{Amit_FTRG, Zinn_QFT, Zinn_RG,  Frisch_Turb, McComb_Turb, Vasilev_RG, Tauber_CritDyn}.
Originally developed for the needs of quantum field theory (QFT), this formalism was subsequently successfully applied to the theory of critical phenomena, critical dynamics, random walks, polymer physics, the theory of developed hydrodynamic turbulence, and, finally, to models of helical magnetohydrodynamic turbulence.
Regarding the latter, we would like to emphasize the most significant contribution of the French school of theoretical physics
(see, e.g., \cite{Brissaud1973, Frisch1975, Pouquet1976, Pouquet1978, FournierFrisch1983}).

The renormalization group (RG) method is an essential component of the statistical field theory formalism discussed here.
It was first applied to the study of statistical properties of randomly stirred fluids (but not yet turbulent) by Forster et al. \cite{ForsterNelson1976}. 
This pioneering work laid the foundation for a vast field of research involving field-theoretical RG methods that continues to this day.
It should be noted that in the context of turbulence theory, this method is employed in various formalisms that are quite different from one another -- such as field-theoretic RG \cite{AdzhemyanVasilev_UFN}, Wilson's RG \cite{ForsterNelson1976, Eyink1996}, iterative averaging over modes near grid scales \cite{YakhotOrszag2016, FournierFrisch1983}, functional RG \cite{Canet2016}, and others -- which can make it challenging for specialists in the field to understand each other's approaches.
In the present work, we systematically use the standard field-theoretic perturbative RG technique \cite{Amit_FTRG, Zinn_RG, Vasilev_RG}, which has a solid foundation in quantum field renormalization theory and well-developed methods for calculating RG functions and critical dimensions, such as dimensional regularization, Minimal Subtraction (MS) scheme, and the $\epsilon$-expansion.
Within this framework, one can directly obtain regularized analogs of divergent Feynman diagrams and, through the ultraviolet (UV) renormalization procedure, arrive at asymptotic $\epsilon$-expansions for universal quantities and the corresponding amplitudes.
Unfortunately, despite the impressive results of multi-loop calculations in critical phenomena and stochastic dynamics, the application of RG theory to fully developed turbulence remains mainly limited to analytical calculations up to two-loop approximations.
Notably, studies on infinite-dimensional turbulence have recently achieved four-loop accuracy \cite{Adzhemyan2008, Adzhemyan2013, Adzhemyan2024, Adzhemyan2024_FourLoop}, offering some hope of finding an exact solution to the Navier-Stokes equation in the limit of infinite dimensions, analogous to the well-known non-linear $\sigma$ model \cite{Bykov2021, Affleck2021, Bykov2022, Bykov2024, Krivorol2025, Zinn_QFT, Zinn_RG}.
On the one hand, reaching two-loop accuracy does not yet give rise to the problem of resuming the obtained asymptotic series, as is typically encountered at this order, the coefficients are even smaller than in one-loop order (see, e.g., the results of calculating the two-loop turbulent Prandtl number \cite{Adzhemyan2005}).
On the other hand, this problem will inevitably arise at higher orders.
After that, a correct resummation in any case requires to know the asymptotics of high orders of the coefficients of the perturbation theory expansion.
To the best of our knowledge, in this area of turbulence theory, progress beyond the simplest Burgers equation \cite{Polyakov1995, GurarieMigdal1996, Balkovsky1997, Fleischer2001, Grafke2015} has not yet been achieved, not to mention the 3D Navier-Stokes or MHD.

In the context of reflection-symmetric magnetohydrodynamics, the RG method was applied in \cite{FournierFrisch1983}, and the corresponding field-theoretical version was considered in \cite{Adzhemyan1985}.
It was shown that in an isotropic case for both infrared (IR) stable fixed points of the model -- the so-called kinetic and magnetic fixed points -- Kolmogorov-type spectra emerge \cite{FournierFrisch1983, Adzhemyan1995}.
Later, authors \cite{Adzhemyan1987} considered the case of helical (gyrotropic in their terminology) MHD without magnetic noise.
The key point is that the violation of reflection symmetry -- even without magnetic fluctuations -- leads to an instability in the linearized MHD equations, a phenomenon analyzed in \cite{Pouquet1978}, resulting in a large-scale homogeneous magnetic field.
This field potentially stabilizes the system in its new ``vacuum state'' through the dynamical spontaneous breaking mechanism \cite{ColemanWeinberg1973} of three-dimensional rotational symmetry, effectively acting as a dynamo mechanism.
From a technical standpoint, the discussed instability generates new polynomial divergent tensor counterterms, analogous to mass renormalization in the $\mathbf{O}(n)$-symmetric $\varphi^4$ theory.
These counterterms arise during the UV renormalization, starting from one-loop diagrams, and are eliminated by applying a specially chosen magnetic field shift.
Additionally, authors \cite{Adzhemyan1987} demonstrated that in such a dynamo regime, the Kolmogorov spectrum holds only for the velocity pair correlator.
At the same time, the magnetic field correlator exhibits a different slope, which, in fact, indicates a fundamental violation of the principle of energy equipartition.

Subsequently, the method was successfully applied to other models of magnetohydrodynamic-like turbulence, such as the well-known kinematic Kraichnan-Kazantsev model \cite{Kazantsev1968, Kraichnan1968, Zeldovich_MFinAstro, Antonov2012, Antonov2012a, Antonov2013}, the kinematic MHD model \cite{Jurcisinova2011, Antonov2015, Jurcisinova2021}, and to various advection-diffusion models similar to MHD to varying degrees, involving various impurities -- including passive scalar impurities \cite{Antonov2014, Antonov2021}, magnetic impurities \cite{Hnatic2016}, and vector impurities \cite{Antonov2015a, Jurcisinova2014, Jurcisinova2016, Antonov2019}. 
For earlier developments on this matter, see \cite{Antonov2006} and the literature cited therein.
It is worth noting here the paradigmatic $A$-model (not to be confused with the
 standard Hohenberg and Halperin notation \cite{HohenbergHalperin1977} for critical dynamics models) of an active vector admixture, a special case of which is chiral MHD.
The combined efforts of the team of authors whose works are presented above have brought the accuracy of calculations in it to two-loop order over the past two decades \cite{Hnatic2024}.

In this paper we undertake a two-loop study of the simplest stochastic model of fully developed homogeneous and isotropic turbulence in an electrically conducting fluid in the turbulent dynamo regime, originally formulated in \cite{Adzhemyan1987}.
Whilst by ``simplest model'', we mean that the turbulent medium is assumed to be incompressible; the expression for the current is taken in the simplest form of Ohm's law for a moving medium; the dynamics of the velocity field does not account for the influence of Coriolis and buoyancy forces; and the magnetic field acts as a passive vector impurity (there is no random magnetic noise).
The present study includes a two-loop calculation of the amplitude of the large-scale mean magnetic field that naturally arises in MHD under strong
non-linearity and an investigation into the features of the energy spectrum of MHD turbulence that emerge in this regime.

The plan of the paper is the following: Sec.~\ref{sec:problem_statement} introduces the stochastic MHD equations in a helical medium, presents their field-theoretical formulation, discusses UV renormalization, and provides an overview of current results and their applicability limits. Sec.~\ref{sec:instability} describes the intrinsic long-wavelength instability in MHD characteristic of helical media and discusses approaches to its removal. 
Sec.~\ref{sec:model_in_dynamo_regime} extends the helical turbulent MHD model to describe the steady-state characterized by the presence of a large-scale magnetic field in the system.
The associated non-trivial calculation technique, the structure of the diagrams, the renormalization, and the absence of additional instabilities are described in detail.
In addition, the presence of a massless Goldstone mode in the model, which represents long-lived power-law corrections to the Alfv\'en waves, is discussed.
Sec.~\ref{sec:two-loop_B_field_calculations} contains detailed two-loop calculations that determine the large-scale mean magnetic field generated by the turbulent dynamo in the system.
Particular attention is paid to the multi-level aspects of the renormalization procedure, as well as to the unification of the results with the RG technique.
Sec.~\ref{sec:RG_in_dynamo_regime} demonstrates how to integrate the ideas from the previous sections with the RG approach to obtain expressions for the energy spectra and critical dimensions in the dynamo regime.
Finally, Sec.~\ref{sec:conclusions} contains some conjectures and concluding remarks, and suggests potential future research prospects.

\section{\label{sec:problem_statement}Preliminaries and State of the Art}

\subsection{\label{sec:chiral_MHD}Stochastic helical magnetohydrodynamics}

The standard non-relativistic, resistive, stochastic MHD equations for a homogeneous fluid without magnetic noise are:
\begin{eqnarray}
&\text{D}_t \boldsymbol{v} = - \boldsymbol{\nabla} \big(p + \boldsymbol{b}^2/2\big) + \nu_0 \boldsymbol{\nabla}^2 \boldsymbol{v} + (\boldsymbol{b} \cdot \boldsymbol{\nabla}) \boldsymbol{b} + \boldsymbol{f}^{v}, \label{eq:MHD_NS}\\ 
&\text{D}_t \boldsymbol{b} = \kappa_0 
\boldsymbol{\nabla}^2 \boldsymbol{b}
+  (\boldsymbol{b} \cdot \boldsymbol{\nabla}) \boldsymbol{v},  \label{eq:MHD_induction} \\
& \boldsymbol{\nabla} \cdot \boldsymbol{v} = 0, \qquad \boldsymbol{\nabla} \cdot \boldsymbol{b} = 0, \qquad \boldsymbol{\nabla} \cdot \boldsymbol{f}^{v} = 0. \label{eq:MHD_transvers_cond}
\end{eqnarray}
Here, $\boldsymbol{v}(t, \boldsymbol{x})$ and $(4\pi \rho_0)^{1/2} \boldsymbol{b}(t, \boldsymbol{x})$ are the stochastic (pulsating) components of the velocity and magnetic induction fields, respectively (the field $\boldsymbol{b}$ is expressed in units of the Alfv\'en's velocity).
The terms $\rho_0 p(t, \boldsymbol{x})$ and $\rho_0 \boldsymbol{f}^{v}(t, \boldsymbol{x})$ represent the pressure and the external random force that drives turbulence, while the operator $\mathrm{D}_t$ denotes material (convective) derivative, defined as $\mathrm{D}_t \coloneqq \partial_t + (\boldsymbol{v}\cdot \boldsymbol{\nabla})$.
Throughout this work, $\boldsymbol{x}$ represents a $d$-dimensional spatial vector, $t$ denotes the time, $\boldsymbol{\nabla}$ is the spatial gradient operator, $\partial_t \coloneqq \partial / \partial t$ denotes the time derivative, $\boldsymbol{\nabla}^2$ is the Laplace operator, and the dot symbol indicate the standard Euclidean inner product
of $d$-dimensional vectors.
The parameters $\nu_0$ and $\kappa_0 = c^2/4\pi \sigma_0 \coloneqq u_0 \nu_0$ represent the kinematic viscosity and magnetic diffusivity, respectively.
The constants $\rho_0$ and $\sigma_0$ denote the density and conductivity of the turbulent medium, while $c$ is the speed of light in vacuum.
For convenience, the diffusivity $\kappa_0$ is also expressed here in terms of the inverse magnetic Prandtl number $u_0$.
Let us note from the outset that, for verification and transparency of intermediate calculations, we retain an arbitrary dimension $d$ in the expressions.
However, ultimately we are interested in the three-dimensional case $d = 3$.
Generalization of the results of the following Sections to dimensions other than three is not trivial, and sometimes not even possible.

The pressure field $p(t,\boldsymbol{x})$ in Eq.~\eqref{eq:MHD_NS} is not a full-fledged dynamic parameter due to the incompressibility conditions and the low Mach number characteristic of developed turbulent flows.
Instead, it acts as a Lagrange multiplier, enforcing the solenoidal condition on the velocity field, and can be determined by solving the following Poisson equation:
\begin{equation}\label{eq:pressure_field}
    \boldsymbol{\nabla}^2 p = - \mathrm{div}\,\mathrm{div}\,
    \left( \boldsymbol{v} \otimes \boldsymbol{v} - \boldsymbol{b} \otimes \boldsymbol{b} \right) - \boldsymbol{\nabla}^2 \left(\boldsymbol{b}^2/2 \right).
\end{equation}
Hereinafter, $\otimes$ denotes the tensor product, and $\mathrm{div}$ is the standard divergence operator.

In the usual field-theoretic formulation of the problem, the system of Eqs. \eqref{eq:MHD_NS} -- \eqref{eq:MHD_transvers_cond} is supplemented by the boundary conditions
\begin{align}\label{eq:MHD_boundary_cond}
    \lim_{\|\boldsymbol{x}\| \rightarrow \infty} \boldsymbol{v}(t, \boldsymbol{x}) = 0, \qquad  \lim_{\|\boldsymbol{x}\| \rightarrow \infty} \boldsymbol{b}(t, \boldsymbol{x}) = 0,
\end{align}
as well as asymptotic initial conditions \begin{align}\label{eq:MHD_initial_cond}
   \lim_{t \rightarrow -\infty} \boldsymbol{v}(t, \boldsymbol{x}) = 0, \qquad \lim_{t \rightarrow -\infty} \boldsymbol{b}(t, \boldsymbol{x}) = 0.
\end{align}
With this formulation, the solution is determined by the long-time balance and does not depend on an arbitrary initial condition at a finite time. 

The force density $\boldsymbol{f}^{v}(t, \boldsymbol{x})$ is a stationary-in-time, homogeneous-in-space translation-invariant random stochastic process with zero-mean Gaussian statistics of the form:
\begin{eqnarray}
\label{eq:force_correlator}
\mathfrak{D}_{i j}^v \coloneqq \langle f_i^{v} (\boldsymbol{x}, t) f_j^{v} (\boldsymbol{0},0) \rangle = \delta(t) \int \frac{\mbox{d}^d k}{(2 \pi)^d} \mathbb{D}_{i j} (\boldsymbol{k}) \mathrm{e}^{i \boldsymbol{k} \cdot \boldsymbol{x}},
\end{eqnarray}
where $\langle \bullet \rangle$ denotes the averaging over the distribution of $\boldsymbol{f}^v$, and $\mathbb{D}_{i j} (\boldsymbol{k})$ is the so-called pumping function, which will be described below.
Translation invariance of \eqref{eq:force_correlator}, as well as Eqs. \eqref{eq:MHD_NS} -- \eqref{eq:MHD_transvers_cond}, leads to the fact that all moments of the corresponding distribution will be translation-invariant automatically. 
This advantageously distinguishes used formulation from the Cauchy problem, for which translation invariance is restored only at $t\rightarrow \infty$, when the system ``forgets'' the initial data.

It is important to note that the absence of a static limit for the Navier-Stokes equation implies that the choice of the random force correlator \eqref{eq:force_correlator} for the corresponding stochastic hydrodynamics is in some sense arbitrary and is considered as part of the problem statement in the stochastic theory of developed turbulence.
In turn, fully developed turbulence in the inertial range does not depend on the details of the random force structure, except for very general aspects.
Note also that, from a technical point of view, some restrictions (power-law asymptotic of the kernel $\mathbb{D}_{i j}$ at $|\boldsymbol{k}| \rightarrow \infty$) on the form of $\langle f_i^{v} f_j^{v} \rangle$ are imposed by the RG method.

The stochastic force itself acts simultaneously in two different roles: on the one hand, it simply models stochasticity, which in reality arises as a consequence of the instability of the laminar flow, and on the other hand, in a problem without magnetic noise, it takes on the entire burden of maintaining the turbulent state through interactions with large-scale eddies.
Accordingly, in developed turbulence, the shape of the correlator of this force should correspond to the expected behavior of the energy pumping into the system.
This pumping is assumed to be concentrated in the IR region.
It compensates for the average energy losses due to viscosity and resistance, thus ensuring a stable state of turbulence.

In this paper, we are interested in the forced MHD turbulence system in a helical media.
Technically, helicity implies that the system's symmetry group is $\mathbf{SO}(3) \simeq \mathbf{O}(3)/\mathbf{Z}_2$, rather than the full spatial rotation group $\mathbf{O}(3)$.
Physically, the group $\mathbf{Z}_2$ represents usually uniaxial reflections, although more exotic cases are possible.
With this in mind, in the inertial range we choose the pumping function in the simplest form as
\begin{align}\label{eq:pump_function}
\mathbb{D}_{ij}(\boldsymbol{k}) = g_0 \nu_0^3 k^{4 - d - 2\epsilon} \theta(\Lambda - k)\theta(k - m)\mathbb{R}_{ij} (\boldsymbol{k}).
\end{align}
Here and throughout the paper $\theta(x)$ is the Heaviside step-function, $g_0 \simeq \Lambda^{2 \epsilon}$ serves as the coupling constant (i.e., the expansion parameter in the ordinary perturbation theory), $\epsilon > 0$ is the model parameter characterizing the type of pumping with the physical value $\epsilon = 2$, and $k \coloneqq |\boldsymbol{k}|$ here and from now on.
The range $m \ll k \ll \Lambda$ defined from \eqref{eq:pump_function} represents a usual inertial range in forced turbulence, with $\Lambda \coloneqq l_{\mathrm{mic}}^{-1}$ and $m \coloneqq l_{\mathrm{mac}}^{-1}$, where $l_{\mathrm{mic}}$ and $l_{\mathrm{mac}}$ are characteristic microscale (e.g. dissipative length) and macroscale (e.g., linear size) of the system, respectively.
In other words, the external scale $m$ contains the main energy, and the dissipation scale $\Lambda$ is where the energy is dissipated into heat.
For further details on the described problem statement and the construction of $\epsilon$-expansion in turbulent models, we refer the reader to Refs. \cite{Adzhemyan1989, Adzhemyan2003,  AdzhemyanVasilev_UFN, Adzhemyan_RGinFullDevTurb, Vasilev_RG}.

The tensor quantity $\mathbb{R}_{ij} (\boldsymbol{k})$ is a legacy of the requirement of fluid incompressibility along with the violation of spatial parity:
\begin{align}\label{eq:tensor_R}
    \mathbb{R}_{ij} (\boldsymbol{k}) = \mathbb{P}_{ij} (\boldsymbol{k}) + \rho \mathbb{H}_{ij} (\boldsymbol{k}),
\end{align}
where $\mathbb{P}_{ij} (\boldsymbol{k}) \coloneqq \delta_{ij} - k_i k_j/k^2$ represents the standard Leray projector, and $\mathbb{H}_{ij}(\boldsymbol{k}) \coloneqq i \varepsilon_{i j m} k_m/k$ is the so-called helical term appearing in the absence of mirror symmetry, $\varepsilon_{i j m}$ is the Levi-Civita symbol, and $\rho$ quantifies the degree of mirror symmetry breaking.
From the Schwarz inequality, it directly follows that $|\rho| \leq 1$.

We assume the basic propositions of the phenomenological K41 theory also apply to MHD turbulence.
In modern form \cite{Vasilev_RG}, these reduce to two hypotheses: (1) in the region $k \gg m$, $\omega \gg W^{1/3} m^{2/3}$, the static Green functions depend solely on the total pumping power $W$ and not on its detailed structure; in particular, for them there exists a limit $m/k \to 0$; 2) in the region $k \ll \Lambda$, $\omega \ll \nu_0 \Lambda^2$, the Green functions (both static and dynamic ) are independent of the viscosity $\nu_0$.
Here, $W$ is the average pump power determined from the corresponding energy budget equation.
Consequently, the model’s independent parameters are $m$, $\nu_0$, $u_0$, and $W$; all others follow from dimensional reasoning (e.g., $\RE = (\Lambda/m)^{4/3}$ with $\Lambda = (W/\nu_0^3)^{1/4}$).

In the practical multiloop calculations, it is convenient to use the simplest IR cutoff regularization, realized by the step function $\theta(k - m)$ in~\eqref{eq:pump_function}.
The crucial point here is that the number of lines in the corresponding diagrams that include the kernel $\mathbb{D}_{ij}(\boldsymbol{k})$ is always the same as the number of loops.
Note that cutoff regularization does not break the global symmetry group of our model, unlike in relativistic field theory.

In formulation above, the quantities actually calculated are various correlation functions -- tensor products of the fields $\boldsymbol{v}$ and $\boldsymbol{b}$ -- as well as response functions, which are variational derivatives of the correlation functions with respect to a non-random external force introduced as an additive term into the right-hand side of the Eqs. \eqref{eq:MHD_NS} -- \eqref{eq:MHD_induction}.
These response functions describe how the physical system reacts to small external excitations.

\subsection{\label{sec:field-theoretic_formulation}Field-theoretic formulation}

The statistical properties of the dynamic stochastic Eqs. \eqref{eq:MHD_NS} -- \eqref{eq:MHD_transvers_cond} can be studied via the functional integral formulation of the Martin-Siggia-Rose (MSR) formalism \cite{MSR1973}, well-established in statistical (Euclidean) field theory \cite{Vasilev_RG, Zinn_QFT, Tauber_CritDyn}.
The corresponding (unrenormalized) De Dominicis-Janssen \cite{Janssen1976, DeDominicis1976} action functional involves classical random fields $\boldsymbol{v}$, $\boldsymbol{b}$, as well as the auxiliary transverse fields $\boldsymbol{v'}$ and $\boldsymbol{b'}$ \cite{Adzhemyan1987}:
\begin{align}\label{eq:MHD_action_without_shift}
\mathcal{S}_0 &= \frac{1}{2}\boldsymbol{v'} \mathfrak{D}^v \boldsymbol{v'} +\boldsymbol{v'}\cdot \left[-\text{D}_t \boldsymbol{v} + {\nu}_0 \boldsymbol{\nabla}^2 \boldsymbol{v} + (\boldsymbol{b} \cdot \boldsymbol{\nabla}) \boldsymbol{b}\right]  \notag \\
&+\boldsymbol{b'}\cdot\left[-\text{D}_t \boldsymbol{b} + u_0 {\nu}_0 \boldsymbol{\nabla}^2 \boldsymbol{b} + (\boldsymbol{b} \cdot \boldsymbol{\nabla}) \boldsymbol{v}\right],
\end{align}
where $\mathfrak{D}^v$ is the force correlator matrix from Eq. \eqref{eq:force_correlator}.
Here and in the expressions that follow, we use a condensed notation in which all integrals over space-time and summations over repeated vector indices are implied.

Following the standard perturbation theory methodology, the action \eqref{eq:MHD_action_without_shift} is compactly rewritten as:
\begin{equation}\label{eq:canonical_action_without_shift}
    \mathcal{S}_0 = -\frac{1}{2}\Phi K \Phi + \mathcal{S}_{\mathrm{int}},
\end{equation}
where the abstract vector $\Phi \coloneqq (\boldsymbol{v}, \boldsymbol{b}, \boldsymbol{v'}, \boldsymbol{b'})$ represents the complete set of model fields, $K$ is a $4 \times 4$ matrix defining the quadratic form derived from \eqref{eq:MHD_action_without_shift}, and $S_{\mathrm{int}}$ denotes the interaction terms.

In the employed functional formalism, all correlation and response functions are recovered by variational differentiation of the generating functional $\mathcal{G}(A)$ of the dressed Green functions with respect to the set of sources $A \coloneqq (\boldsymbol{A}^v, \boldsymbol{A}^b, \boldsymbol{A}^{v'}, \boldsymbol{A}^{b'})$.
The expression for $\mathcal{G}(A)$ is given by integration over the functional space associated with the ``probability distribution'' $\mathrm{e}^{\mathcal{S}}$, while for the functional integral itself we adopt an algebraic definition \cite{FaddeevSlavnov_Gauge} that relates it to the perturbative expansion in powers of the interaction $S_{\mathrm{int}}$:
\begin{align}
&\mathcal{G}(A) = \mathrm{e}^{\frac{1}{2}\frac{\delta}{\delta \Phi}\Delta \frac{\delta}{\delta \Phi}} \left(\mathrm{e}^{S_{\mathrm{int}} + A \Phi}\right)\Big|_{\Phi = 0} = \int \EuScript{D}\Phi \,\mathrm{e}^{\mathcal{S}
+ A\Phi}, 
\label{eq:generation_func_of_Green_funcs0}
\\
&\frac{\delta}{\delta \Phi}\Delta \frac{\delta}{\delta \Phi} \coloneqq \int \mbox{d}x \, \mbox{d}x' \frac{\delta}{\delta \Phi_l(x)} \Delta^{l s}(x -  x')\frac{\delta}{\delta \Phi_s(x')}.
\label{eq:generation_func_of_Green_funcs}
\end{align}
The shorthand in the second equality in \eqref{eq:generation_func_of_Green_funcs} is an example of the condensed notation employed throughout.
The matrix $\Delta \coloneqq K^{-1}$ denotes the propagator matrix that determines the diagrammatic lines, $x \coloneqq (t, \boldsymbol{x})$, $\mbox{d} x : = \mbox{d} t\, \mbox{d}^d x $ is the integration measure, and the indices $l,s$ run over the elements of the vector $\Phi$, implying a scalar product over the corresponding discrete degrees of freedom.
The first equality in \eqref{eq:generation_func_of_Green_funcs0} is the compact part of Wick's theorem \cite{Vasilev_Func}, which underlies the evaluation of non-Gaussian functional integrals (the second equality).
The symbol $\EuScript{D}\Phi$ schematically denotes functional integration over the appropriate space of field configurations.

The functional $\mathcal{G}(A)$, is amenable to the standard Feynman diagram technique for correlation and response functions \cite{Vasilev_RG}.
The correlation functions can be obtained by taking variational differentiations with respect to sources of only ``physical'' fields $\boldsymbol{A}^v$ and $\boldsymbol{A}^b$, and to obtain the response functions, one needs additionally to differentiate with respect to sources of auxiliary fields $\boldsymbol{A}^{v'}$ and $\boldsymbol{A}^{b'}$.
Note that in the formalism used, all correlation functions of only the auxiliary fields are equal to zero (this gives a one-to-one correspondence with the well-known Wyld diagram technique \cite{Wyld1961}).

Along with $\mathcal{G}(A)$, a generating functional of connected Green functions $\mathcal{W}(A)$ is also introduced:
\begin{align}\label{eq:nonrenormalized_generating_functional_for_connected_GF}
    \mathcal{W}(A) \coloneqq \ln{\mathcal{G}(A)}.
\end{align}
Then, the translational invariance also allows one to pass from \eqref{eq:nonrenormalized_generating_functional_for_connected_GF} to the generating functional of one-particle irreducible (1PI) Green functions using the first functional Legendre transformation: 
\begin{align}\label{eq:nonrenormalized_generating_functional_for_1PI_GF}
    \Gamma(\alpha) = \sup \{ \mathcal{W}(A) - \alpha A\}, 
\end{align}
where elements of $\alpha \coloneqq (\boldsymbol{\alpha}^v, \boldsymbol{\alpha}^b, \boldsymbol{\alpha}^{v'}, \boldsymbol{\alpha}^{b'})$ are called mean-fields and the supremum is taken over all configurations of sources $A$ from the functional space.
The various Green functions (dressed, connected, 1PI) are obtained by taking variational derivatives of the corresponding generating functional.

The second equality in \eqref{eq:generation_func_of_Green_funcs0} defines the functional integral as a compact representation of the perturbation theory given by the first equality.
However, the functional integral itself must be a much ``wider'' object.
In this regard, we would like to make a supplementary remark concerning the spontaneous symmetry breaking in functionals of the type $\mathcal{G}(A)$ in Navier-Stokes or MHD turbulence theory.
Because of the $\delta$-correlated pumping, the action \eqref{eq:MHD_action_without_shift} is quasi-invariant\footnote{
That is, it is invariant up to a linear addition $\mathcal{S}_0 \rightarrow \mathcal{S}_0 -(\partial_t \boldsymbol{v'}) \cdot \boldsymbol{w}$.
This invariance and its influence, e.g., on renormalizability was discussed in \cite{Hnatic2019}.
} under the so-called generalized Galilean transformations $\{ \boldsymbol{v}, \boldsymbol{v'} \} \rightarrow \{ \boldsymbol{v}_{w}, \boldsymbol{v'}_{w} \}$, when $x \mapsto x_{w}$:
\begin{align} \label{eq:generalized_Galilei_transform}
\begin{split}
&\boldsymbol{v}_{w}(x) = \boldsymbol{v}(x_w) -  \boldsymbol{w}(t), \qquad \boldsymbol{v}_{w}'(x) = \boldsymbol{v'}(x_w), \\ 
&\boldsymbol{b}_{w}(x) = \boldsymbol{b}(x_w), \qquad \qquad \quad~ \boldsymbol{b}_{w}'(x) = \boldsymbol{b'}(x_w), \\
&x_{w} \coloneqq (t, \boldsymbol{x} + \boldsymbol{s}(t)), \,\,\, \boldsymbol{s}(t)  = 
  \int^{\infty}_{-\infty} \mbox{d} t'\, \theta(t - t') \boldsymbol{w}(t').
\end{split}
\end{align}
and strictly invariant (i.e., $\mathcal{S}_0[\Phi] = \mathcal{S}_0[\Phi_w]$) when $\boldsymbol{w}(t) = \mathrm{const.}$,  corresponding to standard Galilean transformations.
However, as pointed out in \cite{Polyakov1995}, the symmetry in the model is spontaneously broken in the subspace of the velocity field.
Such spontaneous symmetry breaking is caused by boundary conditions in the integral \eqref{eq:generation_func_of_Green_funcs}, specified by  \eqref{eq:MHD_boundary_cond} -- \eqref{eq:MHD_initial_cond}.
Although the functional \eqref{eq:MHD_action_without_shift} is invariant under the transformation \eqref{eq:generalized_Galilei_transform} with $\boldsymbol{w}(t) = \mathrm{const}$, the integration space is not.
If one takes $\boldsymbol{v}(t = -\infty) = \boldsymbol{v}_0$, a different ground state of the system arises, which can be transformed into the original one by the symmetry transformation \eqref{eq:generalized_Galilei_transform}.
In essence, the formulation of the problem \eqref{eq:MHD_NS} -- \eqref{eq:MHD_transvers_cond} in terms of the pulsating components of the respective fields specifies a particular reference frame co-moving with the flow at the relevant laminar velocity.\footnote{
It is also interesting to note that transformations \eqref{eq:generalized_Galilei_transform} can be thought of as the weaker analogs of gauge symmetry.
Consequently, one can explicitly formulate the model \eqref{eq:MHD_action_without_shift} in an arbitrary reference frame by adding a gauge-fixing term, $\mathcal{S}_0 \rightarrow \mathcal{S}_0 + \int \mbox{d} t \, \boldsymbol{u}^2(t)/2\xi$, with some gauge-fixing parameter $\xi$ and $\boldsymbol{u}(t) \propto \int \mbox{d}^d x \, \boldsymbol{v}(t, \boldsymbol{x})$, similar to the approach in gauge theories.
This term fixes only the zero mode of the velocity field but does not affect modes with higher wave numbers.
Additionally, the gauge-fixed theory still possesses a fundamental 
Becchi-Rouet-Stora-Tyutin (BRST) symmetry, although it originates differently from the standard stochastic quantization \cite{Zinn_QFT}: rather than using the MSR determinant, one employs the Faddeev-Popov determinant related to the ``gauge'' \eqref{eq:generalized_Galilei_transform}.
These aspects are commonly addressed in the context of Navier-Stokes \cite{Ivashkevich1997, Berera2007, Berera2009}, but it is evident that they carry over to MHD without any modification.
}
It turns out that in MHD, spontaneous symmetry breaking additionally occurs in the subspace of the magnetic field, only now, rotational symmetry $\mathbf{SO}(3)$ is broken.
Note that the average magnetic field arising in this case cannot be 
eliminated using appropriate Galilean transformation.
We will return to this in Sec. \ref{sec:instability}.

\begin{figure}[t]
    \includegraphics[width=1\linewidth]{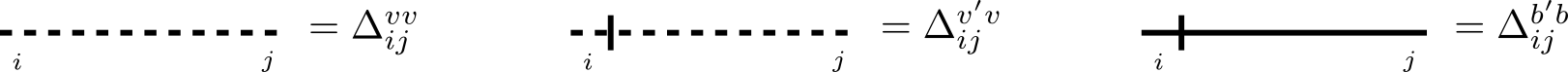}
    \caption{Graphical representation of all independent propagators in \eqref{eq:MHD_action_without_shift}.}
    \label{fig:Unstable_MHD_propagators}
\end{figure}

The non-zero independent elements of the propagator matrix $\Delta$ are shown in Fig. \ref{fig:Unstable_MHD_propagators}. 
In the frequency-momentum $(\omega, \boldsymbol{k})$ representation, these propagators are given by: 
\begin{align}\label{eq:MHD_prop_without_shift}
\Delta^{v' v}_{i j} = \frac{\mathbb{P}_{i j}(\boldsymbol{k})}{\alpha(\omega, k)},\, \Delta^{b' 
 b}_{i j} =
\frac{\mathbb{P}_{i j}(\boldsymbol{k})}{\beta(\omega, k)},\, \Delta^{v v}_{i j} = \frac{\mathbb{D}_{i j}(\boldsymbol{k}) }{\alpha^2(\omega, k)},
\end{align}
where we have introduced the following abbreviations
\begin{align}\label{eq:auxilary_alpha_and_beta}
    \alpha(\omega, k) \coloneqq i \omega + \nu_0 k^2, \quad  \beta(\omega, k) \coloneqq i \omega + u_0\nu_0 k^2.
\end{align}
The remaining elements of $\Delta$ are determined by the relation $\Delta_{i j}^{a b}(\omega, \boldsymbol{k}) = \Delta_{j i}^{b a}(-\omega, -\boldsymbol{k})$ with $a, b\in \Phi$, which reflects the general symmetry properties of $\Delta$.

\begin{figure}[t]
    \includegraphics[width=1\linewidth]{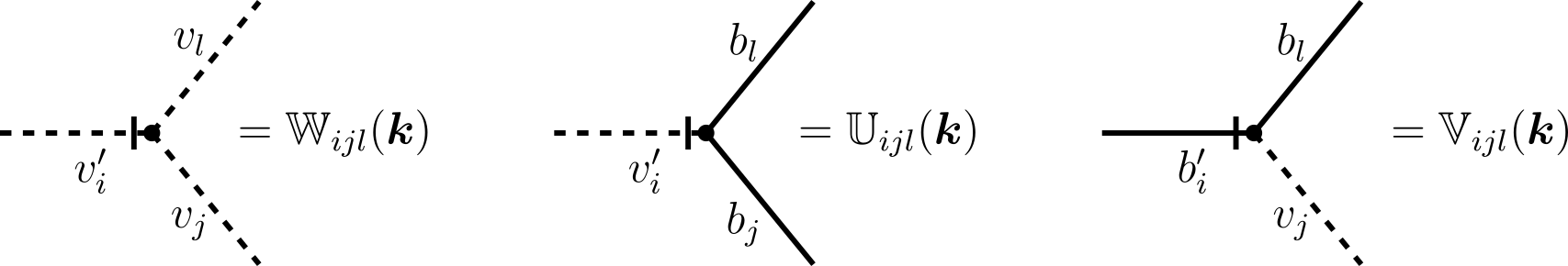}
    \caption{Graphical representation of the structure of all interaction vertices in \eqref{eq:MHD_action_without_shift}.}
    \label{fig:MHD_vertices}
\end{figure}

Exploiting the transversal nature of all fields, the interaction part $\mathcal{S}_{\mathrm{int}}$ in Eq. \eqref{eq:canonical_action_without_shift} can be expressed as:
\begin{align}\label{eq:interaction_part_without_shift}
  \mathcal{S}_{\mathrm{int}} = \frac{1}{2}v'_i \mathbb{U}_{i j l} b_j b_l + \frac{1}{2}v'_i \mathbb{W}_{i j l} v_j v_l + b'_i \mathbb{V}_{i j l} v_j b_l, 
\end{align}
where the vertex structures are defined in the momentum representation by:
\begin{eqnarray}\label{eq:interaction_vertices}
&\mathbb{W}_{i j l}(\boldsymbol{k}) = i(k_j \delta_{i l} + k_l \delta_{i j}), \quad \mathbb{U}_{i j l}(\boldsymbol{k}) = -\mathbb{W}_{i j l}(\boldsymbol{k}), \notag \\
&\mathbb{V}_{i j l}(\boldsymbol{k}) = i(k_j \delta_{i l} - k_l \delta_{i j}).
\end{eqnarray}
The interaction vertex $\mathbb{W}$ is associated with the usual Navier-Stokes equation, $\mathbb{U}$ is responsible for the renormalization of the Lorentz force, and $\mathbb{V}$ is typically referred to as the Ohm’s law vertex.
For convenience, the vertex structures are graphically depicted in Fig. \ref{fig:MHD_vertices}.
The momentum $\boldsymbol{k}$ in \eqref{eq:interaction_vertices} always flows into the vertex through the corresponding auxiliary field: $\boldsymbol{b'}$ for $\mathbb{U}$ and $\boldsymbol{v'}$ for $\mathbb{W}$ and $\mathbb{V}$, respectively.
The factors of $1/2$ in the first two terms of Eq. \eqref{eq:interaction_part_without_shift} account for the symmetry of the vertices $\mathbb{W}$ and $\mathbb{U}$ under the permutation of the corresponding fields.

\subsection{\label{sec:renormalization_without_shift}Ultraviolet renormalization: curl-terms}

By renormalization we have specifically in mind the apparatus of removing ultraviolet divergences of Green functions, originally developed in quantum field theory \cite{Vasilev_RG, Zinn_QFT, Zinn_RG, Amit_FTRG, Collins_Renorm}.
Since its inception in the early 1930s, ultraviolet divergences have haunted quantum field theory.
Although renormalization theory led to remarkable success in quantum electrodynamics, it did not bring satisfaction to its creators. 
The subsequent return of interest in field theory once again raised the question of divergences.
Over time, it became evident that these divergences do not discredit the theory, but, on the contrary, play a positive role, being an effective mechanism for breaking the scale invariance of classical theory.
For further discussion, we introduce several key observations concerning how the renormalization problem is formulated in statistical field theory.
\begin{table*}
\caption{\label{tab:comparasion_of_cutoff_and_dimensional_regularization}
Comparison of calculations performed using dimensional and cutoff regularization with respect to incorporating $\Lambda$-regular terms.
We consider a simple tadpole diagram of the static (time-independent) standard three-dimensional $\varphi^4$ model with one scalar field and action: 
$S = \frac{1}{2}\varphi(-\boldsymbol{\nabla}^2 + m^2)\varphi + \lambda \varphi^4$.
}
\begin{ruledtabular}
\renewcommand{\arraystretch}{1.2}
\begin{tabular}{c|c}
$\Lambda$ cutoff & Dimension regularization\footnote{
The $\Lambda$-regular terms disappear when the answer is analytically continued from the region $0 < d < 2$, where the integral on the left-hand side of the second equality exists, into $d = 3 - \epsilon$, $\epsilon \ll 1$.
} \\
\hline
{\scalebox{.85}{$\displaystyle \qquad \qquad\underset{d = 3}{\raisebox{-0.7ex}{{\includegraphics[scale=1.,angle=90,origin=c]{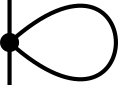}}}}$}} $ = \overline{S}_3 {\displaystyle \int \limits_{0}^{\Lambda}} \mbox{d}k \frac{k^2}{k^2 + m^2} = \frac{\Lambda}{2\pi^2} - \frac{m}{4\pi} + \mathcal{O}\left(\frac{m}{\Lambda}\right) \qquad\qquad ~$
& 
\hspace{-0.2cm}{\scalebox{.85}{$ \displaystyle \quad \underset{d = 3 - \epsilon}{\raisebox{-0.7ex}{{\includegraphics[scale=1.,angle=90,origin=c]{./tadpolephi4.png}}}}$}} $= {\displaystyle \int}\frac{\mbox{d}^d k }{(2\pi)^d} \frac{1}{\boldsymbol{k}^2 + m^2} = \frac{\pi^{d/2}}{(2\pi)^d} \Gamma(1 - d/2) m^{d - 2} = - \frac{m}{4 \pi} + \mathcal{O}(\epsilon)\qquad \qquad$
\end{tabular}
\end{ruledtabular}
\end{table*}

The first step in any renormalization process is UV regularization.
In statistical field theory, a cutoff $\Lambda$ offers a fully natural and physically motivated form of regularization, rather than one artificially imposed.
When distances are measured in units of $\Lambda$, this parameter is typically of the order of unity.
Consequently, it is not, in general, strictly required to take the limit $\Lambda \rightarrow \infty$ or confirm its existence.
As a result, one could conclude that explicit renormalization of the model might be unnecessary in principle.
Nevertheless, it emerges that the capacity to perform such renormalization encodes important implicit information about the underlying model at finite $\Lambda$.
This information is crucial for the application of the RG technique.
More precisely, renormalization becomes most useful when it is multiplicative, meaning that all UV divergences can be reabsorbed into renormalized parameters and fields that are already part of the model.

Technically, UV divergences may be removed recursively via the so-called $R'$ operation (see, e.g., \cite{Vasilev_RG}).
This operation results in the fact that after subtracting all the divergences of the subdiagrams, multi-loop diagrams leave behind only the so-called superficial divergences.
In renormalizable theories the number of types of such divergences is finite.
At the same time, the process of subtracting divergent parts is not unique; it depends on the divergence subtraction scheme and the regularization method, both of which are often chosen for convenience.
In any case, all these schemes are connected by group of UV-finite renormalization transformations -- equivalent from the RG perspective -- and thus remain unaffected by the specific choice of regularization \cite{Collins_Renorm}.
In turbulence theory, multi-loop evaluations are commonly carried out in the MS scheme (which entails subtracting only the divergent parts of the diagrams) and using analytic regularization technique \cite{Speer_GenFeynmanAmplitudes}.
The last is conceptually similar to dimensional regularization as used in critical phenomena, except that in turbulence theory the parameter $\epsilon$ is in no way connected to the spatial dimension.
Instead, it is introduced independently through model-specific representations, such as the type given in the pumping function~\eqref{eq:pump_function}.
Recall that in critical phenomena, the standard small expansion parameter is the deviation of the system’s spatial dimension from the upper critical dimension.
In turbulence theory, a $2 - \epsilon$ expansion requires defining the unrenormalized Green functions of the model as functions of the small parameter $2 - \epsilon$, where $\epsilon \neq 0$.
Simultaneously, the renormalized Green functions must possess a finite limit as $\epsilon \rightarrow 0$.
Also, note that using the analytic regularization technique, we are faced with the problem of non-multiplicativity of the renormalization procedure in some exceptional space dimensions \cite{Vasilev_RG}.
In the case of helical MHD this exceptional spatial dimension is $d = 2$.

A sufficient condition for a diagram to exhibit superficial UV divergences is the non-negativity of its divergence index $w \leq d_I$, where $d_I$ is the total canonical dimension of the diagram $I$.
Index $d_I$ reflects the diagram's invariance under a common scaling transformation of spatial coordinates and time. 
In our case, this transformation is $t \propto x^2$, with all dimensional parameters of the model (including fields) transform accordingly under this scaling.
In turn, one evaluates the divergence index $w$ directly in the logarithmic theory, in which the coupling constant is dimensionless and UV divergences are maximally manifested.
In our case, the theory becomes logarithmic at $\epsilon = 0$ (noting that $g_0 \sim \Lambda^{2\epsilon}$).
If the model’s interaction vertices contain no derivatives, the sufficient condition mentioned above becomes necessary.
Otherwise, the divergence index is lowered by the number of derivatives at each diagram’s vertex.
Such derivatives yield an overall dimensional factor but do not directly influence whether the diagram converges or diverges.

For superficially divergent integrals (multi-loop diagrams after subtracting subdivergences), methods such as, for instance, analytic regularization with a finite cutoff $\Lambda$, ensure, under general assumptions on the integrand and barring IR divergences, that these integrals admit analytic continuation from their strip of convergence to a function analytic throughout the complex $\epsilon$-plane.
Furthermore, upon sending $\Lambda \rightarrow \infty$, the result can be analytically continued to a meromorphic function defined over the entire complex $\epsilon$-plane, with poles appearing at certain isolated points on the real axis.
Such a function, independent of $\Lambda$ and well-defined for all $\epsilon$ values away from its poles, is typically called the formal answer for the integral (diagram).

Regarding UV divergences, the inclusion to integrals of a small parameter $\epsilon$ guarantees convergence for diagrams that for $\epsilon = 0$ originally displayed logarithmic divergences as $\Lambda \rightarrow \infty$.
Nevertheless, such a minor adjustment does not fully remove all divergences.
Diagrams featuring linear or stronger divergences in $\Lambda$ are still divergent.
These leftover power-law divergences in $\Lambda$ correspond to the so-called $\Lambda$-regular contributions from the integrals linked to those diagrams.
These terms appear as polynomials in the IR-relevant parameters upon which the diagram depends.
Consequently, the formal answer reflects the integral with the subtraction of these $\Lambda$-regular terms.
This rest is calculated without a cutoff and embodies the dominant singular contribution to the integral’s exact value.
A simple example illustrating this fact is provided in Table \ref{tab:comparasion_of_cutoff_and_dimensional_regularization}.

Let us now examine the significance of such $\Lambda$-regular contributions in the context of the renormalization theory.
Firstly, these contributions always possess positive dimensions, leaving the dimensionless field-renormalization constant unaffected.
Instead, they alter only the renormalization of IR-relevant parameters $e_0$ by inducing shifts in the initial parameters: $e_0 \rightarrow e'_0 = e_0 + \Delta e_0$, where $\Delta e_0$ includes polynomially in $\Lambda$ divergent terms. 
Such shifts are inconsequential when determining the model’s critical asymptotics \cite{Vasilev_RG}.
Consequently, one generally carries out the so-called $\Lambda$-renormalization at this juncture: the ``shifted'' IR-relevant parameters $e'_0$ are taken as independent variables, and all $\Lambda$-regular terms in the diagrams are discarded.
As a result, Green functions, reparameterized by these ``shifted'' variables $e'_0$, no longer exhibit $\Lambda$-divergences.
Rather, any residual divergences are confined to the relationships between the ``shifted'' and ``unshifted'' parameters, namely $e_0 = e_0(e'_0, \Lambda)$.
At this stage, if $\Lambda$ is finite, Green functions display no divergences in $\epsilon$. 
Nonetheless, ending the process at this step would overlook a crucial aspect of the model -- its multiplicative renormalizability.
To remedy this, one next takes $\Lambda \rightarrow \infty$ for all $\Lambda$-convergent integrals (i.e., those integrals from which any $\Lambda$-regular terms have been removed).
In this procedure, the original dimensional parameters are replaced by the ``shifted'' ones, $e_0 \rightarrow e_0'$, thereby incorporating all regular contributions and the formal answers for the unrenormalized Green functions are then employed.
Within these formal answers, poles in $\epsilon$ arise, requiring the subsequent step: a nontrivial multiplicative $\epsilon$-renormalization of both the fields and parameters.
This stage constitutes the core purpose of RG formalism, facilitating the determination of universal characteristics of critical behavior.
Such universal quantities are typically the most important results derived from any statistical field model and often represent its primary objective.

However, $\Lambda$-renormalization remains straightforward -- entailing a simple substitution $e_0 \rightarrow e_0'$ -- only if the model includes bare parameters capable of consistently ``absorbing'' all $\Lambda$-regular contributions throughout the renormalization process.
This claim is a central point of the current work.
The challenge arises because helical magnetohydrodynamics is among the few models in which a straightforward implementation of this formal scheme fails.
The underlying reason is that some diagrams exhibit power-law dependence on $\Lambda$ that cannot be ``absorbed'' by renormalizing any bare parameters in \eqref{eq:MHD_action_without_shift} simply due to the lack of structure-appropriate parameters.
It is worth emphasizing that this issue is only specific to helical magnetohydrodynamics and does not emerge in non-helical ($\mathbf{O}(3)$-symmetric) MHD.
As we will demonstrate in this paper, while this difficulty does not lead to a fundamental inconsistency, it permits two distinct interpretations of the model's behavior.

In this Section, we focus on the correct two-loop $\Lambda$-renormalization of the model that properly accounts for all $\Lambda$-regular terms.
All necessary results concerning the renormalization of all rest ``standard'' divergences corresponding to $(\ln \Lambda)^p$ (or $1/\epsilon^p$ it the scheme without $\Lambda$) are presented in the forthcoming Section.

For $d > 2$, the renormalizability of \eqref{eq:MHD_action_without_shift} was established in \cite{Adzhemyan1985, Adzhemyan1987}.
The special case $d = 2$ was addressed in \cite{Honkonen1995}; see also \cite{Adzhemyan_RGinFullDevTurb, Vasilev_RG}.
For $d > 2$, the superficial divergences reside exclusively in 1PI functions $\Gamma^{v'v}$, $\Gamma^{b'b}$, and $\Gamma^{v'bb}$.
These divergences are identified by enforcing the dimensionlessness (momentum and frequency separately) of each term in the action \eqref{eq:MHD_action_without_shift}.
The momentum $d^{p}_Q$, frequency $d^{\omega}_Q$ and total $d_Q = d^{p}_Q + 2 d^{\omega}_Q$ canonical dimensions calculated from this requirement for each value of $Q$ in \eqref{eq:MHD_action_without_shift} are shown in table \ref{tab:MHD_canonical_dimensions}.
\begin{table}[t]
\renewcommand{\arraystretch}{1.4}
\caption{\label{tab:MHD_canonical_dimensions}
Canonical (engineering) dimensions of the bare fields and parameters in the action \eqref{eq:MHD_action_without_shift}.}
\begin{ruledtabular}
\begin{tabular}{ccccccc}
Quantity & $\boldsymbol{v}, \boldsymbol{b}$ & $\boldsymbol{v'}, \boldsymbol{b'}$ & $\nu_0$ & $u_0$,  $\rho$ & $g_0$ &  $m, \Lambda$\\
\hline
$d^{p}_Q$  & $-1$ & $d + 1$ & $-2$ & $0$ & $2\epsilon$ & $1$ \\
$d^{\omega}_Q$ & $1$ & $-1$ & $1$ & $0$ & $0$ & $0$ \\ 
$d_Q$ & $1$ & $d - 1$ & $0$ & $0$ & $2\epsilon$ & $1$ \\ 
\end{tabular}
\end{ruledtabular}
\end{table}
Observe that the fields share the same canonical dimensions, which is unusual in simpler models.
Hence, to completely renormalize the model, one need only renormalize divergences in these three vertex functions: $\Gamma^{v'v}$, $\Gamma^{b'b}$, and $\Gamma^{v'bb}$.
The complication is that, in helical magnetohydrodynamics \eqref{eq:MHD_action_without_shift}, these functions can include pseudotensor counterterms that are linear in both $\Lambda$ and the external momentum $\boldsymbol{k}$, with the form $\propto i \rho [\varepsilon_{i j l} k_l] \nu_0\Lambda$ and a real coefficient.
In \cite{Adzhemyan1987}, these terms were dubbed ``rotor'' or ``curl-terms'' because, in coordinate space, they amount to adding a term $\propto \rho[\boldsymbol{\nabla} \times \boldsymbol{b}]\nu_0\Lambda$ into the right hand side of Eq. \eqref{eq:MHD_induction}.

The existence of curl-terms was seemingly first identified in \cite{Pouquet1978}.
This study demonstrated that for $\Gamma^{v v'}$ (i.e., in ordinary helical hydrodynamics), such terms are not generated at least at the one-loop level.
Moreover, in the present paper, we explicitly show that the claim regarding the absence of counterterms of the form $\propto i \rho [\varepsilon_{i j l} k_l] \nu_0\Lambda$ holds to all orders in perturbation theory for every 1PI function whose external momentum enters a vertex that is symmetric under the interchange of fields (see Sec. \ref{sec:absense_of_curl-terms_in_diagram_with_symmetric_vertex}).
Regarding vertex $\Gamma^{b'b}$, then curl-terms can and indeed do emerge there, starting with the simplest one-loop diagram:\vspace{-0.2cm}
\begin{eqnarray}\label{eq:one_loop_curl-term_diagram}
    \Gamma_{i j}^{b'b} = -\beta(\omega, k) \mathbb{P}_{i j}(\boldsymbol{k}) + \raisebox{0ex}{\includegraphics[width=3.0cm]{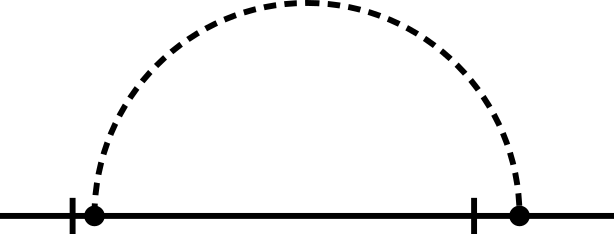}} \,.
\end{eqnarray}
Here, the function $\beta(\omega, k)$ was defined previously in \eqref{eq:auxilary_alpha_and_beta}.
The expansion of the diagram in Eq. \eqref{eq:one_loop_curl-term_diagram} at $m = 0$ and zero external frequency in powers of the external momentum $\boldsymbol{k}$ yields a linear term:\vspace{-0.1cm}
\begin{align}
&\Sigma_{1\,ij}^{[\Lambda]} = \rho \nu_0 \Lambda \big[i \varepsilon_{i j l} k_l\big] \big(g_0 \Lambda^{- 2 \epsilon}\big) c_{1}^{[\Lambda]}(u_0, \epsilon), \label{eq:one_loop_bb'_linear-asymp} \\
&c_{1}^{[\Lambda]}(u_0, \epsilon) = \frac{1}{(1 - 2\epsilon)}\left(\frac{-1}{6\pi^2 (1 + u_0)}\right). \label{eq:one_loop_curl-term}
\end{align}
Hereinafter, the subscript $n$ in quantities like $\Sigma_{n \,ij}^{[\Lambda]}$ indicates the linear-in-$\boldsymbol{k}$ $n$-loop contribution.\footnote{
In Eq. \eqref{eq:one_loop_curl-term}, we have corrected a misprint in the overall sign of the coefficient $c_{1}$ obtained in \cite{Adzhemyan1987}. 
This misprint arose due to the transition from Eq. (13) to Eq. (14), including an erroneous sign change at the antisymmetric vertex $\overline{v}_{135}^p \rightarrow \overline{v}_{153}^p $.
}

Thus, we see that adding only counterterms that coincide in structure with the terms already present in the action \eqref{eq:MHD_action_without_shift} is not enough for correct renormalization of the model.
It contains a dynamically generated new type of divergence that must be eliminated.
In this regard, the discussed model is especially interesting because this elimination can be done without losing the multiplicative renormalizability (see below).

The general theory of renormalization requires incorporating the curl structure into the action \eqref{eq:MHD_action_without_shift} on an equal footing with all the composite operators already present.
Acting formally, let us consider the action \eqref{eq:MHD_action_without_shift} with an added curl term 
\begin{align}\label{eq:MHD_action_with_seed_curl-terms}
    \mathcal{S}_0 + \rho \nu_0 \mathfrak{h}_{0} \boldsymbol{b'} \cdot \left(\boldsymbol{\nabla} \times \boldsymbol{b}\right)
\end{align}
with some bare coefficient $\mathfrak{h}_{0}$ with the dimension of $\Lambda$. 
Now the curl terms arising from the diagrams can be ``absorbed'' into the renormalization of $\mathfrak{h}_{0}$.
By inverting the quadratic-form matrix corresponding to the action \eqref{eq:MHD_action_with_seed_curl-terms} in the Fourier space of divergence-free functions, one finds that propagators $\Delta^{vv}$ and $\Delta^{v'v}$ remain unchanged, while $\Delta^{b'b}$ is replaced by
\begin{align}\label{eq:bb'_prop_with_curl-terms}
     \underline{\Delta}_{i j}^{b'b} = \frac{\beta(\omega, k) \mathbb{P}_{i j}(\boldsymbol{k}) - \rho \nu_0 \mathfrak{h}_{0} k \mathbb{H}_{i j}(\boldsymbol{k})}{\beta^2(\omega, k) - \rho^2 \nu_0^2 \mathfrak{h}_{0}^2 k^2}.
\end{align} 
Unfortunately, this bare propagator exhibits a pole in $\omega$, leading to an instability analogous to tachyonic behavior in QFT. 
On the other hand, the original model presented in Eq. \eqref{eq:MHD_action_without_shift} is also unstable -- it lacks curl terms in the bare propagator, yet such terms emerge perturbatively in the self-energy.
For a more detailed discussion of this instability, refer to Sec. \ref{sec:instability}.
Hence, the only way to renormalize and stabilize either the original model \eqref{eq:MHD_action_without_shift} or the model \eqref{eq:MHD_action_with_seed_curl-terms} is to \textit{eliminate the curl terms entirely}.
\begin{widetext}
The crucial step in derivation of \eqref{eq:bb'_prop_with_curl-terms} is the following calculation:
\begin{align}\label{eq:useful_relation_for_deriving_curl_b'b_propagator}
    \frac{\mathbb{P}_{i s}(\boldsymbol{k})}{\beta(\omega, k)} \left[\delta_{s j} + \frac{\rho \nu_0 \mathfrak{h}_{0} k}{\beta(\omega, k)} \mathbb{H}_{s j}(\boldsymbol{k})\right]^{-1} = \frac{\mathbb{P}_{i s}(\boldsymbol{k})}{\beta^2(\omega, k) - (\rho \nu_0 \mathfrak{h}_{0} k)^2}\bigg[\beta(\omega, k)\delta_{s j} - \rho \nu_0 \mathfrak{h}_{0} k \mathbb{H}_{s j}(\boldsymbol{k})\bigg],
\end{align}
where the property $\mathbb{H}_{i s}(\boldsymbol{k})\mathbb{H}_{s j}(\boldsymbol{k}) = \mathbb{P}_{i j}(\boldsymbol{k})$ and the identity $k\mathbb{H}_{s j}(\boldsymbol{k}) \equiv \big[ i\varepsilon_{s j l}k_l\big]$ were used. \vspace{-0.1cm}
\end{widetext}

There are two possible ways to eliminate the curl terms, each leading to distinct physical behavior of the system.
The first way is to assume that the correct model for describing helical MHD is \eqref{eq:MHD_action_with_seed_curl-terms} rather than \eqref{eq:MHD_action_without_shift}.
From a physical standpoint, this means that the induction equation \eqref{eq:MHD_induction} must initially include the term $\rho \nu_0 \mathfrak{h}_0 (\boldsymbol{\nabla} \times \boldsymbol{b})$ with $\mathfrak{h}_0 \propto g_0$, which acts as a stabilizing factor.
Then, at the level of new diagrammatic technique with propagators $\Delta^{vv}$, $\Delta^{v'v}$ and $\underline{\Delta}^{b'b}$ one can recursively select $\mathfrak{h}_{0}$ to cancel out the curl terms also arising from perturbative expansion of $\Sigma^{b'b}$ in this new theory.
Further, we provide a description of the explicit construction for such $\mathfrak{h}_{0}$.
In this approach, the situation with renormalization is exactly the same as in fluctuation models of the theory of critical behavior.
In practice, this procedure is purely mnemonic: one simply subtracts the curl terms from all diagrams of $\Sigma^{b'b}$ (where they only arise) or, more directly, adopts formal answers for integrals in which such terms never appear.

However, the approach above is only valid if we assume that \eqref{eq:MHD_action_with_seed_curl-terms} is the real physical model. 
Strictly speaking, there is no reason to consider Eq. \eqref{eq:MHD_induction} incorrect; moreover, the existence of an instability in which the fluctuations destroy the initial state $\langle \boldsymbol{b} \rangle = 0$ is a well-established fact (see Sec. \ref{sec:instability}).
Besides this, forced insertion of a stabilizer $\mathfrak{h}_0$ may seem contrived from a physical standpoint, since it is not justified by anything other than the desired result.
To anticipate, let us say that this somewhat contrived procedure makes sense when one can justify the system stabilization in the same or, in a certain sense, a closely related initial state $\langle \boldsymbol{b} \rangle = 0$.
Physically, this applies when the large-scale magnetic field generated by the dynamo effect can be neglected through the decay of $\langle \boldsymbol{b} \rangle = 0$, corresponding to the so-called kinematic MHD.

However, stabilization does not necessarily have to occur in the same state as the initial one.
A very plausible and elegant hypothesis regarding a possible mechanism for such stabilization was proposed in \cite{Adzhemyan1987}.
It is precisely this hypothesis that we will consider in Sec. \ref{sec:instability} and throughout the remainder of this work.
By treating the curl terms as an inherent feature of the theory, let's outline the part of the $\Lambda$-renormalization of the model \eqref{eq:MHD_action_without_shift} that is responsible for eliminating subgraph divergences ($R'$-operation).
The remaining polynomial in $\Lambda$ divergences must then be eliminated ($R$-operation) through a different method.
The idea is that in $\Gamma^{b'b}$ corresponding to the action \eqref{eq:MHD_action_without_shift}, curl terms are already present, not in the propagator, but in the self-energy $\Sigma^{b'b}$.
Therefore, we can extract them from $\Sigma^{b'b}$ (order by order) and incorporate them into the propagator definition representing the Dyson equation for $\Gamma^{b'b}$ as follows \vspace{-0.1cm}
\begin{align}\label{eq:dyson_eq_for_Gammab'b_h0_to_propagator}
    \Gamma_{i j}^{b'b} = -\left(\Delta_{i j}^{b'b}\right)^{-1} + \Sigma_{i j}^{b'b} = -\left( \underline{\Delta}_{i j}^{b'b}\right)^{-1} + \cancel{\Sigma}_{ij}^{b'b},
\end{align}
where $\cancel{\Sigma}_{ij}^{b'b}$ is the self-energy $\Sigma_{ij}^{b'b}$ in the model \eqref{eq:MHD_action_without_shift}, but \textit{without} curl terms.
Thereby we arrive precisely at the the following (unstable) unrenormalized model
\begin{align}\label{eq:MHD_action_with_Gammabb'_curl-terms}
    \mathcal{S}_0 - \rho \nu_0 h_0 \boldsymbol{b'} \cdot \left(\boldsymbol{\nabla} \times \boldsymbol{b}\right),
\end{align}
which corresponds to the action \eqref{eq:MHD_action_with_seed_curl-terms} with formal replacement $\mathfrak{h}_0 \to -h_0$\footnote{
Here we insert into the action exactly what we get from the $\Sigma^{b'b}$ diagrams.
Minus sign for term with $h_0$ comes from the fact that $\mathbb{H}_{ij}(\boldsymbol{k})$ in \eqref{eq:one_loop_bb'_linear-asymp} the convolution goes with the third index of the Levi-Civita symbol. Meanwhile, in $\boldsymbol{b'} \cdot \left(\boldsymbol{\nabla} \times \boldsymbol{b}\right)$ the convolution goes with the second index: $\boldsymbol{b'} \cdot \left(\boldsymbol{\nabla} \times \boldsymbol{b}\right) = b'_i\varepsilon_{i m j}\partial_m b_j$. 
} and in which it is implied that all curl terms are stored in $h_0$ and are absent in diagrams of $\Sigma^{b'b}$. 

\begin{figure*}
    \centering
    \begin{overpic}[percent,grid=false,tics=2,width=0.8\linewidth]{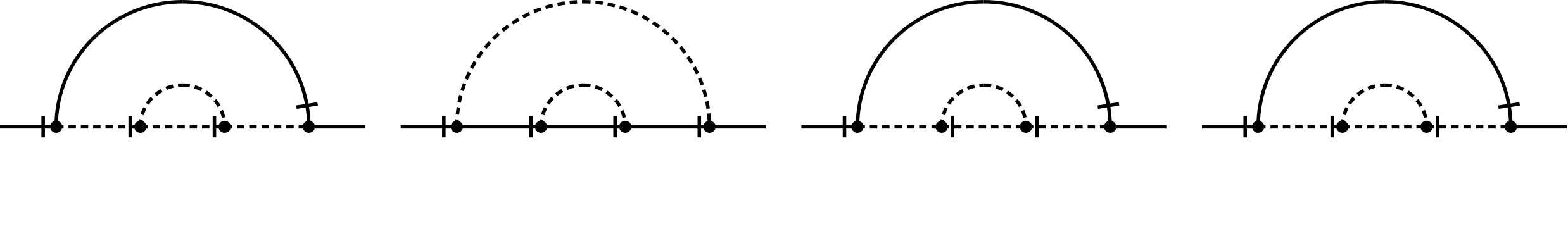}
    \put(0,13){$D_{\mathrm{L}1}^{(2)}$}
    \put(25.5,13){$D_{\mathrm{L}2}^{(2)}$}
    \put(51,13){$D_{\mathrm{L}3}^{(2)}$}
    \put(76.5,13){$D_{\mathrm{L}4}^{(2)}$}
    \end{overpic}
    \begin{overpic}[percent,grid=false,tics=2,width=0.8\linewidth]{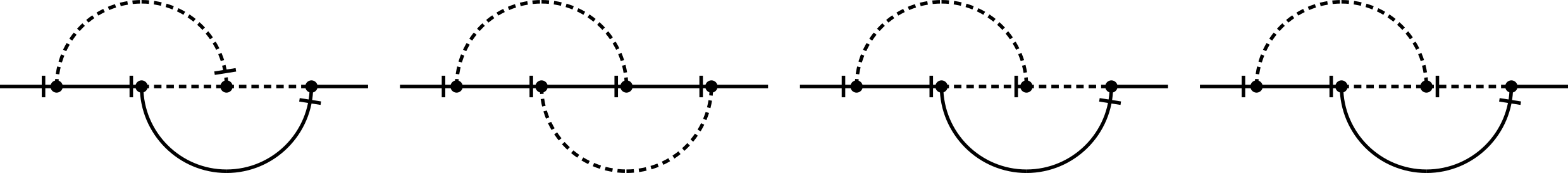}
    \put(0,10){$D_{\mathrm{L}5}^{(2)}$}
    \put(25.5,10){$D_{\mathrm{L}6}^{(2)}$}
    \put(51,10){$D_{\mathrm{L}7}^{(2)}$}
    \put(76.5,10){$D_{\mathrm{L}8}^{(2)}$}
    \end{overpic}
    \caption{Two-loop $\Gamma^{b'b}$  diagrams for the model \eqref{eq:MHD_action_without_shift}.
    The upper four sunrise diagrams are commonly called sunsets or London transport diagrams.
    The lower ones are sometimes called double‐scoop diagrams in the literature.
    The symmetry coefficients of all these diagrams are equal to $1$, except for the sunset $D_{\mathrm{L}4}^{(2)}$, which is equal to $1/2$.}
    \label{fig:two-loop_surface_divergent_diagrams_before_shift}
\end{figure*}
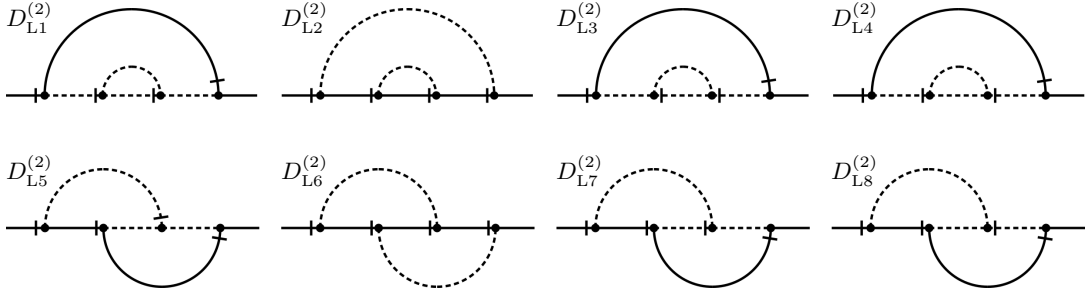

In this manner, we generate a curl operator in the action, which includes a perturbatively generated (infinite) charge $h_0 \propto \Lambda$ to which a counterterm $\delta h$ must now be added.
If one chooses $\delta h$ in such a way as to cancel not only the subgraph divergences arising from the two-loop diagrams $\Sigma^{b'b}$, but also the $\Lambda$-divergences from the additional correction diagrams $\propto h_0$ resulting from the change of the line $\Delta^{b'b} \to \underline{\Delta}^{b'b}$, then the model \eqref{eq:MHD_action_with_Gammabb'_curl-terms} with the implicit subtraction of the curl terms from $\Sigma^{b'b}$ in the sense of Eq. \eqref{eq:dyson_eq_for_Gammab'b_h0_to_propagator} will partially renormalize \eqref{eq:MHD_action_without_shift}.
The general structure of coefficient $h_0$ is as follows:\vspace{-0.1cm}
\begin{align}
\label{eq:h0_structure}
    h_0 = \Lambda \sum \limits_{n = 1}^{\infty} \left(g_0 \Lambda^{-2\epsilon}\right)^n c_n^{[\Lambda]}(u_0, \epsilon)
\end{align}
where the coefficient $c_1^{[\Lambda]}$ is found in Eq.~\eqref{eq:one_loop_curl-term}.
To proceed further and get $c_2^{[\Lambda]}$, we need to take into account perturbatively the difference between the line $\underline{\Delta}^{b'b}$ and $\Delta^{b'b}$.
This perturbative expansion can be schematically represented as the following sum
\begin{align}
\label{eq:schematic_image_of_bb'_propagator_with_curl_terms}
     \underline{\Delta}_{i j}^{b'b} = \, \raisebox{-0.2ex}{\includegraphics[width=2.cm]{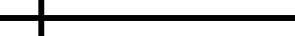}}\, + \,\raisebox{-0.1ex}{\includegraphics[width=2.cm]{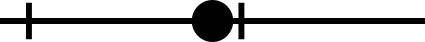}}\, + \mathcal{O}(g_0^2),
\end{align}
where the first term is the original $\Delta^{b'b}$, and the second term is given by
\begin{align}
\label{eq:one-loop_correction_for_bb'_propagator_with_curl_terms}
     \raisebox{-0.1ex}{\includegraphics[width=2.cm]{./h0correctiontobB.png}}\, = \frac{\rho \nu_0 h_{0} k \mathbb{H}_{i j}(\boldsymbol{k})}{\beta^2(\omega, k)}.
\end{align}
Note that correction line \eqref{eq:one-loop_correction_for_bb'_propagator_with_curl_terms} is precisely given by $\Delta_{i l}^{b'b}\left[\rho \nu_0 {h}_{0} k \mathbb{H}_{l s}(\boldsymbol{k})\right]\Delta_{s j}^{b'b}$, where the expression within the square brackets represents the curl contribution from the one-loop diagram in Eq. \eqref{eq:one_loop_bb'_linear-asymp}.
\begin{widetext}
Using the relation \eqref{eq:useful_relation_for_deriving_curl_b'b_propagator} for the derivation of $\left( \underline{\Delta}_{i j}^{b'b}\right)^{-1}$ with $\mathfrak{h}_0 \rightarrow -(h_0 + \delta h)$, the detailed form of Eq. \eqref{eq:dyson_eq_for_Gammab'b_h0_to_propagator} becomes:\vspace{-0.1cm}
\begin{eqnarray}\label{eq:two-loop_Gammabb'_with_curl-terms}
    \Gamma_{i j}^{b'b} = -\beta(\omega, k) \mathbb{P}_{i j}(\boldsymbol{k}) + \rho \nu_0 (h_0 + \delta h) k \mathbb{H}_{i j}(\boldsymbol{k}) + \raisebox{0ex}{\includegraphics[width=2.7cm]{./oneloopCurlTermDiag.png}} + [\text{diagrams from Fig. \ref{fig:two-loop_surface_divergent_diagrams_before_shift}}] + \raisebox{0ex}{\includegraphics[width=2.7cm]{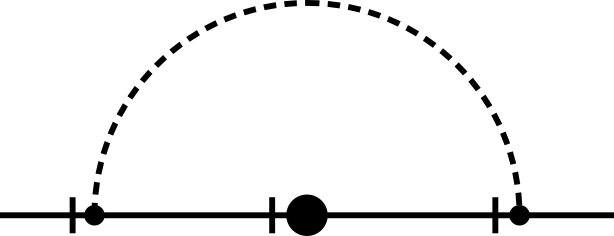}}.\quad~
\end{eqnarray}
Here, it is assumed that the expansion of all two-loop diagrams starts with $k^2$-term (all curl terms are moved to $h_0$).
\end{widetext}

The general picture of the described approach is as follows: the curl contribution \eqref{eq:one_loop_bb'_linear-asymp} from the one-loop diagram in \eqref{eq:two-loop_Gammabb'_with_curl-terms} is attributed to the coefficient $c_1^{[\Lambda]}$ in the $h_0$ term. 
Thus, the propagator $\Delta^{b'b}$ changes to $\underline{\Delta}^{b'b}$ and the action \eqref{eq:MHD_action_without_shift} transforms into \eqref{eq:MHD_action_with_Gammabb'_curl-terms} according to Eq. \eqref{eq:dyson_eq_for_Gammab'b_h0_to_propagator}.
Then, the sum of the curl contributions from the two-loop diagrams (shown in Fig. \ref{fig:two-loop_surface_divergent_diagrams_before_shift}) with extracted subgraph divergencies determines the two-loop coefficient $c_2^{[\Lambda]}$.
In turn, the divergences of the subgraphs along with the last diagram in \eqref{eq:two-loop_Gammabb'_with_curl-terms} are canceled by the counterterm $\delta h$.
The last diagram in \eqref{eq:two-loop_Gammabb'_with_curl-terms} is just a correction diagram generated by the line \eqref{eq:one-loop_correction_for_bb'_propagator_with_curl_terms}.
It exactly duplicates the subgraph divergence in the sunset $D_{\mathrm{L}2}^{(2)}$ in Fig. \ref{fig:two-loop_surface_divergent_diagrams_before_shift}, which is the only diagram containing a $\Lambda$-divergent subgraph. 
Let us emphasize that the technique described above -- adding the first term of the self-energy expansion in $\boldsymbol{k}$ (a local composite operator) to the action -- is similar to the approach used in the effective potential theory. 

Technically, the contributions corresponds to the curl terms are given by the leading (linear) term in the formal asymptotic expansion as $\boldsymbol{k} \rightarrow 0$ (physically, $\boldsymbol{k}$ is taken to lie in the inertial range) of the corresponding diagram evaluated at zero external frequency, with $\boldsymbol{k}$ being the external momentum.
It is convenient to isolate this contribution by keeping the dependence on $\boldsymbol{k}$ solely in the vertex structure \eqref{eq:interaction_vertices} through which it flows, while setting the remaining parts of the diagram to zero external momentum.
In this case, the part of the diagram's tensor structure corresponding to such asymptotics is obtained as a term $\propto \rho$ in the exact tensor structure.
Also, it is convenient to assign a loop momentum to the line containing the kernel \eqref{eq:pump_function}.
The rest two-loop calculations are fairly standard \cite{Adzhemyan2003, Adzhemyan2005}.
A loop momentum and frequency are assigned to the line containing the kernel \eqref{eq:pump_function}.
Then, after integration over frequencies, the resulting integrands $F_l$ become homogeneous functions of degree $1$: 
\begin{align}\label{eq:integrands_homogenity}
    F_l(\lambda p, \lambda q, z) = \lambda F_l(p, q, z), \qquad \lambda > 0,
\end{align}
where the angle variable $z \coloneqq (\boldsymbol{p} \cdot \boldsymbol{q})/(p q)$ is the cosine of the angle between the loop momenta $\boldsymbol{p}$ and $\boldsymbol{q}$.
\begin{widetext}
By close analogy with \eqref{eq:one_loop_bb'_linear-asymp}, the two-loop curl contribution $ \Gamma_{2\, ij}^{[\Lambda]}$ can be expressed as\vspace{-0.1cm}
\begin{eqnarray}
\label{eq:two_loop_bb'_linear-asymp}
    \Sigma_{2\,ij}^{[\Lambda]} = \rho \nu_0 \Lambda \big[i \varepsilon_{i j l} k_l\big] \left(g_0 \Lambda^{- 2 \epsilon}\right){}^2 \,c_{2}^{[\Lambda]}(u_0, \epsilon), \qquad c_{2}^{[\Lambda]}(u_0, \epsilon) \coloneqq \sum \limits_{l = 1}^8 R_{\Lambda}'[I_l], \qquad I_l = \int \limits_{-1}^{1} \mbox{d} z \int \limits_{m/\Lambda}^{1} \frac{\mbox{d} p \, \mbox{d} q}{(pq)^{1 + 2 \epsilon}} F_l(p, q, z), 
\end{eqnarray}
where $R'_{\Lambda}[\bullet]$ denotes the operation of subtracting $\Lambda$-divergent subgraphs in the corresponding integrals (see e.g. \cite{Vasilev_RG} for details) and all two-loop integrands $F_l(p, q, z)$, corresponding to the diagrams in Fig. \ref{fig:two-loop_surface_divergent_diagrams_before_shift}, are provided in the App.~\ref{app:two-loop_Gammab'b_expressions}.
In terms of $\Sigma_{2\,ij}^{[\Lambda]}$, the homogeneity \eqref{eq:integrands_homogenity} implies $\Lambda \partial_{\Lambda}\Sigma_{2\,ij}^{[\Lambda]} = (1 - 4 \epsilon)\Sigma_{2\,ij}^{[\Lambda]}$, leading to the following identity for $I_{l}$: \vspace{-0.1cm} \begin{align}\label{eq:representation_for_diagrams_comes_from_homogeneity}
    I_{l} = \frac{1}{1 - 4\epsilon} \int \limits_{-1}^{1} \mbox{d}z \int \limits_{s}^{1} \left( \frac{\mbox{d}\kappa}{\kappa^{1 + 2\epsilon}} \left(F_l(\kappa, 1, z) - s^{-2\epsilon}F_l(\kappa, s, z)\right) + \left(F_l(1, \kappa, z) - s^{-2\epsilon}F_l(s, \kappa, z)\right) \right), \qquad s \coloneqq m/\Lambda.
\end{align}
Note that the coefficients $c_{2}^{[\Lambda]}$ as well as $c_{1}^{[\Lambda]}$ from \eqref{eq:h0_structure} can also depend on the helicity parameter $\rho$ via $u_0$.
\end{widetext}

Note that in our case, we have an ``overabundance'' of regularizations, namely three parameters $(\Lambda, m, \epsilon)$.
Strictly speaking, proper $\Lambda$-renormalization requires only two parameters, $\Lambda$ and $m$.
However, we are working within the $\epsilon$-expansion anyhow, so the explicit dependency on $\epsilon$ is also useful to us.
The UV cutoff $\Lambda$ is required for the ``existence'' of the curl terms\footnote{
In calculations in the spirit of dimensional regularization, but keeping the parameter $m$, the corresponding integrals, indeed, exist, but give contributions $\sim m^{1 - 2n\epsilon}$ that tend to $0$ as $m \rightarrow 0$.
}, whereas the IR cutoff $m$ is purely a technical device -- an IR regularization needed for multi-loop calculations.
It is crucial that the parameter $m$ appears only at intermediate stages of the calculations and that the final results remain valid in the limit $m \rightarrow 0$.
More precisely, it can be shown that in \eqref{eq:representation_for_diagrams_comes_from_homogeneity}, all two-loop diagrams -- except for the sunset $D_{\mathrm{L}2}^{(2)}$ -- can be evaluated by setting $s = 0$ and $\epsilon = 0$.
In turn, for $D_{\mathrm{L}2}^{(2)}$, one has to isolate the subgraph divergence in the following way
$F_2(\kappa, 1, z) \rightarrow \bigl[F_2(\kappa, 1, z) - F_2(0, 1, z)\bigr] + F_2(0, 1, z)$.
Then the integral with this subtraction can also be evaluated at $s = 0$ and $\epsilon = 0$.
The contribution in $D_{\mathrm{L}2}^{(2)}$ corresponding to the subgraph divergence takes the form
\begin{align}\label{eq:divergent_subgraph_contribution_in_problematic_diagram}
    &\int \limits_{-1}^{1} \mbox{d}z \int \limits_{s}^{1} \frac{\mbox{d}\kappa}{\kappa^{1 + 2\epsilon}} \frac{F_2(0, 1, z)}{1 - 4 \epsilon} = \frac{1 - s^{-2\epsilon}}{36 \pi^4 (1 + u_0)^3 (1 - 4 \epsilon)(-2\epsilon)} \notag\\
    &= C_{\mathrm{sub}}\ln{\left(\frac{m}{\Lambda}\right)} + \mathcal{O}(\epsilon), \quad C_{\mathrm{sub}} \coloneqq \frac{-1}{36 \pi^4 (1 + u_0)^3}.
\end{align}
Meanwhile, a direct calculation of the counterterm diagram from \eqref{eq:two-loop_Gammabb'_with_curl-terms} yields
\begin{align}\label{eq:counterterm_Gammabb'_diagram}
    &\raisebox{0ex}{\includegraphics[width=3.0cm]{./countertermbBDiagram.png}}\, = -\big[i \varepsilon_{i j l}k_l \big] \frac{\rho \nu_0 g_0 (\Lambda^{-2\epsilon} - m^{-2\epsilon})}{6 \pi^2 (1 + u_0)^2 (-2\epsilon)}h_0 \notag \\
    & = \rho \nu_0 g_0^2 \Lambda \big[i \varepsilon_{i j l}k_l \big]\left(\frac{-\ln{\left(\frac{m}{\Lambda}\right)}}{36 \pi^4 (1 + u_0)^3}\right) + \mathcal{O}(\epsilon),
\end{align}
where in the last equality the definition of $h_0$ \eqref{eq:h0_structure} with one-loop coefficient \eqref{eq:one_loop_curl-term} was used.

It is clear that the contribution in \eqref{eq:divergent_subgraph_contribution_in_problematic_diagram}, combined with the definition \eqref{eq:two_loop_bb'_linear-asymp} (restoring all dimensional parameters), precisely doubles the expression in \eqref{eq:counterterm_Gammabb'_diagram}.
It follows that by choosing the counterterm 
\begin{align}\label{eq:deltah_counterterm}
    \delta h = -2 g_0^2 \Lambda \ln{\left(\frac{\Lambda}{m}\right)} C_{\mathrm{sub}} + \mathcal{O}(\epsilon)
\end{align}
we will achieve exactly what we intended, i.e., eliminating the logarithmic contributions in \eqref{eq:two-loop_Gammabb'_with_curl-terms}.\footnote{Note that the structures yielding logarithms of type \eqref{eq:divergent_subgraph_contribution_in_problematic_diagram} are also contained in $D_{\mathrm{L}1}^{(2)}$ and $D_{\mathrm{L}3}^{(2)}$, but they are odd in the variable $z$ and are zeroed out during angular integration.
Such zeroing out of potential singularities in oddness is a general fact reflecting the absence of curl terms in $\Gamma^{v'v}$ and $\Gamma^{v'bb}$.}
On the other hand, it is also clear from this that if we were to consider not the theory \eqref{eq:MHD_action_with_Gammabb'_curl-terms} in the cense of Eq. \eqref{eq:dyson_eq_for_Gammab'b_h0_to_propagator} but the model \eqref{eq:MHD_action_with_seed_curl-terms} with $\mathfrak{h}_0 = h_0$ (i.e., with the insertion of a bare curl term with a coefficient $\propto g_0$ into Eq. \eqref{eq:MHD_induction}), then there would be no need to introduce $\delta h$ at all, since such an insertion of the curl term with an ``incorrect'' sign would exactly cancel all curl contributions from the diagrams in Eq. \eqref{eq:two-loop_Gammabb'_with_curl-terms}.
With this formulation of the problem, the divergence of the subgraph \eqref{eq:divergent_subgraph_contribution_in_problematic_diagram} would be exactly canceled by the diagram \eqref{eq:counterterm_Gammabb'_diagram}.

Finally, choosing $\delta h$ as described in Eq. \eqref{eq:deltah_counterterm} the linear-in-$\boldsymbol{k}$ term of
$\Gamma^{b'b}$ in \eqref{eq:two-loop_Gammabb'_with_curl-terms} takes the form 
\begin{align}\label{eq:Lambda_renormalized_Gammab'b_in_ordinary_model}
   \Gamma_{i j}^{b'b} = \rho \nu_0 h_0 k \mathbb{H}_{i j}(\boldsymbol{k}) + \mathcal{O}(k^2).
\end{align}
Unfortunately, a definitive analytic expression for $c_2^{[\Lambda]}$ in $h_0$ cannot be derived. 
However, by using the data from App.~\ref{app:two-loop_Gammab'b_expressions} together with the formula \eqref{eq:representation_for_diagrams_comes_from_homogeneity}, one can compute it numerically for any specific value of $u_0$.
To the two-loop accuracy, it is enough to set $\epsilon = 0$ in \eqref{eq:representation_for_diagrams_comes_from_homogeneity}.

Let us make a few concluding remarks by comparing the model \eqref{eq:MHD_action_with_Gammabb'_curl-terms} with the well-known scalar theory $\varphi^4$ (see the action in Table \ref{tab:comparasion_of_cutoff_and_dimensional_regularization}).
In the usual situation typical for $\varphi^4$ model, the $\Lambda$-renormalization reduces to a transition from $m_0$ to $m_0' = m_0 - \Delta m$, where $\Delta m$ is generated by $\Lambda$-regular contributions of $\Gamma_2$.
In its structure, the parameter $h_0$  \eqref{eq:h0_structure} is just similar to the mass shift $\Delta m$ in the $\varphi^4$ theory.
In our case, however, it plays the role of the ``mass'' itself, which is then shifted by the parameter $\delta h$.
This happens because, in contrast with $\varphi^4$ theory, we do not have a bare parameter of the type $m_0$ onto which the arising $\Lambda$-regular curl contributions could be attributed.
The only way to ``generate'' them in the action is to act in the sense of Eq. \eqref{eq:dyson_eq_for_Gammab'b_h0_to_propagator}.

Furthermore, note that since the ``mass'' in the model \eqref{eq:MHD_action_with_Gammabb'_curl-terms} appears only at the one-loop level, the divergences contained in $\delta h$ are not primitive (i.e., they are not simple polynomials in $m$).
This apparent contradiction with the well-known fact that the MS scheme we imply here preserves the primitiveness of divergences is explained by the fact that, in the usual situation, the primitiveness of divergences in the MS scheme is ensured inductively, based on the fact that one-loop diagrams -- onto which the counterterm is written -- can have only superficial divergences.
In our case, however, the one-loop diagrams only generate $h_0$, and the counterterm corresponding to it arises from two-loop contributions and is not a simple polynomial in $\Lambda$ and $m$.
Of course, the procedure described here is not a full-fledged $\Lambda$-renormalization.
It has only led to the situation where all $\Lambda$-divergences in the model have been gathered into a structure that is regular in $\Lambda$ (see Eq. \eqref{eq:Lambda_renormalized_Gammab'b_in_ordinary_model}).
The final step to remove this regular divergence will be carried out in Sec. \ref{sec:renormalization_in_dynamo_regime}.

Let us now examine $\epsilon$-renormalization for the theory \eqref{eq:MHD_action_with_Gammabb'_curl-terms} (understood in the sense of Eq. \eqref{eq:dyson_eq_for_Gammab'b_h0_to_propagator}).
As is well known, analytical regularization does not eliminate power-law divergences in $\Lambda$.
In the framework of RG analysis, they can be completely ignored, since they do not affect the critical behavior of the theory.
This is essentially the reason for using for evaluating diagrams the prescriptions of the analytical or dimensional regularization techniques.
In the case of the model with action \eqref{eq:MHD_action_with_Gammabb'_curl-terms}, one can act in the same way, since in it there is already the parameter to which the occurrence of curl terms is ``written off''.
In this scenario, one can perform a multiplicative $\epsilon$-renormalization \cite{Adzhemyan1985, Adzhemyan1987}, where all rest UV divergences in diagrams (logarithmic in $\Lambda$) appear as poles in $\epsilon$.
As a result, introducing the necessary counterterms into \eqref{eq:MHD_action_with_Gammabb'_curl-terms} and simultaneously replacing each bare parameter with its renormalized counterpart according to the prescription
\begin{align}\label{eq:transition_from_non-renormalized_to_basic_action}
    \nu_0 \rightarrow \nu, \quad u_0 \rightarrow u, \quad h_0 \rightarrow h, \quad g_0 \rightarrow g \mu^{2\epsilon}
\end{align}
(with $\rho$ unrenormalized \cite{Adzhemyan1987}) is equivalent to introducing three standard renormalization constants $Z_1$, $Z_2$, and $Z_3$ (see the next Section) plus an additional constant 
\begin{align}\label{eq:Z4_definition}
    Z_4 = Z_{\nu}Z_h
\end{align}
in the curl term in \eqref{eq:MHD_action_with_Gammabb'_curl-terms}, which governs the renormalization of the new parameter $h_0$:
\begin{align}
\label{eq:h0_epsilon_renormalization}
    h_0 +\delta h = h Z_h.
\end{align}
The prescription \eqref{eq:transition_from_non-renormalized_to_basic_action} also involves the so-called renormalization mass $\mu$, an auxiliary dimensional parameter of the theory comparable in scale to $l_{\mathrm{min}}$.
Recall that $h$ is regarded as being of the same order as $g$.

In the MS scheme and under analytic regularization, all renormalization constants $Z_i$ are determined by the principal parts of their Laurent expansions in $\epsilon$:
Hence, a general form for these constants can be written as 
\begin{equation} \label{eq:def_of_Zi_in_MS}
Z_i = 1 + \sum \limits_{n = 1}^{\infty} g^{n} \sum_{j = 1}^{n} \frac{z_{n, j}^{(i)}(\rho, u)}{\epsilon^j}, \qquad i = 1,\ldots,4.
\end{equation}
Note that in MS-like schemes, renormalization of the curl term in the action \eqref{eq:MHD_action_with_Gammabb'_curl-terms} does not affect the renormalization constants computed without it.
To the two-loop precision, the RG constant $Z_4$ is determined by a single counterterm diagram \eqref{eq:counterterm_Gammabb'_diagram}, evaluated without the $\Lambda$ cutoff.
From \eqref{eq:counterterm_Gammabb'_diagram} and \eqref{eq:two-loop_Gammabb'_with_curl-terms}, it directly follows that
\begin{align}\label{eq:two-loop_Z4}
    Z_4 = 1 + \frac{g}{12 \pi^2 (1 + u)^2 \epsilon} + \mathcal{O}(g^2).
\end{align}

\subsection{\label{sec:summary_of_previous_work}Summary of previous work}

In this Section, we summarize the principal findings from the two-loop renormalization group (RG) analysis of the model \eqref{eq:MHD_action_without_shift}, assuming that curl-terms in $\Gamma^{b'b}$ are absent, being exactly canceled by adding bare curl term with $-h_0$ coefficient into the right hand side of \eqref{eq:MHD_induction} (action \eqref{eq:MHD_action_with_seed_curl-terms} with $\mathfrak{h}_0 = h_0$).
As shown in the previous Section, in that case we effectively return to the ``curl-less'' model \eqref{eq:MHD_action_without_shift}, whose action after $\epsilon$-renormalized is given by:
\begin{align}\label{eq:renormalized_MHD_action_without_shift}
\mathcal{S}_R &= \frac{1}{2}\boldsymbol{v'} \mathfrak{D} \boldsymbol{v'} + \boldsymbol{v'} \cdot \left[-\text{D}_t \boldsymbol{v} + {\nu}Z_1 \boldsymbol{\nabla}^2 \boldsymbol{v} + Z_3(\boldsymbol{b} \cdot \boldsymbol{\nabla}) \boldsymbol{b}\right]  \notag \\
&+\boldsymbol{b'} \cdot \left[-\text{D}_t \boldsymbol{b} + u {\nu}Z_2 \boldsymbol{\nabla}^2 \boldsymbol{b} + (\boldsymbol{b} \cdot \boldsymbol{\nabla}) \boldsymbol{v}\right].
\end{align}
Here, $\mathfrak{D}$ is again the correlator \eqref{eq:force_correlator} with all parameters replaced by their renormalized counterparts according to the aforementioned rule \eqref{eq:transition_from_non-renormalized_to_basic_action} and with renormalization constants $Z_1$, $Z_2$, and $Z_3$ calculated in MS-scheme.
Note that for space dimensions $d > 2$, there is no corresponding counterterm for the correlator $\mathfrak{D}_R$, i.e., $g_0 \nu_0^3 = g \nu^3\mu^{2\epsilon}$, and the fields $\boldsymbol{v}$ and $\boldsymbol{v'}$ remain unrenormalized, which is a fundamental feature of turbulence theory in the RG framework (see \cite{AdzhemyanVasilev_UFN, Vasilev_RG}).

From the action \eqref{eq:renormalized_MHD_action_without_shift}, the complete set of renormalization constants for the fields and parameters in the model follows from the relations
\begin{align}
\label{eq:Ren_consts_for_parameters}
\begin{split}
& Z_{\nu} = Z_1, \qquad Z_g = Z_1^{-3}, \qquad Z_u = Z_2 Z_1^{-1}, \\
& Z_{v} = Z_{v'} = 1, \qquad Z_{b} = Z_{b'}^{-1} = Z_3^{1/2}.
\end{split}
\end{align}
Finally, the multiplicative renormalization transformations in \eqref{eq:renormalized_MHD_action_without_shift} take the form:
\begin{align}\label{eq:multiplicative_renorm}
\begin{split}
&\boldsymbol{b} \rightarrow \boldsymbol{b} Z_{b}, \quad
\boldsymbol{b'} \rightarrow \boldsymbol{b'} Z_{b'}, \quad \boldsymbol{v} \rightarrow \boldsymbol{v}, \quad \boldsymbol{v'} \rightarrow \boldsymbol{v'}, \\
&\nu_0 = \nu Z_{\nu}, \quad
u_0 = u Z_u, \quad
g_0 = g \mu^{2\epsilon} Z_g.  
\end{split}
\end{align}

To the one-loop order, the RG constants $Z_i$ were calculated in \cite{Adzhemyan1985}, which addressed a more general version of the problem involving a mixed correlator.
Notably, at this level of accuracy, all the constants proved to be independent of $\rho$, which enabled the findings of \cite{Adzhemyan1985} to be applied in the one-loop analysis of the model \eqref{eq:MHD_action_without_shift} in \cite{Adzhemyan1987}.
From a technical point of view, the RG constants $Z_1$, $Z_2$, and $Z_3$ follow from enforcing UV-finiteness of the corresponding 1PI functions $\Gamma^{v'v}$, $\Gamma^{b'b}$, and $\Gamma^{v'bb}$.
Conveniently, these constants can be determined from simplified models comprising only those vertices and propagators necessary for building the required 1PI function.
This is what was done.
Over time, numerous authors have systematically derived two-loop contributions to $Z_i$, taking into account both mirror-symmetric and helical terms. 
Below, for the benefit of a reader, we have summarized and collected all known results for realistic space dimension $d = 3$ in the Table \ref{tab:2loop_MHD_Ren_consts}.
\begin{table*}
\caption{\label{tab:2loop_MHD_Ren_consts}
Residues at the poles in $\epsilon$ of all renormalization constants $Z_i$ of the model \eqref{eq:MHD_action_without_shift}.}
\begin{ruledtabular}
\renewcommand{\arraystretch}{1.4}
\begin{tabular}{lccc}
Contribution to $Z_i$: &$i = 1$ &$i = 2$ &$i = 3$ \\
\hline\rule{-0.11cm}{0.4cm}
$z_{11}^{(i)}(u)$& $-\frac{1}{40 \pi^2} {~}^{\text{\cite{Adzhemyan1984}}}$ & $-\frac{1}{12\pi^2 u (1 + u)}{~}^{\text{\cite{Adzhemyan1984}}}$ & $-\frac{1}{60 \pi^2 u}{~}^{\text{\cite{Adzhemyan1985}}}$ \\[.2cm]
$z_{21}^{(i)}(u_{\star}, \rho)$& $-1.05868 \times 10^{-5}$ 
\footnote{
This result was originally obtained in \cite{Adzhemyan2003}.
It was later refined in \cite{Jurcisinova2009} by employing more advanced numerical integration techniques.
That refinement entails shifting the parameter $\lambda$ from $\lambda = -1.101$ to $\lambda = -1.0994$, where $\lambda$ is introduced in \eqref{eq:two-loop_kinetic_fixed_point_vaslues}.
Once the value of $\lambda$ is known, the coefficient $z_{21}^{(1)}$ is recovered unambiguously using formulas from \cite{Adzhemyan2003}.
Note that the numerical result $z_{21}^{(1)} = -0.00825$ [Eq. (41)] from \cite{Jurcisinova2009} and in various later works referencing \cite{Jurcisinova2009} includes a typographical error and should be multiplied by the factor $S_3S_2/(2\pi)^6$, where $S_d$ is the area of a unit $d$-dimensional sphere.
Moreover, the independence of $z_{21}^{(1)}$ from $\rho$ essentially indicates (at the two-loop level) that helicity does not affect the Kolmogorov-type spectrum of ordinary (i.e., non-MHD) isotropic turbulence.
}
& $-2.25613 \times 10^{-5} + 3.40506 \times 10^{-7}\,\rho^2$ 
\footnote{
Formally, $z_{21}^{(2)}$ was found in \cite{Jurcisinova2016} and \cite{Hnatic2016}, for non-helical and helical systems, respectively.
However, neither of those works supplies numerical values for it.
For the non-helical case this can theoretically be deduced from the multi-page Appendix in \cite{Jurcisinova2016}.
Yet for the presence of helicity, it is unfeasible because the Appendix in \cite{Hnatic2016} lacks the relevant data.
Accordingly, here we attempt a kind of reverse-engineering approach to extract the pertinent numerical value of $z_{21}^{(2)}(u_{\star}, \rho)$.
Specifically, by using the values of $u^{(0)}(A = 1)$ and $u^{(\rho)}(A = 1)$ from Table I in \cite{Hnatic2016}, along with Eq. (47) and Eqs. (49) -- (53), one can retrieve the claimed value.
}
& $4.30812 \times 10^{-6} + 6.17026 \times 10^{-6}\,\rho^2 {~}^{\text{\cite{Hnatic2024}}}$ \\[.2cm]
$z_{22}^{(i)}(u)$ 
\footnote{
The relationships among the residues of poles of varying orders in $\epsilon$ follow from the UV finiteness of the RG functions \eqref{eq:RG_functions}.
In the MS scheme, this finiteness condition reduces to the RG functions being completely independent of $\epsilon$ \cite{Vasilev_RG}.
}
& $-(z_{11}^{(1)})^2 {~}^{\text{\cite{Adzhemyan2003}}}$ & $-\frac{z_{11}^{(2)}}{2(1 + u)}\left(u z_{11}^{(2)} + (2 + u)z_{11}^{(1)}\right)$ 
\footnote{
In fact, $z_{22}^{(2)}$ was first derived directly from diagrams in the scalar passive admixture model \cite{Adzhemyan2005} and then reproduced in general models \cite{Jurcisinova2016, Hnatic2016}.
Here, for clarity we provide the expression for $z_{22}^{(2)}$ in terms of one-loop coefficients.
}
& $- \frac{z_{11}^{(3)}}{2} \left( z_{11}^{(3)} - 2 z_{11}^{(1)} - z_{11}^{(2)} \right)$ 
\footnote{
A correction to the result \cite{Hnatic2024} has been made here: the factor of $1/2$ from Eq. (40) should be placed on the r.h.s. of Eq. (A10). 
} \\
\end{tabular}
\end{ruledtabular}
\begin{flushleft}
\end{flushleft}
\end{table*}

The ultimate goal of RG analysis is to determine the infrared asymptotic behavior of the model under consideration.
These asymptotic properties are conveniently analyzed through the generating functional of the renormalized connected Green’s functions $\mathcal{W}_R(A) \coloneqq \ln{\mathcal{G}_R(A)}$, where $\mathcal{G}_R(A)$ is the same functional defined in Eq.~\eqref{eq:generation_func_of_Green_funcs}, except that the action \eqref{eq:MHD_action_without_shift} is replaced by its renormalized counterpart \eqref{eq:renormalized_MHD_action_without_shift}.
For the functional $\mathcal{W}_R(A)$, one can write two standard scaling Euler equations, along with the Callan-Symanzik RG equation (see \cite{Vasilev_RG}):
\begin{align}\label{eq:RG_equation_for_W}
    \left[\mathrm{D}_{\mu} + \sum \limits_{g, u}\beta_{\alpha} \partial_{\alpha} - \gamma_{\nu} \mathrm{D}_{\nu} + \sum \limits_{\boldsymbol{b}, \boldsymbol{b'}} \gamma_{\varphi} \mathrm{D}^{\varphi}\right] \mathcal{W}_R = 0.
\end{align}
Here, Eq. \eqref{eq:RG_equation_for_W} states that $\mathcal{W}_R(A)$ is independent of the renormalization mass $\mu$ when the sources $A$ and bare parameters remain fixed.
In Eq. \eqref{eq:RG_equation_for_W}, the operator $\mathrm{D}_{x}$ represents the logarithmic derivative $x \partial_x$ for any variable (field or parameter) $x$ of the theory.
Meanwhile, the functional operators $\mathrm{D}^{\varphi} \coloneqq \int \mbox{d}x A^{\varphi}(x)\delta/\delta A^{\varphi}(x)$, with $\varphi \in {\boldsymbol{b}, \boldsymbol{b'}}$ (the same notation used in Eq. \eqref{eq:generation_func_of_Green_funcs}), act on specific connected functions (i.e., the expansion coefficients $\mathcal{W}$ in powers of $A$) and become $n_{\boldsymbol{b}}$ or $n_{\boldsymbol{b'}}$ (the counts of external $\boldsymbol{b}$ and $\boldsymbol{b'}$ lines).
The coefficients in Eq. \eqref{eq:RG_equation_for_W} are the so-called RG functions: the $\beta$-functions, describing how the theory departs from scale invariance, and the anomalous dimensions, which yield exponents associated with the power-law asymptotics of the model’s correlation functions and parameters.
The RG functions can be written in terms of the renormalization constants $Z_i$.
By definition:
\begin{align}\label{eq:RG_functions_def}
    &\gamma_i \coloneqq \overline{\mathrm{D}}_{\mu} \ln{Z_i} \qquad \beta_{\mathfrak{g}} \coloneqq \overline{\mathrm{D}}_{\mu} \mathfrak{g}, \qquad \mathfrak{g} \coloneqq (g, u).
\end{align}
Here, $\overline{\mathrm{D}}_{\mu} \coloneqq \mu\partial_{\mu}\big|_{e_0}$ is the derivative with respect to $\mu$ while holding all bare parameters $e_0$ fixed.
From \eqref{eq:RG_functions_def}, it follows that for any function $F(g, u)$:
\begin{align}\label{eq:acting_of_Dmu_on_funcs}
    \overline{\mathrm{D}}_{\mu}F(g, u) = \left[\beta_{g}\partial_g + \beta_u\partial_u\right]F(g, u).
\end{align}
By applying Eqs. \eqref{eq:multiplicative_renorm}, \eqref{eq:Ren_consts_for_parameters}, and \eqref{eq:RG_functions_def}, one obtains:
\begin{align}\label{eq:RG_functions}
\begin{split}
    & \gamma_g = -3\gamma_1, \qquad \gamma_{\nu} = \gamma_1 \qquad \gamma_u = \gamma_2 - \gamma_1,\\
    &\gamma_{\boldsymbol{b}} = -\gamma_{\boldsymbol{b'}} =\gamma_3/2, \qquad \gamma_{\boldsymbol{v}} = \gamma_{\boldsymbol{v'}} = 0 \\
    &\beta_g = g(-2\epsilon - \gamma_g), \qquad \beta_u = -u\gamma_u.
\end{split}   
\end{align}

Due to UV finiteness of $\gamma_i$, they can be expressed in terms of the coefficient of the first-order pole in $\epsilon$ in the expansion \eqref{eq:def_of_Zi_in_MS} for the 
renormalization constant $Z_i$: 
\begin{align}
\label{eq:RG_func_through_first_poles}
    \gamma_i \coloneqq \overline{\mathrm{D}}_{\mu} \ln{Z_i} = -2 \left(z_{11}^{(i)}g + 2 z_{21}^{(i)}g^2\right) + \mathcal{O}(g^3).
\end{align}

To determine the IR asymptotic behavior of the correlation functions in the inertial range, one has to identify the coordinates $(g_{\star}, u_{\star})$ of the corresponding IR-stable fixed point, where $\beta_g(g)$ and $\beta_u(g, u)$ both vanish, i.e., $\beta_g(g_{\star}) = \beta_u(g_{\star}, u_{\star}) = 0$.
In the present model without magnetic noise, only the so-called kinetic (or Kolmogorov) fixed point arises.
To two-loop order, its coordinates are given by\footnote{
References regarding the quantities in \eqref{eq:two-loop_kinetic_fixed_point_vaslues} appear in Table \ref{tab:2loop_MHD_Ren_consts}.
}:
\begin{align}\label{eq:two-loop_kinetic_fixed_point}
\begin{split}
    & g_{\star} = g_{\star}^{(1)} \epsilon + g_{\star}^{(2)} \epsilon^2 + \mathcal{O}(\epsilon^3), \\ 
    & u_{\star} = u_{\star}^{(1)} + u_{\star}^{(2)} \epsilon + \mathcal{O}(\epsilon^2),
    \end{split}
\end{align}
where
\begin{align}\label{eq:two-loop_kinetic_fixed_point_vaslues} 
    & u_{\star}^{(1)} = \left(\sqrt{43/3} - 1\right)/2, \quad u^{(2)}_{\star} = 0.0138 + 0.0312 \rho^2 , \notag \\
    &g_{\star}^{(1)} = \frac{40}{3} \pi^2, \quad g_{\star}^{(2)} = \lambda g_*^{(1)}, \quad \lambda = -1.0994. 
\end{align}

The fixed point’s stability is dictated by the positive definiteness of the real part of the matrix $\Omega \coloneqq \bigl(\partial \beta_{\mathfrak{g}_1}/\partial \mathfrak{g}_2\bigr)\big|_{(g_{\star}, u_{\star})}$, $\mathfrak{g}_1, \mathfrak{g}_2 \in \{g, u\}$, evaluated near $\epsilon = 0$. 
Because $\beta_g$ does not depend on $u$, this matrix is triangular, so the signs of $\partial_g \beta_g\big|_{g_{\star}}$ and $\partial_u \beta_u\big|_{(g_{\star}, u_{\star})}$ determine stability.
The first equality in \eqref{eq:two-loop_kinetic_fixed_point} is exactly the fixed point corresponding to the ordinary Navier-Stokes turbulence.
Its positive definiteness is well known \cite{Adzhemyan2003, Jurcisinova2009}.
Additionally, a numerical study of $\partial_u \beta_u\big|_{(g_{\star}, u_{\star})}$ was performed in \cite{Hnatic2016}, demonstrating that 
non-zero helicity further strengthens stability of the kinetic fixed point $(g_{\star}, u_{\star})$.

Note that our model admits two exact identities (i.e., without $\epsilon^2$ or higher-order corrections):
\begin{align}\label{eq:exact_identities_for_gamma}
    \gamma_{1 \star} \coloneqq \gamma_1(g_{\star}) = 2\epsilon/3, \qquad \gamma_{2 \star} \coloneqq \gamma_2(g_{\star}, u_{\star}) =  \gamma_{1 \star}. 
\end{align}
The first relation is standard in RG theory of turbulence; the second follows from the first and from the condition $\beta_u(g_{\star}, u_{\star}) = u_{\star}\bigl(\gamma_{2 \star} - \gamma_{1 \star}\bigr) = 0$.
In addition to \eqref{eq:exact_identities_for_gamma}, there is also a perturbatively calculable anomalous dimension $\gamma_{3 \star}$.
Its two-loop value was found to be \cite{Hnatic2024}:
\begin{align}\label{eq:two-lopp_gamma3}
    \gamma_{3 \star} =  - 0.319 \epsilon + \left(0.0556 - 0.4202 \rho^2\right) \epsilon^2 + \mathcal{O}(\epsilon^3).
\end{align}

To conclude this Section, let us examine the energy spectra derived from the RG in models like \eqref{eq:MHD_action_without_shift} when helicity term is absent.
In general, the magnetic and kinetic energy spectra are related to the corresponding static pair correlation functions $G^{b}(\boldsymbol{k})$ and $G^{v}(\boldsymbol{k})$ via:
\begin{eqnarray}
    &E^{v,\,b}(\boldsymbol{k}) \coloneqq (2 C_d)^{-1} k^{d-1} G^{v,\,b}(\boldsymbol{k}),\qquad \label{eq:magnetic_and_kinetic_spectrums_through_pair_correlators} \\
    &G^{v} \coloneqq \int \frac{\mbox{d}\omega}{2\pi} \langle v_i v_i\rangle, \quad G^{b} \coloneqq \int \frac{\mbox{d}\omega}{2\pi} \langle b_i b_i\rangle. \label{eq:static_magnetic_and_kinetic_correlators}
\end{eqnarray}

There is no magnetic noise in the action functional \eqref{eq:MHD_action_without_shift}, so the IR asymptotics can only be computed for the velocity correlator
\begin{align}\label{eq:K41_spec_for_Ev}
    G^{v}(\boldsymbol{k}) & = \nu^2 \mu^{4\epsilon/3} k^{2 - d - 4\epsilon/3}R(1, g_{\star}, u_{\star}) A^{-2}, \notag \\
    A & = \exp{\left(-\int \limits_0^1 \frac{\mbox{d}s'}{s'}\Big(\gamma_1\left(\mybar{0.8}{1pt}{g}(s')\right) - \gamma_{1\star} \Big)\right)}.
\end{align}
Here, $\mybar{0.8}{1pt}{g}$ denotes the invariant charge governed by the corresponding Gell-Mann-Low equation $\mathrm{D}_s \mybar{0.8}{1pt}{g} = \beta_g(\mybar{0.8}{1pt}{g})$, with the initial condition $\mybar{0.8}{1pt}{g}\big|_{s = 1} = g$.
It is evident that for $d = 3$ and $\epsilon = 2$, yields the standard Kolmogorov spectrum $E^v(\boldsymbol{k}) \sim k^{-5/3}$.

In general, the absence of a magnetic noise precludes a direct discussion of energy equipartition between the magnetic and kinetic components -- i.e., whether their spectra coincide -- within the model \eqref{eq:MHD_action_without_shift}.
Hence, it is worthwhile to briefly mention the exact results (based on identities \eqref{eq:exact_identities_for_gamma}) from \cite{Adzhemyan1995} for the non-helical model that includes magnetic noise and anisotropy.
The action for such a model is 
\begin{align}\label{eq:MHD_with_magnetic_noise}
   \mathcal{S}_R + \frac{1}{2}g_2 \mu^{2 a \epsilon} \nu^3 u^2\boldsymbol{b'}\mathfrak{D}^{b}\boldsymbol{b'}, 
\end{align}
where $\mathfrak{D}^{v}$ in $\mathcal{S}_R$ is evaluated at $\rho = 0$, and $\mathfrak{D}^{b}$ has the same form as $\mathfrak{D}^{v}$ but with $\epsilon$ replaced by $a\epsilon$, also at $\rho = 0$.

RG analysis for the model \eqref{eq:MHD_with_magnetic_noise}
reveals existence of two stable fixed points: the kinetic one, with coordinates $(u_{\star}, g_{\star}, g_{2 \star} = 0)$ -- where $u_{\star}$ and $g_{\star}$ match the values in \eqref{eq:two-loop_kinetic_fixed_point_vaslues} -- and a new, so-called magnetic fixed point.
Remarkably, at the kinetic point of \eqref{eq:MHD_with_magnetic_noise}, energy equipartition indeed occurs, even though $\boldsymbol{b}$ (unlike $\boldsymbol{v}$) is renormalized.
The reason is that the condition $g_{2 \star} = 0$ conflicts with the usual requirement that, in the IR regime, the corresponding scaling functions for Green’s functions with $n_b > n_{b'}$ remain finite.
For such functions, the relevant scaling functions become IR-critical quantities with the dimension $(n_b - n_{b'})\omega_2/2$, where
\begin{align}\label{eq:correction_index_in_MHD}
    \omega_2 = 2\epsilon - 2 a \epsilon - \gamma_{3 \star}
\end{align}
is the correction exponent obtained from the $\beta_{g_2}$ at the fixed point \cite{Adzhemyan1995}.
Note that $\gamma_{3 \star}$ here is taken at $\rho = 0$.

Inside the inertial range $l_{\mathrm{mic}} \ll \ell \ll l_{\mathrm{mac}}$ any quantity $Q$ is characterized, additionally to $d_Q$ (canonical dimension), by new critical dimension $\Delta_Q$, determining the power-law behavior in $k \propto \ell^{-1}$ for fixed $\nu$ and $W$.
The \textit{exact} results concerning the kinetic fixed point in \eqref{eq:MHD_with_magnetic_noise} can be summarized as follows \cite{Adzhemyan1995}.
All Green functions exhibit IR scaling with standard (Kolmogorov) critical dimensions for $t$, $\boldsymbol{v}$, and $\boldsymbol{v'}$, and with critical dimensions
\begin{align}
&\Delta_{\boldsymbol{b}} = 1 - \gamma_{1 \star} + \frac{\gamma_{3 \star} + \omega_2}{2} = 1 - \frac{2\epsilon}{3} + \epsilon(1 - a), \label{eq:b_critical_dimension}\\
&\Delta_{\boldsymbol{b'}} = d - \Delta_{\boldsymbol{b}} = d - 1 + \frac{2\epsilon}{3} - \epsilon(1 - a). \label{eq:b'_critical_dimension}
\end{align}
for the fields $\boldsymbol{b}$ and $\boldsymbol{b'}$, respectively.
In turn, for the magnetic correlator $G^b(\boldsymbol{k})$ one obtains a representation like \eqref{eq:K41_spec_for_Ev}, where the scaling function $R^b \sim (k/\mu)^{\omega_2}$, which according to the formulas for spectra \eqref{eq:magnetic_and_kinetic_spectrums_through_pair_correlators} leads to behavior of the type $E^b(\boldsymbol{k}) \sim k^{4 - d + 2\epsilon(1 - a) - 4\epsilon/3}$.
For $d = 3$, $\epsilon = 2$, and $a = 1$, this matches the Kolmogorov spectrum.\footnote{
For further details on this topic, including IR scaling at a magnetic fixed point and the role of anisotropy, the reader may consult \cite{Adzhemyan1995} and the monographs \cite{Adzhemyan_RGinFullDevTurb, Vasilev_RG}.
}

The final remark here is that even in MHD analyzed without helicity, the introduction of anisotropy radically changes the overall physical picture. 
In the isotropic case, the magnetic field does not affect the velocity field’s dynamics in the inertial range and behaves like a passive admixture.
However, once a certain degree of anisotropy is introduced, the Lorentz force becomes IR-relevant and destabilizes the associated fixed point \cite{Adzhemyan1995}.

\section{\label{sec:instability}Long wavelength instability in helical MHD}

In this Section we examine the curl terms that arise in Sec. \ref{sec:renormalization_without_shift} not in the formal framework of renormalization theory, but in terms of the physical instability they cause.
When discussing instabilities in hydrodynamics, one usually means scenarios where specific perturbations to a stationary flow cause rapid changes that can disrupt the flow structure entirely.
In magnetohydrodynamics, a fluid becomes unstable when the gradients of velocity, pressure, or magnetic field exceed critical values, meaning that the convective transport of momentum, heat, or magnetic flux outpaces the diffusive processes governed by viscosity, thermal conductivity, or resistivity.
Three main types of instabilities are typically identified in MHD \cite{Biskamp_MHD_Turb}: Kelvin-Helmholtz instabilities, triggered by velocity shear; Rayleigh-Taylor instabilities, driven by buoyancy in a stratified medium; and MHD instabilities, arising from electric currents in a magnetized fluid.
All of these instabilities, along with the associated stabilizing mechanisms, are commonly investigated at the level of the underlying (deterministic) equations of motion.
Their onset in the system is taken as the initial stage, beyond which the regular flow transitions into turbulence.

The instability addressed here does not hinge on the flow profile’s structure; rather, it emerges as a fundamental statistical feature of the MHD equations (including stochastic formulations).
Its simplest illustration arises in the theory of the kinematic (mean-field) dynamo.
Within the well-known two-scale approach \cite{Moffatt_MagnFieldGen}, the effective equation for the large-scale magnetic field $\mathbf{B}(t, \boldsymbol{x})$ can be written as: 
\begin{align}
\label{eq:mean-field_dynamo}
\frac{\partial \mathbf{B}}{\partial t} = \boldsymbol{\nabla} \times \boldsymbol{\mathcal{E}} + \kappa_0 \boldsymbol{\nabla}^2\mathbf{B}, \qquad \boldsymbol{\mathcal{E}} \coloneqq  \langle \boldsymbol{v} \times \boldsymbol{b}\rangle,
\end{align}
where the mean electromotive force $\boldsymbol{\mathcal{E}}$ (EMF) is introduced and $\langle \bullet \rangle$ is an assemble averaging.
In this framework, the following gradient expansion for $\boldsymbol{\mathcal{E}}$ is also valid:
\begin{align}\label{eq:gradient_expansion_for_EMF}
\boldsymbol{\mathcal{E}} = \alpha_0 \mathbf{B} - \beta_0 \boldsymbol{\nabla} \times \mathbf{B} + \ldots .
\end{align}
Note that Eq. \eqref{eq:mean-field_dynamo} is derived in a reference frame where $\langle \boldsymbol{v} \rangle = 0$ -- that is, a sensor fixed at the spatial position $\boldsymbol{x}$ in the moving flow observes the instantaneous values of the random variable $\boldsymbol{v}$.
This setup is standard in homogeneous isotropic turbulence, where the transport coefficients $\alpha_0$ and $\beta_0$ are treated as pseudoscalar quantities dependent on the statistics of the velocity field $\boldsymbol{v}$ (viewed here as a prescribed source) or, equivalently, on the correlations of the random force \eqref{eq:force_correlator}.

Combining \eqref{eq:mean-field_dynamo} and \eqref{eq:gradient_expansion_for_EMF} produces the familiar kinematic dynamo equation,
\begin{align}
\label{eq:kinematic_dynamo_eq}
\frac{\partial \mathbf{B}}{\partial t} = \alpha_0 \boldsymbol{\nabla} \times  \mathbf{B} + (\kappa_0 + \beta_0) \boldsymbol{\nabla}^2 \mathbf{B},
\end{align}
which exhibits an instability.
Indeed, in the long-wave regime, Eq. \eqref{eq:kinematic_dynamo_eq} in $\boldsymbol{k}$-representation simplifies to
\begin{align}
\label{eq:IR_limit_for_kinematic_dynamo_eq}
\frac{\partial}{\partial t} \begin{pmatrix}
{B}_1 \\
{B}_2 \\
{B}_3 \\
\end{pmatrix} = i \alpha_0 
\begin{pmatrix}
0 & -k_3 & k_2 \\
k_3 & 0 & -k_1 \\
-k_2 & k_1 & 0 \\
\end{pmatrix} 
\begin{pmatrix}
{B}_1 \\
{B}_2 \\
{B}_3 \\
\end{pmatrix},
\end{align}
where the matrix on the right-hand side is precisely $\mathbb{H}_{ij}(\boldsymbol{k})$ from Eq.~\eqref{eq:tensor_R}.
A Hermitian matrix with zero trace diagonalizes to eigenvalues $\pm k$ and $0$. Depending on the sign of $\alpha_0 k$, one component of $\mathbf{B}$ thus grows exponentially, whereas another decays.
The third eigenmode corresponding to the eigenvalue $0$ is unphysical because its eigenfunction scales with $k$, contradicting the incompressibility condition $(\boldsymbol{k} \cdot \mathbf{B}) = 0$.
Because the growth rate scales as $k$ and dissipation as $k^2$, only the growing mode persists in the long-time limit, thereby giving rise to an instability.
This instability provide a seed that becomes a macroscopic entity in the steady state.

In the context of helical MHD turbulence, this instability causes an exponential amplification of incipient disturbance vortices, thereby triggering a reverse energy cascade and the formation of structures whose characteristic scale greatly exceeds that of the turbulent flow \cite{Moiseev1983a}.
This phenomenon forms the basis for the well-known model that describes large-scale vortex generation in Earth’s atmosphere \cite{Moiseev1983a, Moiseev1983b} and possibly in the atmospheres of other planets \cite{Ivanov1996}.

\subsection{\label{sec:instability_in_bb'}Stability criterion in stochastic MHD}

From the point of view of the formalism used in this paper, the aforementioned instability is simply a violation of the local stability condition of the model \eqref{eq:MHD_action_without_shift} with respect to relatively (infinitely) small fluctuations around its ground state $\langle\boldsymbol{v}\rangle = \langle \boldsymbol{b}\rangle = 0$.
Recall that the necessary stability condition requires
\begin{align}\label{eq:stability_in_stoch_dyn}
    &\langle v_i^{~} v'_j \rangle,\, \langle b_i^{~} b'_j \rangle,\, \Delta_{i j}^{vv'},\, \Delta_{i j}^{bb'} \in \EuScript{A}(\mathbb{C}_+),
\end{align}
where belonging to $\EuScript{A} (\mathbb{C}_+)$ means that these functions are analytic in the upper half-plane $\mathbb{C}_+$ with respect to the variable $\omega$.
The functions $\Delta_{i j}^{vv'}$ and $\Delta_{i j}^{bb'}$ appear in \eqref{eq:MHD_prop_without_shift}, while the exact (also known as dressed in field-theoretic parlance) response functions, $\langle \boldsymbol{v}' \otimes \boldsymbol{v} \rangle = -\left[ \Gamma^{v'v}\right]^{-1}$ and $\langle \boldsymbol{b}'\otimes \boldsymbol{b} \rangle = -\left[ \Gamma^{b'b}\right]^{-1}$ follow from the Dyson equations
\begin{align}
&\Gamma_{ij}^{v'v}(\omega, \boldsymbol{k}) = -\alpha(\omega, k)\mathbb{P}_{ij}(\boldsymbol{k}) + \Sigma_{ij}^{v'v}(\omega, \boldsymbol{k}), \label{eq:Dyson_eq_for_responce_func_vv'}\\
&\Gamma_{ij}^{b'b}(\omega, \boldsymbol{k}) = -\beta(\omega, k)\mathbb{P}_{ij}(\boldsymbol{k}) + \Sigma_{ij}^{b'b}(\omega, \boldsymbol{k}). \label{eq:Dyson_eq_for_responce_func_bb'}
\end{align}
Here, $\Sigma_{ij}^{v'v}$ and $\Sigma_{ij}^{b'b}$ are self-energy contributions containing all loop corrections to $\Gamma^{vv'}$ and $\Gamma^{bb'}$, respectively.

Condition \eqref{eq:stability_in_stoch_dyn} is tantamount to having all the zeroes of  $\Gamma^{vv'}$ and $\Gamma^{bb'}$ 
situated in the upper half of the complex $\omega$-plane.
At first sight, since $\Sigma^{b'b}$ and $\Sigma^{v'v}$ are proportional to $g_0$, it appears that they would be negligible compared to the baseline terms of order $\nu_0 k^2$, and thus incapable of violating the stability condition \eqref{eq:stability_in_stoch_dyn}.
However, we already know that in the expansions of $\Sigma^{b'b}$ and $\Sigma^{v'v}$ around $\omega = k = 0$, there is a $k$-linear term (see Eq. \eqref{eq:one_loop_bb'_linear-asymp}) of the form $\propto \rho g_0 \nu_0 \big[i \varepsilon_{i j l} k_l\big] \Lambda$.
For sufficiently small momentum, such a term dominates over $\nu_0 k^2$.
If, in accordance with the renormalization theory (see Sec. \ref{sec:renormalization_without_shift}), we include this term in the action \eqref{eq:MHD_action_without_shift} in full generality -- equivalent to adding $\nu_0 h_0 \,(\boldsymbol{\nabla} \times  \boldsymbol{b})$ on the right-hand side of Eq. \eqref{eq:MHD_induction} -- we end up with the scenario outlined in the preceding Section.
Near $k = 0$ , this contribution surpasses the dissipation term, causing $\Gamma^{bb'}$ to vanish in the ``wrong'' half-plane at the propagator level \eqref{eq:bb'_prop_with_curl-terms}.

\subsection{\label{sec:stabilization}Stabilization via uniform magnetic field: large-scale turbulent dynamo}

Let us now turn our attention to the second possible mechanism by which the instability may be eliminated.
As noted in Sec. \ref{sec:intro}, the emergence of a large-scale mean magnetic field in a highly turbulent MHD system -- often referred to as a large-scale turbulent dynamo -- is an observable fact.
On the other hand, our model reveals that the initial ground state with zero mean, $\langle \boldsymbol{b} \rangle = 0$, becomes unstable when subjected to large-scale fluctuations.
Bringing these observations together, one naturally concludes that the turbulent dynamo is precisely the process responsible for stabilizing the system \cite{Adzhemyan1987}.
The next step is to clarify how stabilization occurs in practice under the model \eqref{eq:MHD_action_without_shift}.
From a field-theoretic perspective, this appears straightforward: models exhibiting ground-state instability are typically stabilized through the spontaneous breaking of a continuous symmetry, which in turn yields a non-zero vacuum expectation value of the relevant field.
Exactly this approach was proposed and carried out in \cite{Adzhemyan1987} at the one-loop order, yielding a stabilizing field  $\boldsymbol{B} \coloneqq \langle \boldsymbol{b} \rangle$ with magnitude
\begin{align}\label{eq:one-loop_B0}
   B = 16\pi\sqrt{u}(1 + u) |h| \nu/g.
\end{align}
Here, the renormalized $B$ is expressed in terms of $h$, a parameter with dimensions of inverse length introduced at the end of Sec. \ref{sec:renormalization_without_shift}.
The expression \eqref{eq:one-loop_B0} agrees with the result $B = (8/3\pi)\Lambda\nu\sqrt{u}$ presented in \cite{Adzhemyan1987}, aside from the substitution $h \to g\Lambda c_1^{[\Lambda]}(u, 0)$ with $c_1^{[\Lambda]}$ from Eq. \eqref{eq:one_loop_curl-term} (in \cite{Adzhemyan1987} $\Lambda$ simply plays the role of $h$).
For further details, see Sec. \ref{sec:one_loop_approximation_after_shift}.
It also should be noted that the direction of $\boldsymbol{B}$ remains arbitrary, as it should be in spontaneous symmetry breaking.

Let us clarify here in more detail the logic of the approach employed in \cite{Adzhemyan1987}, which greatly diverges from conventional quantum or statistical field theories by introducing a higher level of technical complexity.
The starting points are as follows.
In standard helical MHD \eqref{eq:MHD_action_without_shift}, the instability stems from the presence of curl-terms.
Meanwhile, the soundness of the model \eqref{eq:MHD_action_without_shift} is not questioned, so, to avoid overcomplicating matters, one has to require that once the necessary curl-terms are appended to \eqref{eq:MHD_action_without_shift} -- as demanded by renormalization -- they must somehow cancel out precisely in every order of perturbation theory.
Accepting the instability as a physical reality, we infer it must induce the breakdown of the initial vacuum state, driving the system to a new, ``lower-energy'' stable ground state.
In this stable regime -- the turbulent dynamo -- a large-scale magnetic field materializes by tapping into the energy of small-scale turbulent motion.
Technically, the system moves to this state via spontaneous symmetry breaking driven by radiative corrections, known as the Coleman-Weinberg mechanism \cite{ColemanWeinberg1973, Cheng_GaugeThofElPart}.
Much like ordinary spontaneous symmetry breaking, the Coleman-Weinberg mechanism is typically framed within the effective-potential formalism (see, e.g., \cite{ItzyksonMartin1975, Fukuda1995, Zinn_QFT, DeWitt_GlobalQFT, Kleinert_PathIntegrals, Vasilev_RG}).
The renormalized effective potential (or free energy or first Legendre transform) $\Gamma_{R}(\alpha)$ is the 1PI generating functional of Green functions.
When spontaneous symmetry breaking occurs, $\Gamma_{R}(\alpha)$ remains single-valued (unlike $\mathcal{W}_{R}(A)$), and it is most conveniently treated via the loop expansion \cite{Zinn_QFT, Zinn_RG, Vasilev_RG}:
\begin{equation}\label{eq:loop_expansion_for_Gamma}
   \Gamma_{R}(\alpha) = \mathcal{S}_{R}(\alpha) + \sum \limits_{l = 1}^{\infty}\Gamma^{(l)}(\alpha), 
\end{equation}
where $\alpha$ is the same as in \eqref{eq:nonrenormalized_generating_functional_for_1PI_GF}, and terms $\Gamma^{(l)}(\alpha)$ represent the ``quantum'' loop corrections.
At the lowest tree order, $\Gamma(\alpha) = \mathcal{S}_{R}(\alpha)$, and the stationarity condition becomes the usual requirement of minimal potential energy. 

The mean field $\boldsymbol{B}$ that stabilizes the system is identified with the stationary point of $\Gamma_{R}(\alpha)$.
Different solutions of stationarity equation for $\Gamma_{R}(\alpha)$ then correspond to distinct ``pure phases'' of the system.
In translation-invariant theories, $\Gamma_{R}(\alpha)$ is restricted to homogeneous $\alpha$, and one searches for the stationary point of the resulting numerical (rather than functional) function $\Gamma_{R}(\alpha)$.
However, in dynamic field theories that include both auxiliary, non-physical (non-propagating) fields and interaction vertices each carrying derivatives, the effective-action approach breaks down.

The crux is that we seek extremals of $\Gamma_{R}(\alpha)$ with $\langle \boldsymbol{b} \rangle \neq 0$ while $\langle \boldsymbol{v} \rangle$, $\langle \boldsymbol{v'} \rangle$, and $\langle \boldsymbol{b'} \rangle$ remain zero.
The first is due to there is no curl-term for $\langle \boldsymbol{v'} \otimes \boldsymbol{v} \rangle$ (see Sec. \ref{sec:linear_in_p_asymptotics}), and we work in a ``gauge'' with $\langle \boldsymbol{v} \rangle = 0$ (see the end of Sec. \ref{sec:field-theoretic_formulation}).
For the rest, because the auxiliary fields act merely as Lagrange multipliers, shifting them is irrelevant.

If we simply impose $\langle \boldsymbol{v'} \rangle = \langle \boldsymbol{b'} \rangle = 0$ on the set of auxiliary fields, we end up with a trivial zero.
In stochastic models of this kind, all 1PI functions composed exclusively of the main fields $\boldsymbol{v}$ and $\boldsymbol{b}$ vanish (see, e.g., \cite{Vasilev_RG}).
Even if we take a strict point of view and set the auxiliary fields’ averages to zero in $\delta \Gamma_R(\alpha)/\delta \alpha$, our model’s vertices contain derivatives such that the stationarity conditions hold for \textit{any} spatially homogeneous averages:
\begin{align}\label{eq:stacionarity_cond_for_Gamma}
\frac{\delta}{\delta \alpha}\Gamma_R(\alpha)\Bigg|_{\substack{\langle\boldsymbol{b'}\rangle = 0, \\ \langle \boldsymbol{v'}\rangle = 0.
}} = 0 \quad \forall \text{ constant } \langle\boldsymbol{b}\rangle \text{ and } \langle\boldsymbol{v}\rangle.
\end{align}
Hence, the stationarity condition for $\Gamma_R(\alpha)$ does not constrain these averages in the manner of standard field models.
\footnote{
We also note that certain methods exist for building an effective action solely with auxiliary fields, even in theories whose vertices all include derivatives (e.g., the Kardar-Parisi-Zhang equation \cite{Cooper2016}).
Nevertheless, these examples hinge on Hubbard-Stratonovich-type transformations for four-field interactions.
To our knowledge, nothing comparable has been devised for three-field interactions.}

Therefore, to account for the emergence of the mean field $\boldsymbol{B}$ and its estimation in \cite{Adzhemyan1987}, and its estimation in \cite{Adzhemyan1987}, one may begin by directly examining the expansion of the unrenormalized generating functional for the connected Green functions in the symmetry-broken phase:
\begin{align}\label{eq:W_in_broken_phase}
    \mathcal{W}_{s}(A) = \ln{\int \limits_{\boldsymbol{b}(\infty) = \boldsymbol{B}_0} \EuScript{D}\Phi \,\mathrm{e}^{\mathcal{S}_0 + A\Phi}},
\end{align}
where the index $s$ enumerates the different ``pure phases'' (in our case, the direction of the bare spontaneous field $\boldsymbol{B}_0$),  $\mathcal{S}_0$ is the action \eqref{eq:MHD_action_without_shift}, and $A$ is the same as in \eqref{eq:generation_func_of_Green_funcs}.
In the general case, the relation 
\begin{align}\label{eq:relation_for_W_in_different_phases}
    \mathcal{W}_{s'}(A') =  \mathcal{W}_{s}(A),
\end{align}
applies, indicating that group transformations connect $\mathcal{W}_{s}(A)$ among different pure phases.
For each specific phase $s$, the symmetry is broken: $\mathcal{W}_{s}(A') \neq  \mathcal{W}_{s}(A)$.

For each given pure phase, perturbation theory is built by making the shift $\boldsymbol{b} \rightarrow \boldsymbol{b} + \boldsymbol{B}_0$ in \eqref{eq:W_in_broken_phase}, yielding
\begin{align}\label{eq:W_in_broken_phase_after_shift}
    \mathcal{W}_{s}(A) = \ln{\int \limits_{\boldsymbol{b}(\infty) = 0} \EuScript{D}\Phi \,\mathrm{e}^{\mathcal{S}_0[\boldsymbol{b} + \boldsymbol{B}_0] + A\Phi}\,\mathrm{e}^{  \boldsymbol{A}^{b}\boldsymbol{B}_0}}.
\end{align}
If, by appropriate choice the magnitude of the spontaneous field $B_0$ in the expansion of $\mathcal{W}_{s}(A)$ in \eqref{eq:W_in_broken_phase_after_shift}, one can eliminate the curl-terms that give rise to instability, then $\boldsymbol{B}_0$ is the field sought.

However, both technical and practical considerations motivate a transition to the renormalized functional $\mathcal{W}_{R \,s}(A)$.
It is defined analogously to \eqref{eq:W_in_broken_phase_after_shift}, by shifting the argument $\boldsymbol{b} \rightarrow \boldsymbol{b} + \boldsymbol{B}$ in the corresponding generating functional $\mathcal{W}_{R}(A)$ of the unshifted theory.
Since a shift by a UV-finite quantity in the UV-finite functional does not compromise the UV-finiteness of the object, it is clear that the shift parameter must be renormalized by the same renormalization constant as the field $\boldsymbol{b}$:
\begin{align}\label{eq:B_field_renormalization}
    \boldsymbol{B}_0 = \boldsymbol{B}Z_b.
\end{align}

It is appropriate to recall here that the physical quantities -- that is, those potentially determinable by experiment -- are the renormalized Green functions (including the one-point function $\boldsymbol{B}$).
We stress that the unrenormalized mean field $\boldsymbol{B}_0$ in our model is not externally imposed parameter as $\nu_0$ or $g_0$, which are considered specified; on the contrary, it emerges within the model as a consequence of its intrinsic instability and is expressed through $\Lambda$.
In the renormalized theory, the corresponding absolute value of $\boldsymbol{B}$ is given in terms of the parameter $|h|$, which plays the role of the UV-finite (a consequence of renormalization) source of curl terms.
Indeed, $B$ can be identified with the emergent large-scale mean field in the system.
At the one-loop order, where the instability first appears, one accomplishes this by selecting $B$ from \eqref{eq:one-loop_B0}. 
It is crucial that no more instabilities emerge under this procedure (see Sec. \ref{sec:linear_in_p_asymptotics}).

Let us also emphasize that when we speak about the stability of our system, we mean only the \textit{local} stability.
A more general statement, apparently, cannot be made (at least in the framework of this model).
In addition, we have another important difference from typical problems of the theory of critical phenomena.
According to our assumption, the field $\boldsymbol{B}$ obtained in the manner described above represents the system’s ``true ground state'', below which it cannot fall.
Unlike critical phenomena, where varying a control parameter (e.g., temperature) shifts the system from one stable phase to another, our initial phase is itself unstable, and it collapses into a ``lower-energy'' stable phase, thereafter persisting in that state.
In spirit, this situation is more akin to relativistic field theory or the theory of directed percolation (e.g., the so-called absorbing states, see 
\cite{hinrichsen2000,hinrichsen2008}).

From a symmetry point of view, the vacuum state $\langle \boldsymbol{b}\rangle = 0$ collapses when a nonzero spontaneously oriented mean field $\boldsymbol{B}$ appears, thereby breaking the global symmetry $\mathbf{SO}(3)$ down to $\mathbf{SO}(2)$ (only rotations in the plane perpendicular to $\boldsymbol{B}$).
A ``Goldstone mode'' should then emerge, which is addressed in Sec. \ref{sec:corrections_to_alfven_waves}.

In closing this Section, observe that the above instability, broadly speaking, appears in any model with a helical correlator \eqref{eq:force_correlator} that contains the diagram displayed in \eqref{eq:one_loop_curl-term_diagram}.
Examples include the well-known Kraichnan-Kazantsev model (see, e.g., \cite{Antonov2006}) and the kinematic MHD model discussed in \cite{Jurcisinova2016, Hnatic2016}.
Note that in the Kraichnan-Kazantsev model, the instability cannot be resolved by any of the above approaches, because the corresponding action is invariant under shifts of the magnetic field.
This suggests that the Kraichnan-Kazantsev model is ``too simple'' to capture MHD turbulence accurately.
In the kinematic MHD model, the situation improves somewhat, and in principle one could apply the program outlined above.
However, in this paper, we proceed further by examining a model that reflects the full system of stochastic MHD equations, albeit without magnetic noise for simplicity.
In this approximation, under perturbation theory, not only does the mean field $\langle \boldsymbol{b}\rangle$ vanish, but also all higher-order correlation functions $\langle \boldsymbol{b}\otimes\boldsymbol{b}\rangle$, $\langle \boldsymbol{b}\otimes\boldsymbol{b}\otimes\boldsymbol{b}\rangle$, etc., simply because the relevant propagators are absent.
However, this ``normal'' solution is unstable and becomes stabilized by the emergence of a non-zero mean field $\boldsymbol{B}$.
As $\boldsymbol{B}$ is generated the higher-order correlation functions $\langle \boldsymbol{b}\otimes\boldsymbol{b}\rangle$, etc. also become nonzero.
Conversely, this approximation clearly demonstrates that helicity alone -- not magnetic noise -- plays a critical role in facilitating the dynamo (potentially active only at some earlier stage; see Sec. \ref{sec:conclusions}).
Nevertheless, all the above arguments hold even if magnetic noise is present.
In that scenario, the normal solution features nonzero higher-order correlation functions  $\langle \boldsymbol{b}\otimes\boldsymbol{b}\rangle$, etc., yet $\langle \boldsymbol{b}\rangle = 0$.
The dynamo phenomenon then involves the emergence of a nonzero $\langle \boldsymbol{b}\rangle$.
The only complication in this more general setting is a technical one: the exceedingly large number of diagrams and the associated intricacy of a diagrammatic approach that diverges from the standard form (see Sec. \ref{sec:diagramm_technique_after_shift}).

\section{\label{sec:model_in_dynamo_regime}MHD in a dynamo regime}

We call the model \eqref{eq:MHD_action_without_shift}, supplemented by a potentially stabilizing shift of the magnetic field, as the ``dynamo regime''.
Our primary aim is to demonstrate that second-order perturbation theory does not undermine the hypothesis that the model is stabilized by the appearance of a mean field with magnitude $B$.
For lack of alternatives, we have to carry out the analysis directly using the diagrammatic technique for spontaneous symmetry breaking.
In this case, the stability condition (exact cancellation of the curl terms at the diagram level) determines only the modulus of $\boldsymbol{B}$, leaving its direction arbitrary.
The specific details of the calculations of $B$ are given in Sec. \ref{sec:two-loop_B_field_calculations}.
Here we describe the general properties of the dynamo regime.

\subsection{\label{sec:MHD_after_shift}Field-theoretical formulation of MHD in dynamo regime}

\begin{figure}[t]
    \includegraphics[width=1\linewidth]{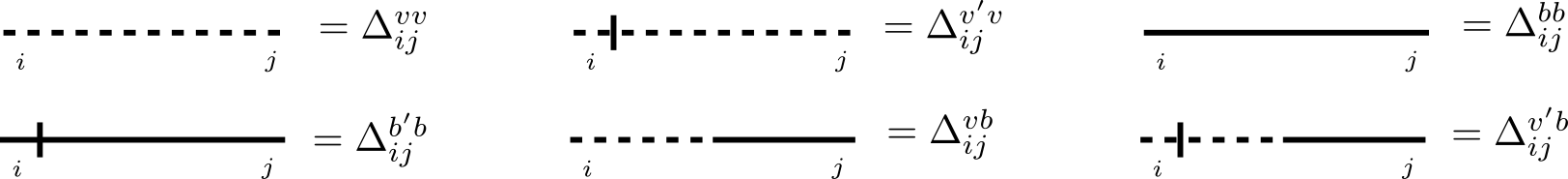}
    \caption{Graphical representation of all independent propagators in \eqref{eq:MHD_action_after_shift}.}
    \label{fig:MHD_propagators_after_shift}
\end{figure}

Once the shift $\boldsymbol{b}(t, \boldsymbol{x}) \to \boldsymbol{b}(t, \boldsymbol{x}) + \boldsymbol{B}_0$ is introduced, where $\boldsymbol{B}_0$ represents a bare spontaneous magnetic field, the action of the model gains extra terms.
For concreteness, we examine this shift in the context of the non-renormalized action \eqref{eq:MHD_action_without_shift}.
For details of renormalization, see later Sec. \ref{sec:two-loop_B_field_calculations}.
Because of the specific structure of \eqref{eq:MHD_action_without_shift}, only its quadratic part is modified:
\begin{align}\label{eq:MHD_action_after_shift}
\mathcal{S}_{\boldsymbol{B}_0} = \mathcal{S}_0 + \boldsymbol{v'}\cdot(\boldsymbol{B}_0 \cdot \boldsymbol{\nabla})\boldsymbol{b} + \boldsymbol{b'}\cdot(\boldsymbol{B}_0 \cdot \boldsymbol{\nabla}) \boldsymbol{v}.
\end{align}
As a result, this shift generates $\boldsymbol{B}_0$-dependent corrections to the existing perturbation diagrams and also gives rise to new diagrams. 
From \eqref{eq:MHD_action_after_shift}, it is evident that only the propagator matrix $\Delta$ changes.
Its new explicit form is 
\begin{align}\label{eq:propagator_matrix_after_shift}
\Delta_{i j} = 
\begin{pmatrix}
\frac{\beta\beta^*}{\xi\xi^*}\mathbb{D}_{ij}  & -i \frac{\gamma \beta^*}{\xi\xi^*}\mathbb{D}_{ij}  & \frac{\beta^*}{\xi^*} \mathbb{P}_{ij} &  i \frac{\gamma}{\xi^*} \mathbb{P}_{i j} \\
i \frac{\gamma \beta}{\xi\xi^*} \mathbb{D}_{ij} &\frac{\gamma^2}{\xi\xi^*}  \mathbb{D}_{ij} & i \frac{\gamma}{\xi^*} \mathbb{P}_{ij} & \frac{\alpha^*}{\xi^*} \mathbb{P}_{ij} \\
\frac{\beta}{\xi} \mathbb{P}_{ij} & -i \frac{\gamma}{\xi} \mathbb{P}_{i j} & 0 & 0 \\
-i \frac{\gamma}{\xi} \mathbb{P}_{ij} &  \frac{\alpha}{\xi} \mathbb{P}_{ij} & 0 & 0 \\
\end{pmatrix} ,
\end{align}
where $\xi(\omega, \boldsymbol{k}) \coloneqq \gamma^2 + \alpha(\omega, k)\beta(\omega, k)$, $\gamma \coloneqq (\boldsymbol{B}_0 \cdot \boldsymbol{k})$, asterisk ${}^*$ indicates complex conjugation, and $\omega$ and $\boldsymbol{k}$ are the implicit arguments of all functions and tensor operators. 
The non-zero, independent matrix elements of \eqref{eq:propagator_matrix_after_shift} are depicted in Fig. \ref{fig:MHD_propagators_after_shift}.
The interaction vertices in \eqref{eq:MHD_action_after_shift} match those in \eqref{eq:MHD_action_without_shift}, so no new diagrammatic structures or divergences arise; only the propagators (lines in diagrams) are modified.
In the original propagators $\Delta^{vv}$, $\Delta^{v'v}$, and $\Delta^{b'b}$ corrections in $B_0$ occur, and entirely new purely $B_0$-lines, namely $\Delta^{vb}$, $\Delta^{v'b}$, $\Delta^{b'v}$, and $\Delta^{bb}$, emerge.
It is easy to see that all new lines and $B_0$-corrections to old ones contain a factor $\gamma^2$.
By the dimension $B_0 \simeq \nu_0 \Lambda$, from which it immediately follows that all $B_0$-graphs constructed from $B_0$-lines do not contain superficial UV divergences. 
In turn, this means that the superficially divergent diagrams in \eqref{eq:MHD_action_after_shift} are the same as in \eqref{eq:MHD_action_without_shift}, presented in the Fig. \ref{fig:two-loop_surface_divergent_diagrams_before_shift}.

Note that the shift $\boldsymbol{b} \to \boldsymbol{b} + \boldsymbol{B}_0$ with constant $\boldsymbol{B}_0$ does not affect the composite operator $\rho \nu_0 h_0 \boldsymbol{b'} \cdot \left(\boldsymbol{\nabla} \times \boldsymbol{b}\right)$, which must necessarily be included into \eqref{eq:MHD_action_after_shift} to ensure a correct $\Lambda$-renormalization.
However, it makes sense to discuss the theory with the insertion of such an operator only in the context of the cancelation (through the choice of $B_0$) of the contributions in generated by it.
This is done in Sec. \ref{sec:two-loop_B_field_calculations}.
If we consider \eqref{eq:MHD_action_after_shift} simply as a model with an arbitrary $B_0$, then in it, along with the IR-relevant contributions $i\rho [\varepsilon_{ijl}k_l] \nu_0 h_0$, $B_0$-structures $i \rho [\varepsilon_{ijl}k_l] j_0$ with certain $j_0$ will appear, as well as some others.
These additional structures do not generate instabilities (see Sections \ref{sec:linear_in_p_asymptotics} -- \ref{sec:corrections_to_alfven_waves}), from which it follows that the curl term $i\rho[\varepsilon_{ijl}k_l] \nu_0 h_0$ can be successfully eliminated by choosing $B_0$.
Ultimately, after the reduction of the curl terms, we again arrive at the model \eqref{eq:MHD_action_after_shift}, but with a specific value of $B_0$.
It is the latter model (with the selected $B_0$) that is considered in this and the next parts of this Section when discussing the diagrammatic technique in the dynamo regime.\footnote{
Note that, from a purely conceptual standpoint and if one so wishes, model \eqref{eq:MHD_action_after_shift} can be viewed in a fully formal manner as a new model featuring an arbitrary bare parameter $B_0$, but with inclusion of the bare curl term $\rho \nu_0 h_0 \boldsymbol{b'} \cdot \left(\boldsymbol{\nabla} \times \boldsymbol{b}\right)$, as discussed in Sec. \ref{sec:renormalization_without_shift}.
}

Observe that in \eqref{eq:MHD_action_after_shift}, one instantly has a bare magnetic correlator $\Delta^{bb}$ and a correlator $\Delta^{vb}$, the latter serving as the tree-level approximation for the electromotive force $\mathcal{E}$ (see Eq. \eqref{eq:gradient_expansion_for_EMF}).
We also note that the function $\xi(\omega, \boldsymbol{k})$, which arises purely formally as a result of the inversion of the quadratic form of interaction, has a clear physical meaning. 
This is exactly the dispersion relation for Alfv\'en waves \cite{Campos1999} (see also Sec. \ref{sec:corrections_to_alfven_waves} and Appendix \ref{app:Alfven_waves_with_exotic_corrections}).

Yet, despite the seeming simplicity of the action \eqref{eq:MHD_action_after_shift}, the diagrammatic technique stemming from the propagators \eqref{eq:propagator_matrix_after_shift} markedly differs from any models the authors are aware of in the realms of statistical or quantum field theory.
In this regard, the model \eqref{eq:MHD_action_after_shift} provides a striking example of ``computing at the edge of feasibility,'' as even a slight increase in complexity would render any standardized method of diagram calculation unworkable -- particularly given the already substantial diagram count at two loops (total number of diagrams reads $488$ after the field shift versus $8$ before).
On one hand, such complications are unsurprising, because we have ``spoiled'' the model by adding a shift; on the other hand, it is, after all, only a constant shift.

\subsection{\label{sec:diagramm_technique_after_shift}Calculation technique in dynamo regime}

In this paragraph, we provide an overview of the computational techniques for the model \eqref{eq:MHD_action_after_shift}.
The main issue is that the diagrams with propagators \eqref{eq:propagator_matrix_after_shift} do not conform to the standard Feynman type.
This is apparent from the fact that the denominators in all propagators involve the function $\xi(\omega, \boldsymbol{k})$, which is a quadratic polynomial in $\omega$ with complex coefficients:
\begin{align}\label{eq:explicit_form_of_xi}
    \scalebox{0.92}{$\displaystyle \xi(\omega, \boldsymbol{k}) = -\omega^2 - i\omega (u_0 + 1) {\nu}_0 k^2 + u_0 {\nu}_0^2 k^4 + (\boldsymbol{B}_0\cdot \boldsymbol{k})^2.$}
\end{align}
Moreover, neither the Schwinger nor Feynman representations, nor any other commonly used ones in quantum or statistical field theory \cite{Weinzierl_FeynInt}, apply to integrals involving the propagators \eqref{eq:propagator_matrix_after_shift}. 
Also, integrands constructed from propagators \eqref{eq:propagator_matrix_after_shift} no longer have the property of homogeneity \eqref{eq:integrands_homogenity}.
Thus, even frequency integration -- normally straightforward and easily handled by standard computer algebra systems (e.g., Maple, Wolfram Mathematica, etc.) -- demands a more careful analysis in this case.
In addition, there is a more technical issue worth noting: in our anisotropic setting, one has to scalarize the tensor integrals that emerge after carrying out the required tensor contractions.

\subsubsection{\label{sec:frequency_integration}Integration over frequencies}

At the one-loop level, the theory of stable polynomials comes to our aid.
More precisely, it turns out that the zeros of $\xi(\omega, \boldsymbol{k})$ or $\xi^*(\omega, \boldsymbol{k})$, viewed as a quadratic polynomial in the frequency $\omega$, lie in the upper or lower half-plane of the complex variable $\omega$ for any given momentum $\boldsymbol{k}$ and positive parameters ${\nu}_0$ and $u_0$.
Polynomials possessing this property are called stable, and the relevant criterion for $\xi(\omega, \boldsymbol{k})$ to belong to this class of polynomials is provided by the Hermite-Biehler (H.-B.) theorem.
In the literature, this theorem is typically stated in various forms.
For our purposes, the most convenient formulation is as follows \cite{Levin_DistZerosEntireFunc}:

Let $g(z)$ and $h(z)$ be nonzero polynomials with real coefficients.
Define $f(z) = g(z) + i h(z)$.
All zeros of $f(z)$ lie in the open lower half-plane $\mathbb{C}{-}$ if and only if the following two conditions are satisfied:
\begin{enumerate}[nosep, left=0pt .. \parindent]
    \item The zeros of $g(z)$ and $h(z)$ are all real, and simple, and between two consecutive zeros of one polynomial, there lies a unique zero of the other polynomial.
    \item There is $x \in \mathbb{R}$ such that $g'(x)h(x) - h'(x)g(x) > 0$.
\end{enumerate}
From the representation $\xi(\omega, \boldsymbol{k}) = g(\omega) + i h(\omega)$, where $g(\omega) \coloneqq -\omega^2 + k^4 {\nu}_0^2 u_0 + (\boldsymbol{B}_0 \cdot \boldsymbol{k})^2$ and $h(\omega) \coloneqq -\omega k^2 {\nu}_0(u_0 + 1)$, it is clear that the conditions of the theorem are satisfied.
\begin{figure}[t]
\begin{overpic}[percent,grid=false,tics=5,width=0.8\linewidth]{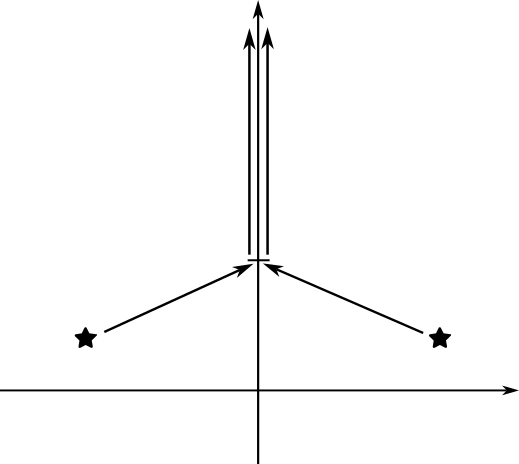}
\put(2,22){$\omega_1(\boldsymbol{k})$}
\put(88,22){$\omega_2(\boldsymbol{k})$}
\put(17,42){$\frac{i}{2}(u_0 + 1)\nu_0 k_{\star}^2 $}
\put(22,32){\rotatebox[origin=c]{30}{$k \rightarrow k_{\star}$}}
\put(43,60){\rotatebox[origin=c]{90}{$k \rightarrow \infty$}}
\put(53,85){$\mathrm{Im}\,\omega$}
\put(90,10){$\mathrm{Re}\,\omega$}
\end{overpic}
\caption{
Behavior of $\omega_{1,2}(\boldsymbol{k})$ for different $k$: for sufficiently small $k < k_{\star}$, where $k_{\star} \coloneqq 2 B_0 |z_k| / \bigl(\nu_0 |1 + u_0|\bigr)$ and $z_k = (\boldsymbol{k} \cdot \boldsymbol{B}_0) / (k B_0)$ is determined by $D(\boldsymbol{k}) = 0$.
Under these conditions, both roots $\omega_{1,2}$ have nonzero real and imaginary parts.
As $k$ grows larger, the roots become purely imaginary and tend to infinity as $k \rightarrow \infty$.
}
\label{fig:poles_in_omega_behaviour}
\end{figure}
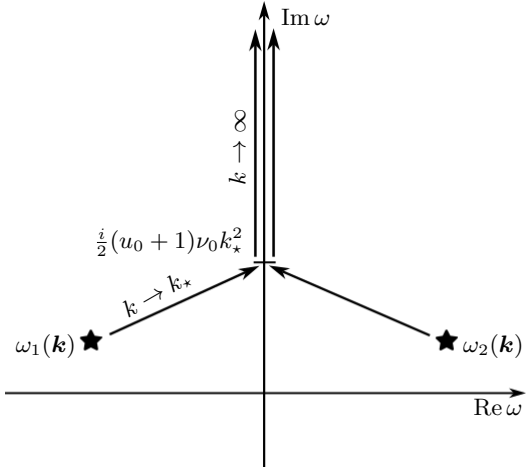
Accordingly, to simplify the calculations, one can close the integration contour in the half-plane containing fewer poles.
In other words, we can carry out the frequency integration formally via the residue theorem -- as if the poles were constant -- for poles of the type
\begin{align}
    &\xi(\omega, \boldsymbol{k}) = -\left(\omega - \omega_1(\boldsymbol{k})\right)\left(\omega - \omega_2(\boldsymbol{k})\right), \label{eq:zeros_of_xi}
\end{align}
where
\begin{align}
    & \omega_{1,2}(\boldsymbol{k}) \coloneqq \frac{i}{2}\left((u_0 + 1)\nu_0 k^2 \pm \sqrt{D(\boldsymbol{k})}\right), \label{eq:omega12(k)} \\
    & D(\boldsymbol{k}) \coloneqq k^4 \nu_0 (u_0 - 1)^2 - 4(\boldsymbol{B}_0 \cdot \boldsymbol{k})^2, \label{eq:minus_discriminant}
\end{align}
while recognizing that each pole’s ``dynamics'' remains confined to a single half-plane in $\omega$.

For two-loop diagrams, however, the situation becomes more complicated.
By virtue of the (H.-B.) theorem, one can always perform the integral over the first frequency variable (say, $\omega_1$).
However, in two-loop diagrams, one encounters terms such as $\xi(\pm\omega_1, \boldsymbol{p})$ or $\xi(\pm\omega_2, \boldsymbol{q})$ along with factors like $\xi(\omega_2 - \omega_1, \boldsymbol{q} - \boldsymbol{p})$, where $\omega_1$ and $\omega_2$ are loop frequencies, and $\boldsymbol{p}$ and $\boldsymbol{q}$ are loop momenta.\footnote{When working with $\xi(\omega, \boldsymbol{k})$, it is useful to note the identities: $\xi(\omega, \boldsymbol{k}) = \xi^{*}(-\omega, \boldsymbol{k})$ and $\xi(\omega, \boldsymbol{k}) = \xi(\omega, -\boldsymbol{k})$.}
Then, performing the integration over $\omega_1$ in the diagram can indeed introduce poles in $\omega_2$, for instance of the form
\begin{align}\label{eq:example_of_dangerous_poles_in_frequency}
   (\omega_2 + \omega_1(\boldsymbol{p}) - \omega_2(\boldsymbol{p} - \boldsymbol{q})),
\end{align}
which may ``move'' between half-planes depending on the sign of $\omega_1(\boldsymbol{p}) - \omega_2(\boldsymbol{p} - \boldsymbol{q})$.
In principle, this issue can be addressed by multiplying the corresponding residue by a step function of the argument $\omega_1(\boldsymbol{p}) - \omega_2(\boldsymbol{p} - \boldsymbol{q})$ of the appropriate sign.
This procedure ensures that the contribution from the pole is counted solely in the half-plane where the integration contour was closed.

Fortunately, this step turns out to be unnecessary.
Direct calculation shows that, at least in every two-loop diagram of $\Gamma^{b'b}$, the contributions from these “dangerous” poles of the type \eqref{eq:example_of_dangerous_poles_in_frequency} cancel out completely.
Evidently, this cancellation stems from another noteworthy property of the diagrams in the model \eqref{eq:MHD_action_after_shift}.
Specifically, one observes that after frequency integration, all one-loop diagrams in $\Gamma^{b'b}$ depend solely on $D(\boldsymbol{p})$ from \eqref{eq:minus_discriminant} and not on $\sqrt{D(\boldsymbol{p})}$.
Remarkably, the same pattern emerges in two-loop diagrams: despite the appearance of more complicated terms such as $\pm \sqrt{D(\boldsymbol{p} - \boldsymbol{q})} \mp \sqrt{D(\boldsymbol{p})}$ in intermediate steps, the final result is always a fractional rational function that does not depend on these square roots, mirroring standard field theory models.

In general, performing symbolic frequency integrations in the two-loop diagrams of the model \eqref{eq:MHD_action_after_shift} is extraordinarily resource-intensive, even for the most advanced computer algebra systems.
We have verified these properties of the diagrammatic technique directly for the $\Gamma^{b'b}$ diagrams, but we believe they apply to at least all two-loop diagrams.
Also, note that attempting to address this problem in the time domain rather than the frequency domain offers no simplification.
Furthermore, direct numerical integration of the diagrams without performing frequency integrals analytically yields no benefit, owing to the multitude of complex-valued functions involved.

We also observe that a rigorous proof that the result of $n$-fold frequency integration in the model \eqref{eq:MHD_action_after_shift} is a fractional rational function of the diagram’s momentum arguments -- along with an investigation of the related function systems for such non-Feynman integrals -- is an interesting mathematical problem in its own right.
Recall that Feynman integrals are generally linked to various generalizations of hypergeometric functions such as Appell, Lauricella, Horn, etc.
\cite{Weinzierl_FeynInt}. 

\subsubsection{\label{sec:scalarization}Scalarization of tensor integrals}

The second technical challenge in diagram computations involves the scalarization of two-loop tensor integrals that arise once the tensor structures of the propagators \eqref{eq:propagator_matrix_after_shift} and the vertex functions \eqref{eq:interaction_vertices} have been convolved for each $s$-point diagram:
\begin{align}\label{eq:tensor_integral_general_form}
    &J_{\boldsymbol{i};\, \boldsymbol{j}}(\{\boldsymbol{k}_s\}) = \int \mbox{d}^d p \,\, p_{i_1}\ldots p_{i_n} \left[\int \mbox{d}^d q \,\, q_{j_1}\ldots q_{j_m} F \right], \notag\\
    &F \coloneqq F\left(p,q, (\boldsymbol{p} \cdot \boldsymbol{q}), (\boldsymbol{p} \cdot \boldsymbol{B}_0), (\boldsymbol{q} \cdot \boldsymbol{B}_0),\{\boldsymbol{k}_s\} \right),
\end{align}
where, for clarity, we use the following multi-index notations: $\boldsymbol{i} \coloneqq (i_1, \ldots, i_n)$, $\boldsymbol{j} \coloneqq (j_1, \ldots, j_m)$, and $\{\boldsymbol{k}_s\} \coloneqq (\boldsymbol{k}_1, \ldots, \boldsymbol{k}_s)$.

For standard Feynman integrals, scalarization typically goes hand in hand with reducing the number of scalar integrals required for the computation.
Such a procedure \cite{Davydychev1991, Tarasov1996, Tarasov1997} (see also \cite{Weinzierl_FeynInt}) relies on the Schwinger representation of the respective integrals, along with the method of performing calculations in arbitrary dimensional space.
In our case, however, this approach proves entirely inadequate because of the distinct structure of our integrands.

In principle, there is another approach to converting tensor integrals to scalar form -- a procedure in the spirit of Passarino-Veltman \cite{Passarino1979}.
Typically, this method entails the following steps:
\begin{enumerate}[nosep, left=0pt .. \parindent]
    \item[(a)] The tensor integral is expressed as a sum of independent tensor structures (built from the external momenta and the metric tensor) multiplied by scalar factors;
    \item[(b)] By examining various convolutions, one obtains a system of linear equations that defines these scalar quantities as integrals with scalar numerators;
    \item[(c)] The scalar numerators are expressed in terms of the denominators, yielding a representation involving the original scalar integrals.
\end{enumerate}
Step (c) is what reduces the number of scalar integrals needed for the calculation.
It relies on the fact that any one-loop Feynman graph admits a Baikov representation \cite{Weinzierl_FeynInt}, making it possible to express any scalar product involving the loop momentum and external momenta as a linear combination of inverse propagators plus a term independent of the loop momentum.
This is the point where the fundamental limitation of the Passarino-Veltman procedure lies, as multi-loop Feynman integrals may contain so-called irreducible scalar products.

On the other hand, for our problem, step (c) cannot be done (while the first two steps can).
In practical terms, this means we lack a system of basis integrals, so for each tensor integral, we must construct an ad hoc representation as described in step (b).
Naturally, this procedure is highly resource-intensive and unsuitable for genuinely large multi-loop problems; however, for our purposes its level of complexity is acceptable.

In this paper, we focus exclusively on the curl terms.
Therefore, we can handle the scalarization of the relevant integrals by setting all external momenta to zero in the general formula \eqref{eq:tensor_integral_general_form}.
A description of the tensor reduction procedure used here -- in particular, decomposing tensor integrals into linear combinations of symmetric tensors of the same rank, built from the metric and external vectors -- together with its application to a specific two-loop diagram of $\Gamma^{b'b}$, is provided in Apps.~\ref{app:scalarization_of_typical_integral} and \ref{app:tensor_structure_of_specific_diagram}.

\subsection{\label{sec:linear_in_p_asymptotics} General structure of the linearized equations \\ of motion in the presence of \texorpdfstring{$B_0$}{B\_0}
}

As previously mentioned and as will be shown in the next Section, curl structures leading to instability can occur only in $\Gamma^{b'b}$, while in $\Gamma^{v'v}$ and $\Gamma^{v'bb}$ they are forbidden by the symmetry of the corresponding interaction vertex.
After passing to model \eqref{eq:MHD_action_after_shift}, the situation becomes even more involved; during the tensor reduction process, besides curl terms, new momentum-linear structures also appear.
Moreover, additional propagators emerge, yielding corresponding two-point vertices that represent off-diagonal modifications to the equations of motion (medium polarization).
Note that here and until the end of this Section we consider simply the model \eqref{eq:MHD_action_after_shift} for an arbitrary $\boldsymbol{B}_0$.
For a generic two-point vertex $\Gamma_{ij}$, the most general form of such a structure -- built from the Levi-Civita symbol, the external momentum $\boldsymbol{k}$, and the field $\boldsymbol{B}_0$ -- is given by
\begin{align}\label{eq:general_structure_of_linear_asymptotic_of_Gamma}
    &\Gamma_{ij}(0, \boldsymbol{k})\Big|_{k\rightarrow 0} = i \rho h_0\epsilon_{i j s} k_s + i\rho \mu_{01} \left[\boldsymbol{k}\times \boldsymbol{\hat{B}}_0\right]_i \hat{B}_{0j} \notag \\
    &+ i \rho \mu_{02} \left[\boldsymbol{k}\times \boldsymbol{\hat{B}}_0\right]_j \hat{B}_{0i} + i \rho \mu_{03} (\boldsymbol{k} \cdot \boldsymbol{\hat{B}}_0)\epsilon_{i j s} \hat{B}_{0s},
\end{align}
where $\boldsymbol{\hat{B}}_0 \coloneqq \boldsymbol{B}_0/B_0$ is a unit vector in the $\boldsymbol{B}_0$ direction.
Let us emphasize that other linear terms in $\boldsymbol{k}$ cannot arise in two-point functions $\Gamma_{ij}$ (they simply cannot be constructed at all).
All the terms shown in \eqref{eq:general_structure_of_linear_asymptotic_of_Gamma} arise already at the two-loop level (see, e.g., App.~\ref{app:tensor_structure_of_specific_diagram}).
In \cite{Adzhemyan1987}, the term with coefficient $\mu_{01}$ was called exotic.
The structures involving $\mu_{02}$ and $\mu_{03}$, which emerge at the two-loop level, are referred here to as bizarre and drift terms.
In the coordinate representation, the exotic term corresponds to $\left(\boldsymbol{\nabla} \times (\boldsymbol{\hat{B}}_0 \cdot \boldsymbol{\phi}) \boldsymbol{\hat{B}}_0\right)$, i.e., to the projection of the physical field $\boldsymbol{\phi}$ ($\boldsymbol{v}$ or $\boldsymbol{b}$) onto the spontaneous magnetic field $\boldsymbol{\hat{B}}_0$.
Meanwhile, the bizarre term yields to the projection of the auxiliary field $\boldsymbol{\phi'}$ ($\boldsymbol{v'}$ or $\boldsymbol{b'}$) in the direction of $\boldsymbol{\hat{B}}_0$.
The drift term generally corresponds to the mixed product $\left(\boldsymbol{\hat{B}}_0 \cdot \boldsymbol{\nabla}\right) \left(\boldsymbol{\hat{B}}_0 \times \boldsymbol{\phi}\right)$.

Under our assumption, the curl terms involving $h_0$ in $\Gamma^{b'b}$ are eliminated through shifting the magnetic field at all orders of perturbation theory.
However, it is still of interest whether they might appear in other 1PI functions.
Strictly speaking, this question is mostly academic because only the one-loop curl terms in $\Gamma^{v'v}$ would present a genuine threat.
In general, the appearance of curl terms from higher-loop contributions would only provide corrections (within perturbation theory) to the stable solution of the corresponding linearized equations of motion obtained in \cite{Adzhemyan1987} (see also Sec. \ref{sec:corrections_to_alfven_waves}). 
Nevertheless, as demonstrated in \cite{Pouquet1978} and \cite{Adzhemyan1987}, at the one-loop level, $\Gamma^{v'v}$ contains no curl terms.
Naturally, if such terms did occur in $\Gamma^{v'v}$, it would undermine the entire rationale of stabilizing the model by shifting the magnetic field.
Indeed, it is entirely impossible to eliminate curl terms by shifting the velocity field, since the setup of our problem already entails moving to a frame comoving with the fluid by subtracting the laminar component of the velocity field (see Sec. \ref{sec:field-theoretic_formulation}).\footnote{In this regard, we note the erroneous work \cite{Kim2010}, where the aforementioned curl terms were obtained.
The findings of \cite{Kim2010} contradict previous works \cite{Pouquet1978} and \cite{Adzhemyan1987} (although \cite{Kim2010} cites both, it incorrectly states that they dealt with antisymmetric magnetic noise). They also conflict with the results of Sec. \ref{sec:absense_of_curl-terms_in_diagram_with_symmetric_vertex}. The crucial error in \cite{Kim2010} consists in an incorrect evaluation of the tensor structure of the self-energy $\Sigma^{v'v}$.
Notably, Eq. (14) in \cite{Kim2010} is identically zero.}

Let us examine the general form of the linearized equations of motion corresponding to the MSR action \eqref{eq:MHD_action_after_shift}:
\begin{eqnarray}
\label{eq:linearized_equations_of_motion_in_polarized_media}
\left\{
\begin{aligned}
&\partial_t \boldsymbol{v} = \left(\nu_0 k^2 + \cancel{\Sigma}^{v'v}\right)\boldsymbol{v} + \left(i(\boldsymbol{B}_0 \cdot \boldsymbol{k}) + \cancel{\Sigma}^{v'b}\right)\boldsymbol{b}, \\
&\partial_t \boldsymbol{b} = \left(i(\boldsymbol{B}_0 \cdot \boldsymbol{k}) + \Sigma^{b'v}\right)\boldsymbol{v} + \left(u_0\nu_0 k^2 + \Sigma^{b'b}\right) \boldsymbol{b}.
\end{aligned}
\right.
\end{eqnarray}
Here, the symbol $\cancel{\Sigma}^{a b}$, with $a, b \in \{v, v', b, b'\}$ indicates the absence of curl contributions in the respective correction to self-energy matrix $\Sigma^{a b}$.

We now inquire whether a stable leading-order approximation exists to solve this system within perturbation theory.
From the perspective of instability, in the matrix $\Sigma^{a b}$ only the linear-in-$\boldsymbol{k}$ terms of the form $i \rho \varepsilon_{i j l} k_l \nu_0 \Lambda$ pose a threat.
Thus, for the leading-order solution of \eqref{eq:linearized_equations_of_motion_in_polarized_media}, we are interested solely in the diagonal elements of $\Sigma^{a b}$, because for small $\boldsymbol{k}$ and any finite $g_0$, the linear-in-$\boldsymbol{k}$ contributions to $\Sigma^{v'v}$ and $\Sigma^{b'b}$ always dominate the bare propagator terms $\nu_0 k^2$ and $u_0\nu_0 k^2$.
By contrast, for the off-diagonal elements of $\Sigma^{a b}$, the bare propagator contributions are already linear in $\boldsymbol{k}$ and dominate any $g_0 k$ terms in $\Sigma^{v'b}$ and $\Sigma^{b'v}$.
Hence, to the first order, they can be neglected.
Note that this is fortuitous for our analysis, because, in contrast to $\Sigma^{v'b}$, a curl term can be produced in $\Sigma^{b'v}$, and the entire scheme would fail if it had to be included at the leading order.

In any case, if any of the $\boldsymbol{B}_0$-structures from \eqref{eq:general_structure_of_linear_asymptotic_of_Gamma} in $\Sigma^{v'v}$ or $\Sigma^{b'b}$ were to trigger an instability analogous to the curl terms in $\Sigma^{b'b}$, then the notion of stabilizing the system by introducing a large-scale field $\boldsymbol{B}_0$ would have to be discarded.
However, as shown in Section \ref{sec:corrections_to_alfven_waves}, none of the newly emerging $\boldsymbol{B}_0$-dependent momentum-linear structures produce additional instabilities (at least for $d = 3$).
Thus, we confirm the consistency of the assumption that the large-scale field $\boldsymbol{B}_0$ may serve as a necessary means of stabilizing (at least locally) the system.

\subsection{\label{sec:absense_of_curl-terms_in_diagram_with_symmetric_vertex}Absence of curl-terms in \texorpdfstring{$\Gamma^{v'v}$}{} and \texorpdfstring{$\Gamma^{v'bb}$}{}}

\begin{figure}[t]
    \includegraphics[width=0.8\linewidth]{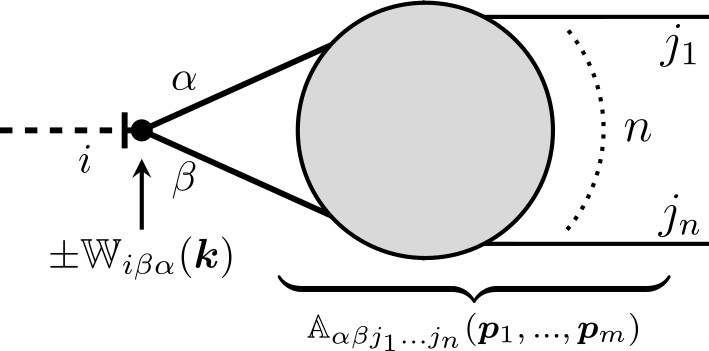}
    \caption{The general structure of the linear in external momentum $\boldsymbol{k}$ asymptotics of $(n + 1)$-point 1PI $m$-loop diagram, in which $\boldsymbol{k}$ flows in through a symmetric vertex.}  \label{fig:arbitrary_diagram_with_symmetric_vertex}
\end{figure}

Let us show that in the general case, no curl terms can appear in the 1PI functions $\Gamma^{v'v}$ and $\Gamma^{v'bb}$.
Consequently, the $\Lambda$-renormalization of the $\mathrm{curl}$ operator is fully governed by its contributions to $\Gamma^{b'b}$, with $\Gamma^{v'v}$ and $\Gamma^{v'bb}$ playing no role in that process.
More precisely, let us consider the leading (linear) contribution to the $k \rightarrow 0$ asymptotics of an arbitrary $(n + 1)$-point, $m$-loop 1PI diagram in which the external momentum $\boldsymbol{k}$ enters through the vertex $\mathbb{W}_{i j l}(\boldsymbol{k})$ or $\mathbb{U}_{i j l}(\boldsymbol{k})$ (noting that these differ only by a sign).
Moreover, for $n > 1$, we assume that the diagram is symmetric under the permutation $j_s \leftrightarrow j_l$ of the external lines (as in the case of $\Gamma^{v'bb}$).
The general layout of such a diagram is depicted in Fig. \ref{fig:arbitrary_diagram_with_symmetric_vertex}.
Henceforth, we label the external indices of the diagram with Latin letters and the indices used for internal convolutions with Greek letters.

Because the external momentum $\boldsymbol{k}$ enters the diagram via the vertex $\mathbb{W}_{i \beta \alpha}$ (we consider this vertex for definiteness), the linear asymptotics of the diagram arises by setting $\boldsymbol{k} = 0$ in the remaining part of it.
Then the rest tensor structure (apart from the vertex $\mathbb{W}_{i \beta \alpha}$) is described by the tensor $\mathbb{A}_{\alpha\beta j_1\ldots j_n}$.
The discussed linear asymptotics must be a pseudotensor (there is no purely momentum-linear transverse true tensor), i.e., it must incorporate the Levi-Civita symbol, which appears in propagators of the form $\mathbb{D}_{i j}$ in \eqref{eq:propagator_matrix_after_shift}.
From this, it follows that the tensor $\mathbb{A}_{\alpha\beta j_1\ldots j_n}$ has the most general form
\begin{align}\label{eq:general_form_of_tensor_A}
    \mathbb{A}_{\alpha\beta \boldsymbol{j}} = \varepsilon_{\alpha \beta \gamma} C^{1}_{\gamma  \boldsymbol{j}} + \sum \limits_{s = 1}^n \left(\varepsilon_{\alpha j_s \gamma}C^{2s}_{\gamma \beta} +  \varepsilon_{j_s \beta \gamma}C^{3s}_{\gamma \alpha} \right),
\end{align}
where, for brevity, we again employ previously introduced multi-index $\boldsymbol{j} \coloneqq (j_1, \ldots, j_n)$  and suppress momentum dependence.

A few remarks are in order regarding \eqref{eq:general_form_of_tensor_A}.
It is straightforward to see that in our situation we can immediately set $C^{1}_{\gamma  \boldsymbol{j}} = 0$, because this term vanishes upon convolution with the symmetric vertex $\mathbb{W}_{i \beta \alpha}$.
We include this term here only for illustration, as it is precisely the one that would generate curl and exotic contributions in diagrams if $\mathbb{W}_{i \beta \alpha}$ were replaced by an antisymmetric under $\beta \leftrightarrow \alpha$ vertex $\mathbb{V}_{i \beta \alpha}$.
Therefore, besides our assumption that $\mathbb{A}_{\alpha\beta \boldsymbol{j}}$ is symmetric in the entire index set $\boldsymbol{j}$, we can, without loss of generality, treat it as symmetric under $\alpha \leftrightarrow \beta$.
Note also that the requirement of symmetry under permutations of the set $\boldsymbol{j}$ in \eqref{eq:general_form_of_tensor_A} excludes terms in which the Levi-Civita symbol incorporates two or more indices from $\boldsymbol{j}$ (e.g., $\varepsilon_{\alpha j_s j_r}C_{\beta \cancel{\boldsymbol{j}}}$), where $\cancel{\boldsymbol{j}}$ indicates the omission of indices $j_s$ and $j_r$.
Moreover, one might initially suppose that terms like $\varepsilon_{\alpha \mu \nu}X_{\mu \nu \beta \boldsymbol{j}}$ or $\varepsilon_{\gamma \mu \nu}Y_{\gamma \mu \nu \alpha \beta \boldsymbol{j}}$ could appear in \eqref{eq:general_form_of_tensor_A}, but this is not the case. 
For such terms to exist, the coefficients $X_{\mu \nu \beta \boldsymbol{j}}$ or $Y_{\gamma \mu \nu \alpha \beta \boldsymbol{j}}$ would have to be antisymmetric under $\mu \leftrightarrow \nu$ and totally antisymmetric in $\gamma, \mu, \nu $, respectively.
However, convolving an antisymmetric object with an $\varepsilon$-tensor over two or more indices ``eliminates'' that $\varepsilon$-tensor, leaving a symmetric structure.\footnote{
Further, none of the tensor coefficients in $\mathbb{A}_{\alpha\beta \boldsymbol{j}}$ can contain any powers of mixed products $\left(\varepsilon_{\gamma \mu \nu} (p_{r})_\gamma (p_{s})_\mu (p_{l})_\nu\right)^k$, $r,s,l = 1, \ldots, m$, $r\neq s \neq l$.
During tensor reduction, the $\varepsilon$-symbols are inevitably rearranged into a chained convolution that ultimately yields contributions of order $k^2$ or higher in the diagram’s asymptotic expansion, by virtue of the relation
$\varepsilon_{\alpha l k} \varepsilon_{\alpha r s} = \delta_{l r}\delta_{k s} - \delta_{l s} \delta_{k r}$.
}

Accordingly, and given that our assumptions impose symmetry on the remaining indices $\alpha$, $\beta$ and $\boldsymbol{j}$, these constructions cannot give rise to a Levi-Civita -- proportional linear term in the $k \rightarrow 0$ asymptotics.
Hence, the structure \eqref{eq:general_form_of_tensor_A} can include only $\varepsilon$-tensors that carry two free indices.

From these considerations, the final tensor structure $\mathbb{T}_{i \boldsymbol{j}}(\{\boldsymbol{p}_m\})$ of the diagram follows
\begin{align}\label{eq:general_tensor_structure_of_leading_in_k_asymptotics_of_symmetric_diagram}
    \mathbb{T}_{i \boldsymbol{j}}(\{\boldsymbol{p}_m\}) &= \mathbb{W}_{i \beta \alpha}(\boldsymbol{k}) \mathbb{A}_{\alpha\beta \boldsymbol{j}}(\{\boldsymbol{p}_m\}) \notag\\
    &= \sum \limits_{s = 1}^n \Big( i \varepsilon_{\alpha j_s \mu}k_{\alpha}T^s_{\mu i} + i \varepsilon_{i j_s \mu}k_{\alpha}T^s_{\mu \alpha} \Big),
\end{align}
where $T^s_{\gamma \delta}(\{\boldsymbol{p}_m\}) \coloneqq C^{2s}_{\gamma \delta}(\{\boldsymbol{p}_m\}) - C^{3s}_{\gamma \delta}(\{\boldsymbol{p}_m\})$.
As discussed in Appendix \ref{app:scalarization_of_typical_integral}, upon tensor reduction the two-index object $T^s_{\gamma \delta}$ inevitably transforms to $T^s_{\gamma \delta} \rightarrow \delta_{\gamma \delta} T_1^s + \hat{B}_{0\gamma}\hat{B}_{0\delta} T_2^s$, involving some scalar coefficients $T_1^s$ and $T_2^s$.
Hence, for the linear dependence on
 $\boldsymbol{k}$ in any diagram of the type shown in Fig. \ref{fig:arbitrary_diagram_with_symmetric_vertex}, we find
\begin{align}\label{eq:general_form_of_linera_in_k_contribution_of_symmetric_diagram}
    &\int \prod \limits_{r = 1}^{m} \mbox{d}^d p_r\, \mathbb{T}_{i \boldsymbol{j}}F = \sum \limits_{s = 1}^n \bigg(i\varepsilon_{\alpha j_s \gamma}k_{\alpha} \delta_{\gamma i}F_1 \notag +i\varepsilon_{i j_s \gamma}k_{\alpha} \delta_{\gamma \alpha}F_1 \\
    &+ i\varepsilon_{\alpha j_s \gamma}k_{\alpha}\hat{B}_{0i}\hat{B}_{0\gamma}F_2 +
    i\varepsilon_{i j_s \gamma}k_{\alpha} \hat{B}_{0\alpha}\hat{B}_{0\gamma}F_2 \bigg),
\end{align}
where $F$ is the scalar function obtained after frequency integration, and $F_1$, $F_2$ are certain scalar integrals.
The vanishing of the first two terms on the right-hand side of the first equality in \eqref{eq:general_form_of_linera_in_k_contribution_of_symmetric_diagram} signifies the absence of curl (and exotic) terms in the diagram in question.
The nonzero contributions that remain are those associated with $\mu_{02}$ (bizarre) and $\mu_{03}$ (drift) from \eqref{eq:general_structure_of_linear_asymptotic_of_Gamma}.
Note again that if $\mathbb{V}_{i \beta \alpha}$ were used instead of $\mathbb{W}_{i \beta \alpha}$, the term $C^{1}_{\gamma  \boldsymbol{j}}$ in $\mathbb{T}_{i \boldsymbol{j}}$ from \eqref{eq:general_tensor_structure_of_leading_in_k_asymptotics_of_symmetric_diagram} would not vanish and would generate curl and exotic contributions.
Finally, as a particular example of this result, we see that no curl terms arise in ordinary hydrodynamic turbulence (the vertex $\mathbb{W}_{i \beta \alpha}$ in that case corresponds to the usual Navier-Stokes equation).

\subsection{\label{sec:corrections_to_alfven_waves}Power-law corrections to Alfv\'en waves}

The system of linearized equations \eqref{eq:linearized_equations_of_motion_in_polarized_media} was examined in \cite{Adzhemyan1987, Hnatic2019} in connection with the study of the role of exotic terms appearing in \eqref{eq:general_structure_of_linear_asymptotic_of_Gamma} already at the one-loop level in the dynamo regime.
It was demonstrated that they, does not lead to instability (unlike the curl contribution); rather, it favors the development of distinct long-lived pulses of the form $\sim t\, \mathrm{e}^{- \kappa_0  t}$, where $\kappa_0  \coloneqq (1 + u_0) \nu_0 k^2 > 0$, in Alfv\'en waves.
It is precisely the Alfv\'en waves polarized orthogonally to the plane $\boldsymbol{B}_0 - \boldsymbol{k}$ that play in our case the role of Goldstone modes, striving, as it were, to restore the isotropy they break.
Unfortunately, the solution of \eqref{eq:linearized_equations_of_motion_in_polarized_media} is explicitly known solely for ideal MHD \cite{Adzhemyan1987}, where occurs an irremovable weak polynomial in $t$ instability.
As was also noted in \cite{Adzhemyan1987}, the inclusion of viscosity corrects the situation, generating exponential damping.
Due to its physical significance, we provide here an explicitly exact stable solution of the linearized system \eqref{eq:linearized_equations_of_motion_in_polarized_media} in the general resistive MHD case and discuss the potential inclusion of two-loop structures within perturbation theory.

We examine wave-type solutions of the form
\begin{align}\label{eq:wave_type_solutions}
    \boldsymbol{v}(t, \boldsymbol{x}) = \boldsymbol{v}(t)\mathrm{e}^{i\boldsymbol{k}\cdot \boldsymbol{x}}, \qquad \boldsymbol{b}(t, \boldsymbol{x}) = \boldsymbol{b}(t)\mathrm{e}^{i\boldsymbol{k}\cdot \boldsymbol{x}},
\end{align}
for which it is helpful to employ the following orthonormal basis \cite{Adzhemyan1987}:
\begin{align}\label{eq:orthonormal_basis}
    \boldsymbol{e}_1 \coloneqq \frac{\boldsymbol{k}}{k}, \quad \boldsymbol{e}_2 \coloneqq \frac{\boldsymbol{\hat{B}}_0 - \boldsymbol{e}_1\cos{\delta}}{\sin{\delta}} \quad \boldsymbol{e}_3 \coloneqq \frac{[\boldsymbol{e}_1 \times \boldsymbol{\hat{B}}_0]}{\sin{\delta}}, 
\end{align}
where $\boldsymbol{\hat{B}}_0$ is from Eq. \eqref{eq:general_structure_of_linear_asymptotic_of_Gamma}, and $\delta$ is the angle between $\boldsymbol{k}$ and $\boldsymbol{B}_0$.
From the transversality conditions \eqref{eq:MHD_transvers_cond}, it follows that in the basis \eqref{eq:orthonormal_basis}, the fields $\boldsymbol{v}(t)$ and $\boldsymbol{b}(t)$ each possesses two non-zero components: $ \boldsymbol{v}(t) = v_2(t)\boldsymbol{e}_2 + v_3(t)\boldsymbol{e}_3$ and $\boldsymbol{b}(t) = b_2(t)\boldsymbol{e}_2 + b_3(t)\boldsymbol{e}_3$.

With these preparations, using \eqref{eq:wave_type_solutions} and \eqref{eq:orthonormal_basis}, we recast the system \eqref{eq:linearized_equations_of_motion_in_polarized_media} into the Euclidean form of the Schr\"odinger equation
\begin{align}\label{eq:linearized_equations_of_motion_in_matrix_form}
    \partial_t\Psi = \left(A_0 + a A_1\right)\Psi,
\end{align}
where $\Psi \coloneqq (b_2, b_3, v_2, v_3)$ and $a = g_0 \Lambda^{-2\epsilon}$ is a natural dimensionless expansion parameter.
In addition to the loopless terms, the matrix $A_0$ incorporates the diagonal one-loop contributions to $\Sigma^{v'v}$ and $\Sigma^{b'b}$ (see the text after Eq. \eqref{eq:linearized_equations_of_motion_in_polarized_media}).
Among these, only the exotic contribution in $\Sigma^{b'b}$ remains nonzero.
This is a specific instance of the result in the previous Section (the corresponding curl term is assumed to be removed via an appropriate choice of $\boldsymbol{B}_0$).
Meanwhile, the matrix $A_1$ gathers all structures that arise at the two-loop level and further, including the two-loop correction to the exotic term in $\Sigma^{b'b}$, as well as all one-loop off-diagonal terms.
Note that, according to Sec. \ref{sec:absense_of_curl-terms_in_diagram_with_symmetric_vertex}, the form of the matrix $A_0 + a A_1$  remains unaltered under the inclusion of higher-loop contributions; such contributions merely provide corrections to the elements already present in $A_1$.

Treating $A_1$ as a small perturbation, one can straightforwardly derive the solution of \eqref{eq:linearized_equations_of_motion_in_matrix_form} to the two-loop order in the form
\begin{align}\label{eq:perturbation_theory_solution_for_linearized_system}
    \Psi(t) = \mathrm{e}^{A_0 t}\left(1 + a \int \limits_0^t \mbox{d}s\,\, \mathrm{e}^{-A_0 s}A_1\mathrm{e}^{A_0 s}\right)\boldsymbol{c},
\end{align}
where the vector $\boldsymbol{c} \coloneqq (c_1, c_2, c_3,c_4)$ is an arbitrary constant wave amplitude.

The matrix $A_0$ from \eqref{eq:linearized_equations_of_motion_in_polarized_media} takes the explicit form
\begin{align}\label{eq:matrix_A0}
A_0 =
\begin{pmatrix}
i \gamma & 0 & -u_0 \nu_0 k^2 & 0 \\
0 & i \gamma & 2i \rho \lambda & -u_0 \nu_0 k^2 \\
-\nu_0 k^2 & 0 & i \gamma & 0 \\
0 & -\nu_0 k^2 & 0 & i \gamma \\
\end{pmatrix},
\end{align}
where $\gamma = k B_0 \cos{\delta}$ is as defined in \eqref{eq:propagator_matrix_after_shift}, and $2\lambda \coloneqq \mu_{01} k \sin{\delta}^2$.
At one-loop approximation, $\mu_{01} = g\big/16\pi(1 + u_0)\sqrt{u_0}$ from \cite{Adzhemyan1987}.
For brevity and because it is not urgently required, we do not present the explicit form of $A_1$ here.
If needed, it can be constructed following the procedure outlined above.

Hence, the resistive one-loop solution of \eqref{eq:linearized_equations_of_motion_in_matrix_form} containing corrections polynomial in $t$ takes the form $\Psi^{(1)}(t) = \mathrm{e}^{A_0 t}\,\boldsymbol{c}$.
Its components are detailed in App. \ref{app:Alfven_waves_with_exotic_corrections}.
One observes that this solution decays as $\sim t \mathrm{e}^{-\kappa_0 t}$, in accordance
with the prediction of \cite{Adzhemyan1987}.

Turning to the two-loop solution \eqref{eq:perturbation_theory_solution_for_linearized_system}, the principal question is whether the perturbation $A_1$ could introduce any extra instability that undermines the exponential decay at $t \rightarrow \infty$.
Within perturbation theory, the answer is no.
In fact, since all components of $\mathrm{e}^{A_0 s}\,\boldsymbol{c}$ decay like $\mathrm{e}^{-\kappa_0 s}$ (it is evident from their explicit form provided in App. \ref{app:Alfven_waves_with_exotic_corrections}), it is straightforward to see that the components of the vector $\mathrm{e}^{-A_0 s}A_1\mathrm{e}^{A_0 s}\boldsymbol{c}$ with constant matrix $A_1$ can exhibit only polynomial dependence on $s$.
Integrating such a polynomial in $s$ increases its degree by $1$, but does not disrupt the exponential decay caused by prefactor $\mathrm{e}^{A_0 t}$ in \eqref{eq:perturbation_theory_solution_for_linearized_system}.

\section{\label{sec:two-loop_B_field_calculations}Two-loop calculation of the spontaneous magnetic field \texorpdfstring{$B$}{B}}

In this Section, we detail the two-loop calculation of the spontaneous field magnitude $B$.
As noted, the magnetic field may be shifted either before renormalization -- yielding an expression for $B_0$ in terms of unrenormalized parameters -- or after renormalization (both after $\Lambda$- and $\epsilon$-renormalization), resulting in a similar relation for renormalized (physical) $B$.
This introduces some arbitrariness, so it is worthwhile to discuss all these stages.

Note that MHD calculations in the dynamo regime are considerably more complicated than in kinematic one.
In addition to the reasons described in Sec. \ref{sec:diagramm_technique_after_shift}, many two-loop expressions become extremely unwieldy, while the renormalization procedure requires significant manual effort.
Ultimately, implementing the renormalization prescriptions to ensure UV-finiteness and regularity in $\Lambda$ and $B_0$ not only illustrates a local renormalizable field theory with nontrivial $\Lambda$-renormalization and an unconventional diagram technique, but also provides a valuable check on the final results.

\subsection{\label{sec:Non-renormalized_two-loop_diagrams} Integrals of unrenormalized perturbation theory}

Let us start with the unrenormalized action $\mathcal{S}_{\boldsymbol{B}_0}$ from \eqref{eq:MHD_action_after_shift}, where $\boldsymbol{B}_0$ yet remains unfixed. 
For such a shifted theory, we refer back to the linearized equations of motion \eqref{eq:linearized_equations_of_motion_in_polarized_media}.
As detailed in Sections \ref{sec:model_in_dynamo_regime} -- \ref{sec:corrections_to_alfven_waves}, the 1PI function $\Gamma^{v'v}$, $\Gamma^{v'b}$, $\Gamma^{b'v}$, and $\Gamma^{b'b}$ contain various structures linear in $\boldsymbol{k}$.
Notably, only the curl structure in the corresponding self-energy $\Sigma^{b'b}$ is ``problematic''.
In turn, as discussed in Sec. \ref{sec:diagramm_technique_after_shift}, the integrals corresponds to 1PI diagrams in the model \eqref{eq:MHD_action_after_shift} can be divided into two categories.
The first, called $\Lambda$-type integrals, contain only $\Lambda$-regular terms (superficial divergences).
The second, referred to as $B_0$-type integrals, consist of diagrams with subtracted
$\Lambda$-regular parts as well as diagrams that arise after the field shift (which, by definition, do not exhibit superficial divergences).
This classification implies that the curl structure in $\Sigma^{b'b}$ comprises two parts: $i \rho \nu_0 [\varepsilon_{i j l} k_l] \big(\mathcal{Q}^{[\Lambda]} + \mathcal{Q}^{[B_0]}\big)$, where 
$\mathcal{Q}^{[\Lambda]} = \mathcal{Q}^{[\Lambda]}(\Lambda, m, \epsilon)$ arises from $\Lambda$-type diagrams and $\mathcal{Q}^{[B_0]} = \mathcal{Q}^{[B_0]}(\Lambda_0, \Lambda, m, \epsilon)$ from the $B_0$-type diagrams.
Importantly, $\mathcal{Q}^{[\Lambda]}$ is entirely determined by the diagrams of the unshifted model $\mathcal{S}_0$.
Thus, the general form of the formal condition for canceling the curl contributions becomes 
\begin{eqnarray}\label{eq:equation_for_B_general_form}
    \Lambda H_1\left(\frac{g_0}{\Lambda^{2\epsilon}}, \frac{m}{\Lambda}, \epsilon\right) + \Lambda_0 H_2\left(\frac{g_0}{\Lambda_0^{2\epsilon}}, \frac{\Lambda}{\Lambda_0}, \frac{m}{\Lambda_0}, \epsilon\right) = 0, \quad
\end{eqnarray}
where we have introduced the parameter $\Lambda_0 \coloneqq B_0/\nu_0$ (which has the same dimension as $\Lambda$) and, for dimensional reasons, expressed $\mathcal{Q}^{[\Lambda]} = \Lambda H_1(g_0\Lambda^{-2\epsilon}, m/\Lambda, \epsilon)$ and $\mathcal{Q}^{[B_0]} = \Lambda_0 H_2(g_0\Lambda_0^{-2\epsilon}, \Lambda/\Lambda_0, m/\Lambda_0, \epsilon)$.

The equality in Eq. \eqref{eq:equation_for_B_general_form} is formal, since it only makes sense to discuss its renormalized analogue.
Recall that $B_0$ does not have a clear physical meaning, unlike $B$.
Nevertheless, its solution to Eq. \eqref{eq:equation_for_B_general_form} must be positive and this requirement is essentially the only potential obstacle to validating the concept of stabilizing the system through the emergence of the spontaneous field.

Before delving into renormalization, let us first outline the general structure of the resulting diagrams corresponding to the curl term in $\Gamma^{b'b}$.
The constructions discussed here are applicable to any $n$-loop integral $\mathcal{J}_n$.
However, for our purposes, we only need to consider two-loop integrals $\mathcal{J}_2$.
Corresponding one-loop integrals are discussed in Sec. \ref{sec:one_loop_approximation_after_shift}.
As before the shift, two-loop diagrams are categorized into two topological types: sunsets and double-scoop diagrams (see Fig. \ref{fig:two-loop_surface_divergent_diagrams_before_shift}).
However, the number of diagrams has increased significantly: there are now $232$ sunset diagrams and $256$ double-scoop diagrams, up from $4$ of each type previously.
In total, there are altogether $488$ two-loop diagrams contributing to $\Sigma^{b'b}$.
A more detailed classification of these diagrams is given in the Sec. \ref{sec:Two-loop_diagrams_classification}.

After integrating over frequencies, performing tensor reduction, and extracting the curl structure $i \rho [\varepsilon_{i j l} k_l] \nu_0 g_0^2$, the remaining scalar two-loop integral $\mathcal{J}_2$ over three-dimensional momenta contributing for generality to both of coefficients $\mathcal{Q}^{[\Lambda]}$ and $\mathcal{Q}^{[B_0]}$ (i.e., corresponding to some superficial divergent diagram) is given by the general expression 
\begin{align}\label{eq:general_form_for_diagram_contribution_after_shift}
    \mathcal{J}_2 = \int \limits_{m}^{\Lambda} \frac{\mbox{d}^3 p}{p^{1 + 2\epsilon}} \frac{\mbox{d}^3 q}{q^{1 + 2\epsilon}} \,\, \mathscr{F},
\end{align}
where $\mathscr{F} \coloneqq \mathscr{F}\left(p, q, (\boldsymbol{p} \cdot \boldsymbol{q}), (\boldsymbol{p} \cdot \boldsymbol{B}_0), (\boldsymbol{q} \cdot \boldsymbol{B}_0) \right)$.
Here we once again utilize the fact that the loop momenta $\boldsymbol{p}$ and $\boldsymbol{q}$ can always be assigned to lines containing the kernel $\mathbb{D}_{ij}$ from \eqref{eq:pump_function}, as it is proportional to the coupling constant $g_0$.
The parameter $m$ ensures proper IR-regularization of individual diagrams.

The presence of a superficial divergence in $\mathcal{J}_2$ indicates that the function $\mathscr{F}$ at $B_0 = 0$ contains a non-zero part denoted as $\mathscr{F}_0 \coloneqq \mathscr{F}_0(p, q, (\boldsymbol{p} \cdot \boldsymbol{q}))$.
Then the replacement $\mathscr{F} \rightarrow [\mathscr{F} - \mathscr{F}_0] + \mathscr{F}_0$ in \eqref{eq:general_form_for_diagram_contribution_after_shift}, as well as the replacement of variables $\boldsymbol{p} \rightarrow \boldsymbol{p}/\Lambda_0$, $\boldsymbol{q} \rightarrow \boldsymbol{q}/\Lambda_0$ in the integral in brackets and $\boldsymbol{p} \rightarrow \boldsymbol{p}/\Lambda$, $\boldsymbol{q} \rightarrow \boldsymbol{q}/\Lambda$ in the rest integral with $\mathscr{F}_0$, leads to the representation
\begin{align}\label{eq:separated_Lambda_and_B0_integrals}
    &\mathcal{J}_2 = \Lambda^{1 - 4 \epsilon}\int \limits_{m/\Lambda}^{1} \frac{\mbox{d}^3 p }{p^{1 + 2\epsilon}} \frac{\mbox{d}^3 q }{q^{1 + 2\epsilon}} \,\, \mathscr{F}_0 \notag \\
    &+ \Lambda_0^{1 - 4 \epsilon}\int \limits_{m/\Lambda_0}^{\Lambda/\Lambda_0} \frac{\mbox{d}^3 p}{p^{1 + 2\epsilon}} \frac{\mbox{d}^3 q}{q^{1 + 2\epsilon}}\left[\mathscr{F} - \mathscr{F}_0\right],
\end{align}
where in both integrands the indicated replacement is implied.
Next, the first integral in \eqref{eq:separated_Lambda_and_B0_integrals} denoted as $ \mathcal{J}^{[\Lambda]}$, is transformed into:
\begin{align}\label{eq:Lambda-integral}
    \mathcal{J}^{[\Lambda]}_2 = \Lambda^{1 - 4 \epsilon}\int \limits_{-1}^1 \mbox{d} z \int \limits_{m/\Lambda}^{1} \frac{\mbox{d} p }{p^{1 + 2\epsilon}} \frac{\mbox{d} q }{q^{1 + 2\epsilon}} \,\, \mathscr{F}_0,
\end{align}
where the angular variable $z$ is the same as in \eqref{eq:integrands_homogenity}, and the function $\mathscr{F}_0$ was multiplied by the density of the resulting measure: $\mathscr{F}_0 \rightarrow 8\pi^2 k^{2}q^{2}\mathscr{F}_0$.
This expression exactly matches the definition of $I_i$ provided in Eq. \eqref{eq:two_loop_bb'_linear-asymp}.
The second integral in \eqref{eq:separated_Lambda_and_B0_integrals} denoted as $ \mathcal{J}^{[B_0]}$, depends on $\boldsymbol{\hat{B}}_0$ and is transformed into: 
\begin{align}\label{eq:B-integral}
    \mathcal{J}^{[B_0]}_2 = \Lambda_0^{1 - 4 \epsilon}\int \mbox{d} \Omega \int \limits_{m/\Lambda_0}^{\Lambda/\Lambda_0} \frac{\mbox{d} p}{p^{1 + 2\epsilon}} \frac{\mbox{d} q}{q^{1 + 2\epsilon}}\left[\mathscr{F} - \mathscr{F}_0\right],
\end{align}
where the angular measure $\mbox{d} \Omega$ taking into account the asphericity introduced by $\boldsymbol{\hat{B}}_0$ is defined below.
It is precisely $\mathcal{J}^{[B_0]}_2$ integrals correspond to all the new diagrams that arise after shifting the magnetic field ($B_0$-diagrams).
For finite $m$ and $\Lambda$, the representation \eqref{eq:separated_Lambda_and_B0_integrals} is always valid.
However, within $\Lambda \to \infty$ and/or $m \rightarrow 0$, the splitting into $\Lambda$ and $B_0$-integrals is possible only after renormalization, which is discussed in Sec. \ref{sec:renormalization_in_dynamo_regime}.

The angular measure $\mbox{d} \Omega$ in Eq. \eqref{eq:B-integral} in three dimensions can be derived geometrically.
Let us choose a Cartesian coordinate system $(q_1, q_2, q_3)$ in $\boldsymbol{q}$-space, aligning one axis, say $q_3$, with the vector $\boldsymbol{\hat{B}}_0$.
Also in this space there is an external vector $\boldsymbol{p}$ (without loss of generality, we assume that the integral over $\boldsymbol{q}$ is calculated first).
Similarly, in $\boldsymbol{p}$-space we can also choose a coordinate system $(p_1, p_2, p_3)$ coaxial with $\boldsymbol{\hat{B}}_0$.
The coordinate systems $(q_1, q_2, q_3)$ and $(p_1, p_2, p_3)$ are related by rotation by a constant angle.
This rotation can always be performed, thereby superimposing one system onto the other.
Following these steps, we introduce spherical coordinates for both $\boldsymbol{p}$ and $\boldsymbol{q}$.
In these coordinates, the scalar products are expressed as
$(\boldsymbol{p} \cdot \boldsymbol{\hat{B}}_0) = p \cos{\theta_p}$, $(\boldsymbol{q} \cdot \boldsymbol{\hat{B}}_0) = q \cos{\theta_q}$, and $(\boldsymbol{p} \cdot \boldsymbol{q}) = p q z$, where $z = \cos{\theta_p}\cos{\theta_q} + \sin{\theta_p}\sin{\theta_q}\cos{(\phi_q - \phi_p)}$.
Noting thus that the integrand in Eq. \eqref{eq:B-integral} depend only on $\cos{(\phi_q - \phi_p)}$, one can reduce the number of angular integrals by one using the identity\vspace{-0.1cm}
\begin{equation}\label{eq:identity_reduces_number_of_integrals}
    \int \limits_{0}^{2\pi} \mbox{d} \phi_q\mbox{d} \phi_p f = 4 \int \limits_{0}^{\pi} \mbox{d} \vartheta (\pi - \vartheta) f = 4\pi \int \limits_{0}^{\pi} \mbox{d} \vartheta f,
\end{equation}
where $f = f\left(\cos{2\vartheta}\right)$, is any integrable function $2\vartheta \coloneqq (\phi_q - \phi_p)$.
This identity allows us to express $\mbox{d} \Omega$ in Eq. \eqref{eq:B-integral} essentially as the usual spherical measure
\begin{eqnarray}\label{eq:angular_measure}
    \mbox{d} \Omega \coloneqq \sin{\theta_p}\sin{\theta_q} \mbox{d} \theta_p \mbox{d} \theta_q\mbox{d} \vartheta, ~~ 0 \leqslant \theta_p, \theta_q, \vartheta \leqslant \pi.
\end{eqnarray}
The factor $4\pi$ and the $p^2 q^2$ term from the spherical coordinates Jacobian are incorporated into the integrand, resulting in the substitution $\left[\mathscr{F} - \mathscr{F}_0\right] \rightarrow 4\pi p^2q^2\left[\mathscr{F} - \mathscr{F}_0\right]$.

The sums of the integrals in Eqs. \eqref{eq:Lambda-integral} and \eqref{eq:B-integral} exactly determine the two-loop contributions to the coefficients $F_1$ and $F_2$ in Eq. \eqref{eq:equation_for_B_general_form}, forming in linear-in-$\boldsymbol{k}$ part of $\Sigma^{b'b}$ a two-loop curl term $\Sigma_{2\,ij}^{\mathrm{curl}}$:
\begin{align}\label{eq:two-loop_Lambda_and_B_curl_coefficients}
    &\Sigma_{2\,ij}^{\mathrm{curl}} = \rho \nu_0 [i\varepsilon_{i j l} k_l]g_0^2  \left(\sum \limits_{s = 1}^{8}\mathcal{J}^{[\Lambda]}_{2s} + \sum \limits_{s = 1}^{488} \mathcal{J}^{[B_0]}_{2s}\right).
\end{align}
Here, the first sum runs over contributions from $8$ superficially divergent diagrams depicted in Fig. \ref{fig:two-loop_surface_divergent_diagrams_before_shift}, while the second includes contributions from all $488$ two-loop $\Sigma^{b'b}$ diagrams.


\subsection{\label{sec:Two-loop_diagrams_classification} Classification of the \texorpdfstring{$\Sigma^{b'b}$}{} two-loop diagrams}

As just noted, at the two-loop level, $488$ diagrams contribute to $\Sigma^{b'b}$, but we only require the leading linear term in their expansion in
$\boldsymbol{k}$ that generates the curl term.
Moreover, when solving Eq. \eqref{eq:equation_for_B_general_form}, it is crucial to isolate the contributions present at $B_0 = 0$ (i.e., those computed in Sec. \ref{sec:renormalization_without_shift}), which poses additional challenges.

\begin{figure*}
    \centering
    \begin{overpic}[percent,grid=false,tics=2,width=1\linewidth]{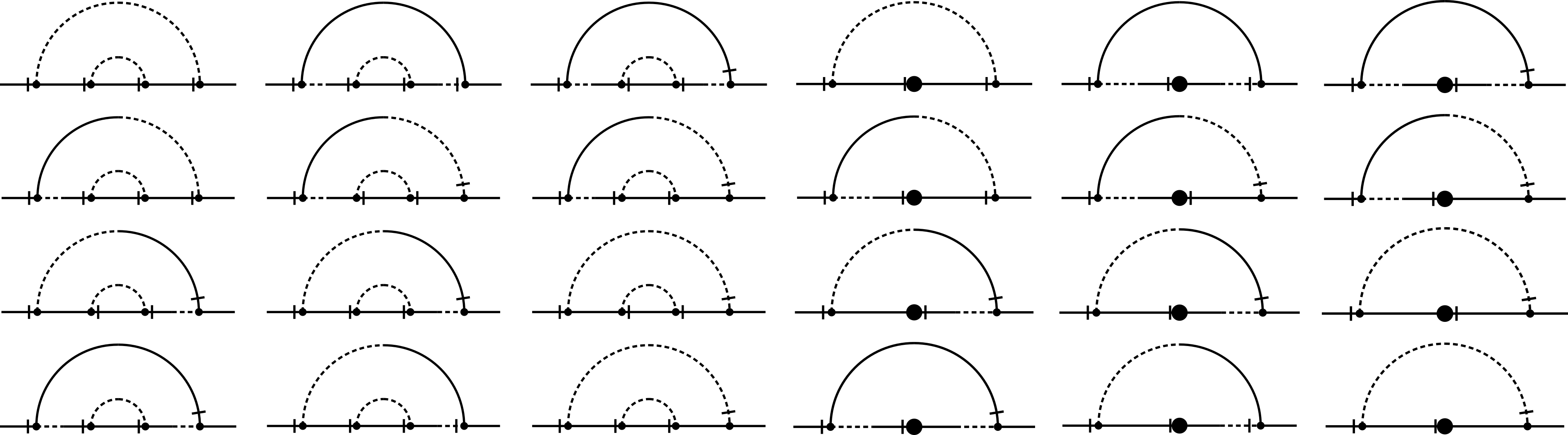}
    \put(0,27){\scalebox{.8}{$\displaystyle D_{\mathrm{S}1}^{(2)}$}}
    \put(17,27){\scalebox{.8}{$\displaystyle D_{\mathrm{S}2}^{(2)}$}}
    \put(34,27){\scalebox{.8}{$\displaystyle D_{\mathrm{S}3}^{(2)}$}}
    \put(50.5,27){\scalebox{.8}{$\displaystyle D_{\mathrm{add}1}^{(2)}$}}
    \put(67.5,27){\scalebox{.8}{$\displaystyle D_{\mathrm{add}2}^{(2)}$}}
    \put(84.5,27){\scalebox{.8}{$\displaystyle D_{\mathrm{add}3}^{(2)}$}}
    \put(0,19.5){\scalebox{.8}{$\displaystyle D_{\mathrm{S}4}^{(2)}$}}
    \put(17,19.5){\scalebox{.8}{$\displaystyle D_{\mathrm{S}5}^{(2)}$}}
    \put(34,19.5){\scalebox{.8}{$\displaystyle D_{\mathrm{S}6}^{(2)}$}}
    \put(50.5,19.5){\scalebox{.8}{$\displaystyle D_{\mathrm{add}4}^{(2)}$}}
    \put(67.5,19.5){\scalebox{.8}{$\displaystyle D_{\mathrm{add}5}^{(2)}$}}
    \put(84.5,19.){\scalebox{.8}{$\displaystyle D_{\mathrm{add}6}^{(2)}$}}
    \put(0,12){\scalebox{.8}{$\displaystyle D_{\mathrm{S}7}^{(2)}$}}
    \put(17,12){\scalebox{.8}{$\displaystyle D_{\mathrm{S}8}^{(2)}$}}
    \put(34,12){\scalebox{.8}{$\displaystyle D_{\mathrm{S}9}^{(2)}$}}
    \put(50.5,12){\scalebox{.8}{$\displaystyle D_{\mathrm{add}7}^{(2)}$}}
    \put(67.5,12){\scalebox{.8}{$\displaystyle D_{\mathrm{add}8}^{(2)}$}}
    \put(84.5,12){\scalebox{.8}{$\displaystyle D_{\mathrm{add}9}^{(2)}$}}
    \put(0,5){\scalebox{.8}{$\displaystyle D_{\mathrm{S}10}^{(2)}$}}
    \put(17,5){\scalebox{.8}{$\displaystyle D_{\mathrm{S}11}^{(2)}$}}
    \put(34,5){\scalebox{.8}{$\displaystyle D_{\mathrm{S}12}^{(2)}$}}
    \put(50.5,5){\scalebox{.8}{$\displaystyle D_{\mathrm{add}10}^{(2)}$}}
    \put(67.5,5){\scalebox{.8}{$\displaystyle D_{\mathrm{add}11}^{(2)}$}}
    \put(84.5,5){\scalebox{.8}{$\displaystyle D_{\mathrm{add}12}^{(2)}$}}
    \end{overpic}
    \caption{
    Two-loop type S diagrams along with the $h_0$-correction diagrams.
    Here, all symmetry coefficients are equal to $1$.
    }    \label{fig:diagrams_with_divergent_subgraph_and_their_counterterms}
\end{figure*}

Thus, the corresponding diagrams can be classified into non-overlapping types based on their behavior as $m \to 0$ and $\Lambda \to \infty$:
\begin{itemize}[nosep, left=0pt .. \parindent]
    \item Type \verb|F| (Finite). 
    These diagrams contribute only to $B_0$-integrals without divergences, computed directly at $m = 0$ and remaining finite as $\Lambda \to \infty$.
    In total, there are $63$ sunset diagrams and $243$ double-scoop diagrams.
    \item Type \verb|IR| (Infrared).
    These sunset diagrams also contribute solely to the $B_0$-integrals but exhibit singularities as $m \to 0$. 
    There are $72$ diagrams whose individual IR divergences cancel in the sum.
    \item Type \verb|P| (Poles).
    These diagrams lack superficial divergences but contain $\epsilon$-divergent subgraphs (from $\Gamma^{v'v}$ and $\Gamma^{v'bb}$) that do not diverge as $m \to 0$.
    This group includes $82$ sunset and $9$ double-scoop diagrams.
    \item Type \verb|L| (Lambda).
    This type includes $4$ double-scoop and $3$ sunset diagrams (present in the unshifted theory) that immediately generate power-law contributions in  $\Lambda$ and $B_0$.
    All Type \verb|L| diagrams are shown in Fig. \ref{fig:two-loop_surface_divergent_diagrams_before_shift}, which also includes ``special'' diagram $D_{\mathrm{L}2}^{(2)}$ placed in the next class.
    Some Type \verb|L| diagrams (labeled $D_{\mathrm{L}1}^{(2)}$ and $D_{\mathrm{L}3}^{(2)}$) contain $\epsilon$-poles and contribute to the $\epsilon$-renormalization process.
    \item Type \verb|S| (Subgraph).
    This class comprises $12$ sunset diagrams with a one-loop $\Gamma^{b'b}$ subgraph that, besides the standard $\epsilon$-divergences, also possesses a power-law $\Lambda$-divergence.
    One diagram, $D_{\mathrm{L}2}^{(2)} \coloneqq D_{\mathrm{S}1}^{(2)}$, comes from the pre-shift theory and contributes to both $\Lambda$- and $B_0$-integrals, while the remaining $11$ appear after the shift.
    All such diagrams are shown in Fig. \ref{fig:diagrams_with_divergent_subgraph_and_their_counterterms}.
\end{itemize}

In accordance with this classification, Appendix \ref{app:supplementary_info_about_two-loop_Gammab'b_diagrams} contains human-readable files with all explicit two-loop expressions contributing to the curl term in $\Sigma^{b'b}$.
Within these files the sunset topology is referred to as topological type \verb|I1| and the double-scoop topology as type \verb|I2|.

\begin{widetext}
Symbolically, the two-loop curl contributions to $\Sigma^{b'b}$ decomposed into sums of terms corresponding to the aforementioned types for each topology, can be represented as
\begin{eqnarray}\label{eq:two_loop_curl_b'b_self-energy_after_shift}
     &\Sigma_{2\, ij}^{\mathrm{curl}} = [\text{all 488 two-loop diagrams from App. \ref{app:supplementary_info_about_two-loop_Gammab'b_diagrams}}] = \rho \nu_0 [i\varepsilon_{i j l} k_l] \nu_0 g_0^2 \left( \mathcal{J}_2^{\mathrm{F}} +  \mathcal{J}_2^{\mathrm{IR}} + \mathcal{J}_2^{\mathrm{P}} + \mathcal{J}_2^{\mathrm{L}} + \mathcal{J}_2^{\mathrm{S}} \right), \notag \\
     &\mathcal{J}_2^{\mathrm{F}} \coloneqq \sum \limits_{s \in \Pi_1} J_{2s}, \quad \mathcal{J}_2^{\mathrm{IR}} \coloneqq \sum \limits_{s \in \Pi_2} J_{2s}, \quad \mathcal{J}_2^{\mathrm{P}} \coloneqq  \sum \limits_{s \in \Pi_3} J_{2s}, \quad \mathcal{J}_2^{\mathrm{L}} \coloneqq \sum \limits_{s \in \Pi_4} J_{2s}, \quad \mathcal{J}_2^{\mathrm{S}} \coloneqq \sum \limits_{s \in \Pi_5} J_{2s}, 
\end{eqnarray}
where the sum over set $\Pi_i$ denotes the sum over all diagrams of a certain topology (\verb|I1| or \verb|I2|) and a certain type (\verb|F|, \verb|IR|, \verb|P|, \verb|L|, or \verb|S|) from the corresponding files in App. \ref{app:supplementary_info_about_two-loop_Gammab'b_diagrams}.
Concretely, $\Pi_i$ are defined as: $\Pi_1 \coloneqq \{\verb|I1-F|,\, \verb|I2-F|\}$, $\Pi_2 \coloneqq \{\verb|I1-IR|\}$, $\Pi_3 \coloneqq \{\verb|I1-P|,\, \verb|I2-P|\}$, $\Pi_4 \coloneqq \{\verb|I1-L|,\, \verb|I2-L|\}$ and $\Pi_5 \coloneqq \{\verb|I1-S|\}$. 
\end{widetext}

Each term in the second equality in Eq. \eqref{eq:two-loop_integral_after_rescaling_by_B0} must be treated separately, with the integral $J_{2s}$ in each case defined according to representation
\begin{align}\label{eq:two-loop_integral_after_rescaling_by_B0}
    &J_{2s} = \Lambda_0^{1 - 4 \epsilon}\int \mbox{d} \Omega \int \limits_{m/\Lambda_0}^{\Lambda/\Lambda_0} \frac{\mbox{d} p\, \mbox{d} q}{(p q)^{1 + 2\epsilon}} \,\, \mathscr{F}_0(p, q, \Omega) \notag \\
    &+ \Lambda_0^{1 - 4 \epsilon}\int \mbox{d} \Omega \int \limits_{m/\Lambda_0}^{\Lambda/\Lambda_0} \frac{\mbox{d} p\, \mbox{d} q}{(pq)^{1 + 2\epsilon}} \, \left[\mathscr{F} - \mathscr{F}_0\right](p, q, \Omega),
\end{align}
which is derived from the general formula \eqref{eq:general_form_for_diagram_contribution_after_shift} by first scaling $\boldsymbol{p} \rightarrow \boldsymbol{p}/\Lambda_0$, $\boldsymbol{q} \rightarrow \boldsymbol{q}/\Lambda_0$ and then by explicitly identifying the part that remains finite at $B_0 = 0$ (in those integrals where such a part is present).\footnote{Since $\mathscr{F}_0$ always depends on only one angular variable $z = \cos{\theta_p}\cos{\theta_q} + \sin{\theta_p}\sin{\theta_q}\cos{2\vartheta}$ and the function $\mathscr{F}_0$ (unlike $\mathscr{F}$) always has the property \eqref{eq:integrands_homogenity}, it is easy to see that the first integral in \eqref{eq:Lambda-integral} exactly coincides with \eqref{eq:two-loop_integral_after_rescaling_by_B0}.}
Hereinafter, for clarity, the argument $\Omega$ in all functions denotes three angular variables: $\Omega \coloneqq (\theta_p, \theta_q, \vartheta)$.
Let us emphasize that the dependence on $\vartheta$ in integrands comes only through the argument $\cos{2\vartheta}$.
The abundance of various singularities in integrals \eqref{eq:two-loop_integral_after_rescaling_by_B0} entails the necessity of dividing the integration domain into subdomains where only singularities of a certain type are localized and considering integrals over these subdomains separately.
Such partition is given by the following formula: 
\begin{align}\label{eq:masterform} 
    &\mathcal{I} = \int \frac{\mbox{d} \Omega}{\Lambda_0^{-1 + 4 \epsilon}} \left[\rule{0cm}{1cm}\right.\int\limits_{\frac{m}{\Lambda_0}}^{1}\mbox{d} p \int\limits_{\frac{m}{\Lambda_0}}^{1}\mbox{d} q + \int\limits_{\frac{m}{\Lambda_0}}^{1} \mbox{d} p \int\limits_{1}^{\frac{\Lambda}{\Lambda_0}} \mbox{d} q + \int\limits_{1}^{\frac{\Lambda}{\Lambda_0}} \mbox{d} p \int\limits_{\frac{m}{\Lambda_0}}^{1} \mbox{d} q \notag \\
    &+\int\limits_{1}^{\frac{\Lambda}{\Lambda_0}} \mbox{d} p \int\limits_{1}^{\frac{\Lambda}{\Lambda_0}} \mbox{d} q \left.\rule{0cm}{1cm}\right]\frac{\mathscr{F}(p, q, \Omega)}{(p q)^{1 + 2\epsilon}} \coloneqq \mathcal{I}_1 + \mathcal{I}_2 + \mathcal{I}_3 + \mathcal{I}_4.
\end{align}
It is implicitly assumed here that $\Lambda/\Lambda_0 > 1$.
However, this assumption is not essential.
The integral $\mathcal{I}$ in Eq.~\eqref{eq:masterform} can be either some individual integral of the type $J_{2s}$, or their partial sum, e.g., $\mathcal{J}_2^{\mathrm{IR}}$.
The integral $\mathcal{I}_1$ may depend only on the ratio $m/\Lambda_0$ and $\epsilon$, while $\mathcal{I}_2$, $\mathcal{I}_3$, and $\mathcal{I}_4$ depend on $\Lambda/\Lambda_0$, $m/\Lambda_0$ and $\epsilon$. 
It is more convenient to conduct further practical discussion of singularities in any integral $\mathcal{I}$ in terms of integrals $\mathcal{I}_i$.

As already noted, for integrals contributing to $\mathcal{J}_2^{\mathrm{F}}$, the corresponding integrals $\mathcal{I}_i^{\mathrm{F}}$ are free of complications and can be computed numerically for a given $u_0$.
In turn, the integrals contributing to $\mathcal{J}_2^{\mathrm{IR}}$ exhibit issues in the limit $m \to 0$.
Technically, these singularities manifest as the following asymptotic behavior of the integrand $\mathscr{F}$ in $p$:
\begin{equation}\label{eq:expansionIR}
\frac{\mathscr{F}(p, q)}{(p q)^{1 + 2\epsilon}} = \frac{1}{(p q)^{2\epsilon}}\left(\frac{\mathscr{C}_{1} (q)}{p} + \mathcal{O}(p)\right),
\end{equation}
where $\mathscr{C}_{1}(q)$ is regular in $q$.
Here and further in similar expressions we omit for brevity the dependence of all quantities on angular variables $\Omega$.
Since the integrand $\mathscr{F}$ shows no further singularities, isolating these divergences requires subtracting the $p\to 0$ asymptotic \eqref{eq:expansionIR} from the corresponding integrals $\mathcal{I}_1^{\mathrm{IR}}$ and $\mathcal{I}_2^{\mathrm{IR}}$.
After this one can see that the sum of all $72$ functions $\mathscr{C}_{1} (q, \Omega)/p$ remaining after subtractions is equal to zero (see App. \ref{app:supplementary_info_about_two-loop_Gammab'b_diagrams}).
The treatment of the remaining three types of integrals, which is considerably more labor-intensive, is presented in the next Section.

\subsection{\label{sec:renormalization_in_dynamo_regime}Two-loop renormalization in presence of \texorpdfstring{$B$}{B}}

\subsubsection{\label{sec:one_loop_approximation_after_shift}Exact one-loop approximation}

Before proceeding to two-loop estimates, we have to first refine the one-loop result for the curl term part of self-energy correction $\Sigma^{b'b}$.
In \cite{Adzhemyan1987}, only the result accurate up to $\mathcal{O}(\epsilon^0)$ is given.
Since the curl terms in $\Sigma^{b'b}$ originate at the one-loop level, they start to correct the propagator lines \eqref{eq:propagator_matrix_after_shift} only in higher loops.
Consequently, at the one-loop level, the relevant analysis can be conducted directly using diagrams constructed from the propagator lines \eqref{eq:propagator_matrix_after_shift}.
At this level, $\Sigma^{b'b}$ comprises four diagrams.
One of these is the ``old'' diagram given in Eq. \eqref{eq:one_loop_curl-term_diagram} (with updated propagator lines) that carries the $\Lambda$-divergence as shown in Eqs. \eqref{eq:one_loop_bb'_linear-asymp} -- \eqref{eq:one_loop_curl-term}, along with a UV-finite $B_0$-contribution.
The other three are purely $B_0$-diagrams.
Graphically, the one-loop expression for $\Sigma^{b'b}$ takes the form:
\begin{align}\label{eq:one-loop_diagrams_after_shift}
     &\Sigma_1^{b'b}  = ~\raisebox{0ex}{\includegraphics[width=3.0cm]{./oneloopCurlTermDiag.png}} ~+~ \raisebox{0ex}{\includegraphics[width=3.0cm]{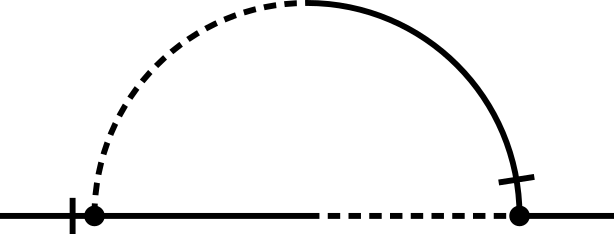}} \notag \\
     &+ ~ \raisebox{0ex}{\includegraphics[width=3.0cm]{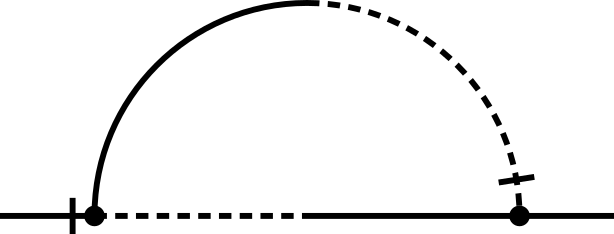}} ~+~ \raisebox{0ex}{\includegraphics[width=3.0cm]{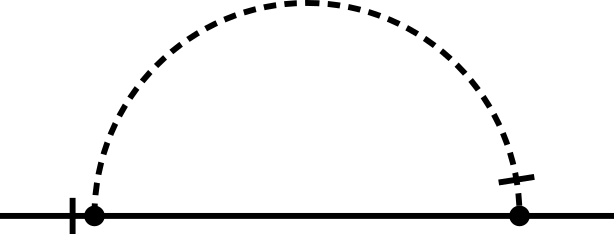}}.
\end{align}
As before, we calculate the linear-in-$\boldsymbol{k}$ part of $\Sigma_1^{b'b}$, denoted as $\Sigma_{1}^{\mathrm{exact}}$, at zero external frequency and $m = 0$.
It consists of two contributions -- curl and exotic:
\begin{align}\label{eq:two-terms_in_exact_one-loop_linear_in_k_self-energy}
    &\Sigma_{1\, ij}^{\mathrm{exact}} = \Sigma_{1\, ij}^{\mathrm{curl}} + \Sigma_{1\, ij}^{\mathrm{exotic}}, 
\end{align}
where the $\Sigma_{1}^{\mathrm{curl}}$ and $\Sigma_{1}^{\mathrm{exotic}}$ contributions at the arbitrary space dimension $d$ have the forms
\begin{eqnarray}
    &\Sigma_{1\, ij}^{\mathrm{curl}} = \rho \nu_0 g_0 \left[i\varepsilon_{i j l} k_l\right] \hspace{-0.07cm}\left(\hspace{-0.05cm}\frac{-4 C(d)\Lambda^{1 - 2\epsilon}}{3(1 + u_0)(1 - 2\epsilon)} + \mathcal{J}_{1 \,\mathrm{curl}}^{[B_0]} \hspace{-0.05cm}\right)\hspace{-0.09cm}, \quad~~\, \label{eq:one-loop_сurl_self-energy}\\
    &\Sigma_{1\, ij}^{\mathrm{exotic}} = \rho \nu_0 g_0\left[i\varepsilon_{i m l} k_l \hat B_{0j} \hat B_{0m} \right] \qquad \qquad \qquad \qquad \quad~~ \notag\\
    &\times\left(\frac{-2 (d - 3)(d - 1) C(d)\Lambda^{1 - 2\epsilon}}{3(1 + u_0)(1 - 2\epsilon)} + \mathcal{J}_{1\,\mathrm{exotic}}^{[B_0]} \right). \quad \label{eq:one-loop_exotic_self-energy}
\end{eqnarray}
Here, $C(d) \coloneqq S_{d - 1}/(2 \pi)^d(d - 1)$, $S_{d} \coloneqq 2\pi^{d/2}\big/\Gamma(d/2)$ is the surface area of a unit sphere in $d$-dimensional space, $\Gamma(z)$ is the Euler gamma function.
The parameter $\Lambda_0$ is defined in \eqref{eq:equation_for_B_general_form}, and the unit vector $\hat{\boldsymbol{B}}_0$ follows from \eqref{eq:general_structure_of_linear_asymptotic_of_Gamma}. 
The one-loop integrals $\mathcal{J}_{1 \,\mathrm{curl}}^{[B_0]}$ and $\mathcal{J}_{1\,\mathrm{exotic}}^{[B_0]}$ have structures similar to \eqref{eq:B-integral}:
\begin{eqnarray}
    &\mathcal{J}_{1 \,\mathrm{curl}}^{[B_0]} = \frac{C(d)}{\Lambda_0^{ 2\epsilon - 1}}\hspace{-0.1cm}\int \limits_{-1}^{1} \mbox{d} z \hspace{-0.15cm}\int \limits_{0}^{\Lambda/\Lambda_0}\hspace{-0.15cm} \frac{\mbox{d} p}{p^{2\epsilon}}\frac{z^2 (1 - z^2) (3 p^2 u_0 + z^2)}{\left(1 + u_0\right) \left(p^2 u_0+z^2\right)^2}, \quad~~ \label{eq:one-loop_сurl_B0-integral}\\ 
    &\mathcal{J}_{1\,\mathrm{exotic}}^{[B_0]} = \frac{C(d)}{\Lambda_0^{ 2\epsilon - 1}}\hspace{-0.1cm}\int \limits_{-1}^{1} \mbox{d} z \hspace{-0.1cm}\int \limits_{0}^{\Lambda/\Lambda_0}\hspace{-0.1cm} \frac{\mbox{d} p}{p^{2\epsilon}} \Bigg[\frac{z^2 \left(d z^2 - 1\right)}{(1 + u_0)^2} \qquad \qquad \quad \notag \\ 
    &\times \frac{\left(u_0 \left(2 p^2 u_0 + 3 p^2 + z^2\right) + 2 z^2\right)}{\left(p^2
   u_0 + z^2\right)^2} \Bigg]. \label{eq:one-loop_exotic_B0-integral}
\end{eqnarray}

First, let us emphasize that Eq.~\eqref{eq:one-loop_exotic_self-energy} clearly shows that the system can be stabilized by transitioning to a state with a nonzero mean magnetic field \textit{only} at $d = 3$.
Specifically, for dimensions other than $d = 3$, the exotic contributions cause extra instability that, along with the curl-term instability, cannot be eliminated by merely adjusting the magnitude of spontaneous magnetic field.
Also, it is instructive to consider the calculation of one loop stabilizing mean field firstly in terms of unrenormalized perturbation theory.
Its magnitude is determined by setting $\Sigma_{1\, ij}^{\mathrm{curl}}$ to zero.
At $d = 3$, we obtain:
\begin{eqnarray}\label{eq:one-loop_exact_calculation_for_B0}
    \Lambda \left(\frac{g_0}{\Lambda^{ 2\epsilon}}\right) c_1^{[\Lambda]}(u_0, \epsilon)  + \Lambda_0 \left(\frac{g_0}{\Lambda_0^{ 2\epsilon}}\right) \mathfrak{F}\left(\frac{\Lambda}{\Lambda_0}, u_0, \epsilon\right) = 0, \qquad
\end{eqnarray}
where $c_1^{[\Lambda]}(u_0, \epsilon)$ from \eqref{eq:one_loop_curl-term} coincides with the first term in the brackets in \eqref{eq:one-loop_сurl_B0-integral} at $d = 3$ and the function $\mathfrak{F}(\lambda, u_0, \epsilon)$ is precisely the integral \eqref{eq:one-loop_сurl_B0-integral} divided by $\Lambda_0^{1 - 2\epsilon}$, interpreted as a function of the upper limit $\lambda = \Lambda/\Lambda_0$.
In principle, $\mathfrak{F}(\lambda, u_0, \epsilon)$ can be expressed explicitly in terms of hypergeometric functions of the dimensionless argument $\lambda \sqrt{u_0}$.
However, for our purposes, it suffices to note that by substituting the ansatz $B_0 = c \Lambda \nu_0 \sqrt{u_0}$ into $\Lambda_0$, we obtain a dimensionless equation for $c$:
\begin{align}\label{eq:one-loop_dimensionless_equation_for_B0}
   c_1^{[\Lambda]}(u_0, \epsilon) + c\,\mathfrak{F}(1/c, u_0, \epsilon) = 0. 
\end{align}
It is straightforward to show that, for any $\epsilon \in [0,2]$, excluding a few singular points (discussed further), this equation has a solution only in the limit $c \to \infty$.
In other words, the absolute value $|c_1^{[\Lambda]}(u_0, \epsilon)|$ is precisely a horizontal asymptote for $c\,\mathfrak{F}(1/c, u_0, \epsilon)$.
Thus, within unrenormalized one-loop calculations, $B_0$ formally equal to infinity but is in fact an \textit{exact} result.
In quantum field theory, a divergent unrenormalized vacuum average is a well-known phenomenon.
As noted in Sec.~\ref{sec:stabilization}, the renormalized field $B$ has direct physical meaning.

Technically, one-loop renormalization involves replacing all bare parameters in Eq.~\eqref{eq:transition_from_non-renormalized_to_basic_action} with the corresponding UV-finite renormalized ones, where the one-loop $h_0$ is identified with $\Lambda(g_0\Lambda^{-2\epsilon})c_1^{[\Lambda]}(u_0, \epsilon)$.
Furthermore, we extend the integrals \eqref{eq:one-loop_сurl_B0-integral} and \eqref{eq:one-loop_exotic_B0-integral} to an infinite upper limit
(since $\Lambda/B \to \infty$ for renormalized $B$, whereas $\Lambda/B_0$ remains constant as $\Lambda \to \infty$).
These integrals can then be evaluated analytically for arbitrary $u$ and $\epsilon$, yielding
\begin{align}
&\mathcal{J}_{1 \,\mathrm{curl}}^{[B]} = \frac{\pi u^{-\frac{1}{2} + \epsilon}C(d) B^{1 - 2\epsilon} \nu^{-1 + 2\epsilon}}{(1 + u)(2 - \epsilon)\cos{(\pi \epsilon)}}, \label{eq:one-loop_сurl_B-integral}\\
&\mathcal{J}_{1\,\mathrm{exotic}}^{[B]} = (d - 2 + \epsilon(1 - d)) \mathcal{J}_{1 \,\mathrm{curl}}^{[B]}. \label{eq:one-loop_exotic_B-integral}
\end{align}
At $\epsilon = 0$, the condition for the renormalized field $B$ becomes:
\begin{align}\label{eq:determining_one_loop_B}
    h + g\left(\frac{B}{\nu}\right) c_1^{[B]}(u, 0) = 0
\end{align}
which solution reproduces precisely \eqref{eq:one-loop_B0}.
By construction, $h < 0$ and $h \propto g$.
Note that, by construction, $h < 0$ and $h \propto g$.
Analogously to $c_1^{[\Lambda]}$, in \eqref{eq:determining_one_loop_B} we have introduced an exact one-loop $B$-coefficient:
\begin{align}\label{eq:one-loop_three_dimensonal_j_coeffisient}
    c_1^{[B]}(u, \epsilon) = \frac{u^{-\frac{1}{2} + \epsilon}}{(2 - \epsilon)\cos{(\pi \epsilon)}}\left(\frac{1}{8\pi(1 + u)}\right).
\end{align}

The obtained exact one-loop expressions allow us to address briefly the analytic continuation of the one-point Green function $B(\epsilon)$ from the vicinity of $\epsilon = 0$ to the ``physical'' region $\epsilon \geqslant 2$.
As is typical in turbulence theory, individual renormalized diagrams contain IR singularities at finite values of $\epsilon$, appearing in the form of poles.
In this case, the poles at $\epsilon = 1/2$, $\epsilon = 1$ and $\epsilon = 2$ arise.
The pole at $\epsilon = 2$ is standard and reflects the power-law representation of the $\delta$-shaped pumping \eqref{eq:pump_function}, commonly canceled by the factor $(2 - \epsilon)$ in the amplitude $g \nu^3 \mu^{2\epsilon}$ (see, e.g., \cite{Adzhemyan1989, Adzhemyan2003}).
The singularities of individual diagrams at $\epsilon = 1$ cancel out in the sum (see Eqs. \eqref{eq:one-loop_сurl_B-integral} and \eqref{eq:one-loop_exotic_B-integral}), which is a consequence of the Galilean symmetry of the theory \cite{Adzhemyan2003}.
Concerning the pole at $\epsilon = 1/2$, one can only speculate on whether a path exists in the complex $\epsilon$-plane that would allow its analytic bypass.
These considerations suggest that the function  $B(\epsilon)$ (assumed analytic on a connected domain including vicinity of points $\epsilon = 1$ and $\epsilon = 2$) can be approximated by its asymptotic $\epsilon$-expansion.
As always, this assumption can only be validated by comparing the final results with experiment.


\subsubsection{\label{sec:Lambda-renormalization}\pdfmath{\Lambda}-renormalization}

Now we can return to the two-loop renormalization of the curl diagrams in $\Sigma^{b'b}$.
Conceptually, the procedure of $\Lambda$-renormalization for the model after the shift $\boldsymbol{b} \to \boldsymbol{b} + \boldsymbol{B}_0$ is analogous to that described in Sec. \ref{sec:renormalization_without_shift}.
We begin by examining the self-energy $\Sigma^{b'b}$, isolating the $\Lambda$-divergent contributions, and associating them with the corresponding propagator term in the Dyson equation for $\Gamma^{b'b}$.
As a result, these contributions are removed from $\Sigma^{b'b}$.
This yields an equation similar to \eqref{eq:dyson_eq_for_Gammab'b_h0_to_propagator}, but with all quantities taken from the shifted theory \eqref{eq:MHD_action_after_shift}.
Analogous to the transition from \eqref{eq:MHD_action_without_shift} to \eqref{eq:MHD_action_with_Gammabb'_curl-terms}, the presence of $\Lambda$-regular contributions in the $\underline{\Delta}^{b'b}$ is equivalent to adding a curl contribution, $\rho \nu_0 {h}_{0} \boldsymbol{b'} \cdot \left(\boldsymbol{\nabla} \times \boldsymbol{b}\right)$, to the action \eqref{eq:MHD_action_after_shift} (implying simultaneous subtraction of the curl-terms from the $\Sigma^{b'b}$-diagrams):
\begin{align}\label{eq:MHD_action_with_seed_curl-terms_after_shift}
    \mathcal{S}_{\boldsymbol{B}_0} - \rho \nu_0 {h}_{0} \boldsymbol{b'} \cdot \left(\boldsymbol{\nabla} \times \boldsymbol{b}\right).
\end{align}
Here, ${h}_{0}$ remains as in \eqref{eq:MHD_action_with_Gammabb'_curl-terms}, since the field shift does not introduce additional superficial divergences.
As before, a counterterm $\delta h$ has to be added to $h_0$ in \eqref{eq:MHD_action_with_seed_curl-terms_after_shift} to cancel both the subgraph divergences in diagrams corresponding to the lines of the original theory \eqref{eq:MHD_action_after_shift} and the new divergences arising from correction diagrams.
The corresponding propagator $\underline{\Delta}^{b'b}$ in \eqref{eq:MHD_action_with_seed_curl-terms_after_shift} takes the form 
\begin{eqnarray}\label{eq:bb'_prop_with_curl-terms_after_shift}
     &\underline{\Delta}_{i j}^{b'b} = \frac{\alpha(\omega, k)\xi(\omega, k)}{\eta(\omega, k)} \mathbb{P}_{i j}(\boldsymbol{k}) + \frac{\alpha^2(\omega, k)\rho \nu_0 {h}_{0} k}{\eta(\omega, k)} \mathbb{H}_{i j}(\boldsymbol{k}), \notag \\
     & \eta(\omega, k) \coloneqq \xi^2(\omega, k) - \alpha^2(\omega, k)\rho^2 \nu_0^2 {h}_{0}^2 k^2.
\end{eqnarray}

Note that incorporating the curl contribution $\rho \nu_0 {h}_{0} \boldsymbol{b'} \cdot \left(\boldsymbol{\nabla} \times \boldsymbol{b}\right)$ into the action \eqref{eq:MHD_action_after_shift} modifies every element of the propagator matrix \eqref{eq:propagator_matrix_after_shift}, not just $\Delta^{b'b}$.
Thus, the corresponding $h_0$ corrections -- analogous to \eqref{eq:schematic_image_of_bb'_propagator_with_curl_terms} and leading to the correction diagrams in Fig. \ref{fig:diagrams_with_divergent_subgraph_and_their_counterterms} -- are best obtained by formally applying the prescriptions of the $R'$-operation.
This involves contracting the $\Lambda$-divergent subgraphs in the two-loop diagrams to a point, thereby generating the required lines in correction diagrams.
This approach is justified by our earlier finding that introducing a curl term with the opposite sign converts these correction diagrams into counterterm diagrams that completely cancel the curl contributions (recall that in this formulation, the curl terms are not subtracted from the $\Sigma^{b'b}$-diagrams).

Similar to \eqref{eq:two-loop_Gammabb'_with_curl-terms}, the Dyson equation for $\Gamma^{b'b}$  in the shifted theory \eqref{eq:MHD_action_with_seed_curl-terms_after_shift} is given by
\begin{align}\label{eq:eq:two-loop_Gammabb'_with_curl-terms_after_shift}
    \Gamma_{i j}^{b'b} =& -\left(\underline{\Delta}_{i j}^{b'b} \right)^{-1} + \Sigma_{1\,ij}^{\mathrm{curl}} + \Sigma_{2\,i j}^{\mathrm{curl}} \notag \\
    &+ [D_{\mathrm{add}}^{(2)} \text{ correction diagrams from Fig. \ref{fig:diagrams_with_divergent_subgraph_and_their_counterterms}}],
\end{align}
where the propagator $\underline{\Delta}_{i j}^{b'b}$ from Eq. \eqref{eq:bb'_prop_with_curl-terms_after_shift} assumes the replacement $h_0 \to h_0 + \delta h$ and $\Sigma_{1\,ij}^{\mathrm{curl}}$ is defined in \eqref{eq:one-loop_сurl_self-energy}.

Extracting the curl part from $\Gamma_{i j}^{b'b}$ in \eqref{eq:eq:two-loop_Gammabb'_with_curl-terms_after_shift} leads to an expression analogous to Eq. \eqref{eq:Lambda_renormalized_Gammab'b_in_ordinary_model}, but with an additional $B_0$-curl term:
\begin{eqnarray}
\label{eq:Lambda_renormalized_Gammab'b_after_shift}
   \Gamma_{i j}^{b'b} = \rho\nu_0\Bigl(h_0 + j_0\Bigr) k \mathbb{H}_{i j}(\boldsymbol{k}) + \mathcal{O}(k).
\end{eqnarray}
Here, the $h_0$ term originates from the propagator,
while $j_0$ remains in the self-energy. 
Recall that, following the emergence of $B_0$, in $\Gamma^{b'b}$ other linear-in-$\boldsymbol{k}$ terms appear (e.g., exotic terms etc.) which are not ``dangerous''.
Using this equation, one can carry out a complete $\Lambda$-renormalization of the model (unlike Eq. \eqref{eq:Lambda_renormalized_Gammab'b_in_ordinary_model}) by treating it as an equation for $B_0$, which is thereby defined at $B_0$ already at $m = 0$ with finite $\Lambda$.

Proceeding to the $j_0$ calculation, note that it, together with $h_0$, is extracted from the self-energy in Eq. \eqref{eq:eq:two-loop_Gammabb'_with_curl-terms_after_shift}:
\begin{align}\label{eq:total_two-loop_purely_power-law_curl_term}
    \rho \nu_0 \big[i\varepsilon_{i j l}k_l \big]\bigl(h_0 + j_0\bigr) = \Sigma_{1\,ij}^{\mathrm{curl}} + R'_{\Lambda}\left[\Sigma_{2\,ij}^{\mathrm{curl}} \right],
\end{align}
where $R'_{\Lambda}\left[\bullet\right]$ is as in Eq. \eqref{eq:two_loop_bb'_linear-asymp}.
The subgraph divergences removed from $\Sigma_{2\,ij}^{\mathrm{curl}}$ by $R'_{\Lambda}$, together with contributions from the correction diagrams, are canceled in Eq. \eqref{eq:eq:two-loop_Gammabb'_with_curl-terms_after_shift} by an appropriate choice of the counterterm $\delta h$.

The parameter $j_0$ generally has the same structure as $h_0$ and is defined perturbatively by
\begin{align}\label{eq:j0_structure}
    j_0 = \Lambda_0 \sum \limits_{n = 1}^{\infty} \left(g_0 \Lambda_0^{-2\epsilon}\right)^n c_n^{[B_0]}(u_0, \epsilon),
\end{align}
with $\Lambda_0 = B_0/\nu_0$ as introduced in Eq. \eqref{eq:equation_for_B_general_form} and the coefficient $c_1^{[B_0]}$ is given exactly by Eq. \eqref{eq:one-loop_three_dimensonal_j_coeffisient} (in renormalized variables).
In turn, $c_2^{[B_0]}$ is defined in the next Section using the following quantity
\begin{align}\label{eq:two-loop_B0_coeffisient}
    &C_2^{[B_0]}(\Lambda/\Lambda_0, u_0, \epsilon) \coloneqq \sum \limits_{s = 1}^{488}R_{\Lambda}\left[J_{2s}\right] \notag \\
    &= \mathcal{J}_2^{\mathrm{F}} +  \mathcal{J}_2^{\mathrm{IR}} + \mathcal{J}_2^{\mathrm{P}} + R_{\Lambda}\left[\mathcal{J}_2^{\mathrm{L}}\right] + R_{\Lambda}\left[\mathcal{J}_2^{\mathrm{S}}\right],
\end{align}
where the $R_{\Lambda}$-operation combines $R'_{\Lambda}$, which removes subgraph divergences, with the subtraction of remaining superficial divergences.
For the $\mathcal{J}_2^{\mathrm{L}}$ integrals, $R_{\Lambda}$ only subtracts superficial divergences, while for the $\mathcal{J}_2^{\mathrm{S}}$ integrals, the action of $R_{\Lambda}$ is nontrivial. 

Let us first consider the integrals $\mathcal{J}_2^{\mathrm{L}}$.
Their superficial divergences can be isolated using Eq. \eqref{eq:two-loop_integral_after_rescaling_by_B0}.
At the $\Lambda$-renormalization level (with $\Lambda/\Lambda_0 < \infty$), dividing these integrals into $\Lambda$- and $B_0$-parts parts does not introduce extra issues as $m \to 0$ (unlike the case for $\mathcal{J}_2^{\mathrm{S}}$, discussed later).
Hence, subtracting the $\Lambda$-regular contributions from $\mathcal{J}_2^{\mathrm{L}}$ is straightforward.
The sum $\mathcal{J}_2^{\mathrm{L}}$ comprises integrals of types \verb|I1-L| and \verb|I2-L| (see Eq. \eqref{eq:two_loop_curl_b'b_self-energy_after_shift}).
For our purposes, it is useful to examine the UV behavior of the integrands with respect to $q$ flowing in the subgraph.\footnote{The momentums in the diagrams are always arranged so that $q$ flows in the subgraph, and $p$ on the outer loop.}
For \verb|I1-L| integrals with subtracted $\Lambda$-parts we have
\begin{equation}\label{eq:expansionUVLambda}
    \frac{\mathscr{F}^{\mathrm{S}}(p, q)}{(p q)^{1 + 2\epsilon}} = \frac{1}{(pq)^{2\epsilon}}\left(\mathscr{C}_{2}(p) + \frac{\mathscr{C}_{3}(p)}{q} + \mathcal{O}(q^{-2}) \right),   
\end{equation}
with regular $\mathscr{C}_{2}(p)$ and $\mathscr{C}_{3}(p)$.
Although $\mathscr{C}_{2}(p)$ would typically cause a power-law $\Lambda$ divergence, the angular integration over $\mbox{d} \Omega$ nullifies this term.
Note that this is not the case for integrals $\mathcal{J}_2^{\mathrm{S}}$.
This cancelation, noted in Sec. \ref{sec:renormalization_without_shift}, is linked to the absence of power-law (curl) divergences in the subgraphs of $\Gamma^{v'v}$ and $\Gamma^{v'bb}$.
For \verb|I2-L| integrals, the UV asymptotic behavior in $q$ is
\begin{equation}\label{eq:expansionUV}
\frac{\mathscr{F}(p, q)}{(p q)^{1 + 2\epsilon}} = \frac{1}{p^{2\epsilon}}\left(\frac{\mathscr{C}_{4}(p)}{q^{1 + 2\epsilon}} + \mathcal{O}(q^{-2})\right),    
\end{equation}
with $\mathscr{C}_{4}(p)$ regular.
Together with Eq. \eqref{eq:expansionIR}, these asymptotics characterize all two-loop $\Sigma^{b'b}$ curl integrals, including those in $\mathcal{J}_2^{\mathrm{S}}$.
Finally, for both \verb|I1-L| and \verb|I2-L| integrals, poles in $\epsilon$ appear as $\Lambda/\Lambda_0 \to \infty$.
Their elimination is the subject of the $\epsilon$-renormalization discussed in the next Section.
At finite $\Lambda/\Lambda_0$ these issues can be ignored by setting $\epsilon = 0$.

From Eqs. \eqref{eq:two-loop_B0_coeffisient}, \eqref{eq:total_two-loop_purely_power-law_curl_term}, and \eqref{eq:two_loop_curl_b'b_self-energy_after_shift} it follows that
\begin{eqnarray}\label{eq:action_of_the_R'-operation_on_integrals_of_type_S}
    \mathcal{J}_2^{\mathrm{S}} &= R_{\Lambda}\left[\mathcal{J}_2^{\mathrm{S}}\right] + \mathcal{J}_2^{\mathrm{S}[\Lambda]} + \mathcal{J}_2^{\left[D^{(2)}_{\mathrm{S}1}\right]}, \qquad
\end{eqnarray}
where $\mathcal{J}_2^{\mathrm{S}[\Lambda]}$ represents the sum of contributions from $\Lambda$-divergent subgraphs (including diagram $D^{(2)}_{\mathrm{S}1}$) and $\mathcal{J}_2^{\left[D^{(2)}_{\mathrm{S}1}\right]}$ denotes the superficial divergence of diagram $D^{(2)}_{\mathrm{S}1}$, after the corresponding subgraph divergence has been subtracted.
This contribution is specified below.

The ``special'' diagram $D^{(2)}_{\mathrm{S}1}$ contains both a subgraph divergence and a superficial $\Lambda$-divergence.
In the integrand of $D^{(2)}_{\mathrm{S}1}$ at $B_0 = 0$, both divergences are present.
After subtracting the $B_0$-independent part as in Eq. \eqref{eq:two-loop_integral_after_rescaling_by_B0}, only the subgraph divergence remains in the $B_0$-dependent part, and the subtracted part matches that obtained in Sec. \ref{sec:renormalization_without_shift} for $h_0$.
Thus, splitting this diagram into $\Lambda$- and $B_0$-parts yields an expression of the form
\begin{eqnarray}\label{eq:problematic_diagram_after_shift_all_contributions}
   C_{\mathrm{sub}} \Lambda\ln{\left(\frac{m}{\Lambda}\right)} + C_{\mathrm{surf}}\Lambda +  \Lambda_0\mathcal{F}_{\epsilon}^{\left[D^{(2)}_{\mathrm{S}1}\right]}\left(\frac{\Lambda}{\Lambda_0}\right), \quad
\end{eqnarray}
where $C_{\mathrm{sub}}$ is defined in Eq. \eqref{eq:divergent_subgraph_contribution_in_problematic_diagram}, $C_{\mathrm{surf}}$ comes from the integral $I_2$ in Eq. \eqref{eq:two_loop_bb'_linear-asymp} with integrand $[F_2(\kappa, 1, z) - F_2(0, 1, z)]$ (the explicit form of the function $F_2$ is given in App. \ref{app:two-loop_Gammab'b_expressions}), and the rest $\mathcal{F}_{\epsilon}^{[D^{(2)}_{\mathrm{S}1}]}$ represents the $B_0$-type contribution of $D^{(2)}_{\mathrm{S}1}$.
The surface term $C_{\mathrm{surf}} \Lambda$ is absorbed into the definition of $h_0$, leaving only the contribution $C_{\mathrm{sub}}\Lambda\ln{\left(m/\Lambda\right)}$, which is handled by the $R'_{\Lambda}$-operation.
The remaining $B_0$-type part, $\mathcal{F}_{\epsilon}^{[D^{(2)}_{\mathrm{S}1}]}$, is then combined with the other $11$ pure $B_0$-type integrals in $\mathcal{J}_2^{\mathrm{S}}$.

Let us now describe how to obtain  $\mathcal{J}_2^{\mathrm{S}[\Lambda]}$.
In general, the total integrand $\mathscr{F}^{\mathrm{S}}/(pq)^{1 + 2\epsilon}$ of $\mathcal{J}_2^{\mathrm{S}}$ after subtraction of the surface term $C_{\mathrm{surf}}\Lambda$, exhibits the following behavior: 
\begin{itemize}
    \item[(a)] $p\to 0$: analogous to Eq. \eqref{eq:expansionIR};
    \item[(b)] $q \to 0$: regular;
    \item[(c)] $p \to \infty$: regular (the superficial divergence has been subtracted);
    \item[(d)] $q \to \infty$: as in Eq. \eqref{eq:expansionUVLambda}.
\end{itemize}
As in Eq. \eqref{eq:masterform}, $\mathcal{J}_2^{\mathrm{S}}$ is divided into four parts: $\mathcal{I}_1^{\mathrm{S}}$, $\mathcal{I}_2^{\mathrm{S}}$, $\mathcal{I}_3^{\mathrm{S}}$, and $\mathcal{I}_4^{\mathrm{S}}$.
It is clear that $\mathcal{I}_{3}^{\mathrm{S}}$ is free of divergences. 
For the other three integrals, we explicitly isolate singularities by adding and subtracting the appropriate asymptotic forms, then sum the results.

Starting with $\mathcal{I}_1^{\mathrm{S}}$, we subtract the asymptotic of type (a), yielding
\begin{align}\label{eq:subtracted_asumpotics_from_I1}
    C_{\mathrm{sub}} \Lambda_{0}\ln{\left(\frac{\Lambda_{0}}{m}\right)} + \Lambda_0\mathcal{F}_{\epsilon}^{(1)}\left(\frac{\Lambda}{\Lambda_0}\right),
\end{align}
where $\mathcal{F}_{\epsilon}^{(1)}$ remains finite as $m = 0$, $\epsilon = 0$ and $\Lambda/\Lambda_{0} \to \infty$.
In $\mathcal{I}_{4}^{\mathrm{S}}$, subtracting the asymptotic of type (d) gives
\begin{align}\label{eq:subtracted_asumpotics_from_I4}
    (\Lambda_0 - \Lambda)\mathcal{A}_1 + \Lambda_0\mathcal{F}_{\epsilon}^{(2)}\left(\frac{\Lambda}{\Lambda_0} \right).
\end{align}
Here, $\mathcal{A}_1 \Lambda_0$ is assigned to the finite part, while $- \mathcal{A}_1\Lambda$ accounts for the subgraph divergence.
The rest, $\mathcal{F}_{\epsilon}^{(2)}$,  stays finite as $m \to 0$  and produces the standard UV poles in $\epsilon$ when $\Lambda/\Lambda_{0} \to \infty$ (the second term in Eq. \eqref{eq:expansionUVLambda} is responsible for this behavior).
At the $\Lambda$-renormalization stage, we can ignore this issue.
The remaining part, $\mathcal{I}_{2}^{\mathrm{S}}$ exhibits divergences both at large $q$ and at small $p$.
Therefore, one must first subtract the asymptotic behavior (d) from the original expression -- which itself diverges as $p \to 0$ -- and then subtract the small-$p$ asymptotics from the resulting difference.
The combined subtracted asymptotic contributions have the form\vspace{-0.1cm}
\begin{eqnarray}\label{eq:subtracted_asumpotics_from_I2}
    (\Lambda_0 - \Lambda)\left(\mathcal{A}_{2} - C_{\mathrm{sub}}\ln{\left(\frac{\Lambda_0}{m}\right)}\right) + \Lambda_0\mathcal{F}_{\epsilon}^{(3)}\left(\frac{\Lambda}{\Lambda_0}\right),\qquad
\end{eqnarray}
where $\mathcal{F}_{\epsilon}^{(3)}$, as well as $\mathcal{F}_{\epsilon}^{(2)}$ from Eq. \eqref{eq:subtracted_asumpotics_from_I4}, remains finite at $m = 0$ and generates an $\epsilon$-pole as $\Lambda/\Lambda_0 \to \infty$.
Within two-loop accuracy, all asymptotics are computed at $\epsilon = 0$ (explicitly highlighted first terms in Eqs. \eqref{eq:subtracted_asumpotics_from_I1} -- \eqref{eq:subtracted_asumpotics_from_I2}), while the remaining integrals retain their dependence on $\epsilon$ until $\epsilon$-renormalization.

Finally, taking into account the contributions from Eqs. \eqref{eq:problematic_diagram_after_shift_all_contributions}, \eqref{eq:subtracted_asumpotics_from_I1}, \eqref{eq:subtracted_asumpotics_from_I4}, \eqref{eq:subtracted_asumpotics_from_I2}, and the finite integral $\mathcal{I}_3^{\mathrm{S}}$, we obtain \vspace{-0.1cm}
\begin{align}\label{eq:divergent_and_convergent_parts_of_S_type_integrals}
    \mathcal{J}^{\mathrm{S}}_{2} & = \mathcal{J}^{\mathrm{S[\Lambda]}}_{2} + \Lambda_0(\mathcal{A}_1 + \mathcal{A}_2) + \mathcal{I}_3^{\mathrm{S}} \notag \\
    &+ \Lambda_0
    \sum \limits_{s = 1}^{3}\mathcal{F}_{\epsilon}^{(s)}\left(\frac{\Lambda}{\Lambda_0}\right) + C_{\mathrm{surf}}\Lambda, \quad
\end{align}
with the explicit form of $\mathcal{J}^{\mathrm{S}}_{2\,\Lambda}$ is given by
\begin{eqnarray}
\label{eq:lsubdivergence}
&\mathcal{J}^{\mathrm{S}[\Lambda]}_{2} = \frac{\Lambda  \ln{\left(\frac{\Lambda}{\Lambda_0}\right)}}{36 \pi^4 \left(u_0 + 1\right)^3} - \frac{\Lambda\left(u_0 - 3 u_0 \ln{\left(u_0\right)} + 3\right)}{216 \pi^4 u_0 \left(u_0 + 1\right)^3}.\qquad
\end{eqnarray}
Here, the logarithmic terms proportional to $ \Lambda_0\ln{\left(\Lambda_0/m\right)}$  in Eqs. \eqref{eq:subtracted_asumpotics_from_I4} and \eqref{eq:subtracted_asumpotics_from_I2} cancel, with the remaining logarithms are collected in the first term in $\mathcal{J}^{\mathrm{S}[\Lambda]}_{2}$.
This cancelation of logarithms is expected, since the full integrand $\mathscr{F}^{\mathrm{S}}/(pq)^{1 + 2\epsilon}$ (as a sum of all diagrams) is regular near $p = 0$ and $q = 0$; however, isolating the UV divergences necessarily disrupts this regularity at $p \to 0$, which is why a detailed subtraction of all asymptotics is required.
Note that although the final expression for $\mathcal{J}^{\mathrm{S}[\Lambda]}_{2}$ appears well-behaved, many intermediate expressions are not.
For this reason, we only provide the sum $\mathcal{A}_1 + \mathcal{A}_2$ as it appears in $\mathcal{J}^{\mathrm{S}[\Lambda]}_{2}$, rather than the individual coefficients.

Returning to the correction diagrams in Fig. \ref{fig:diagrams_with_divergent_subgraph_and_their_counterterms}, the corresponding integrals exhibit similar issues as those in $\mathcal{J}^{\mathrm{S}[\Lambda]}_{2}$.
However, since these integrals are single-valued with respect to the momentum variable, the analysis is somewhat simplified. 
The sum of the correction diagrams $\mathcal{J}^{\mathrm{S}}_{\mathrm{add}}$ is given by\vspace{-0.1cm}
\begin{eqnarray}\label{eq:sum_of_counterterms}
    &\mathcal{J}^{\mathrm{S}}_{\mathrm{add}} = \rho \nu_0\big[i \varepsilon_{i j l}k_l \big]h_0 \left(g_0 \Lambda_0^{- 2\epsilon}\right) \int \limits_{-1}^{1} \mbox{d} z_p \hspace{-0.1cm}\int \limits_{0}^{\Lambda/\Lambda_0}\hspace{-0.1cm} \frac{\mbox{d} p}{p^{2\epsilon}} \mathscr{F}^{\mathrm{S}}_{\mathrm{add}}, \qquad\\
    &\mathscr{F}^{\mathrm{S}}_{\mathrm{add}} = \frac{p \left(z_p^2 - 1\right) \left(p^4 u_0^3 - 3 p^2 u_0 \left(u_0 + 1\right) z_p^2 + z_p^4\right)}{8 \pi ^2
   \left(u_0 + 1\right)^2 \left(p^2 u_0 + z_p^2\right)^3}. \label{eq:integrand_of_counterterms_sum} \qquad
\end{eqnarray}
This integrand is regular as $p \to 0$ but has a logarithmic divergence as $p \to \infty$, corresponding to the diagram $D_{\mathrm{add}\,1}^{(2)}$.
However, an immediate subtraction of the logarithmic asymptotics at $p \to \infty$ would spoil the regularity at the lower limit.
Therefore, the integral $\mathcal{J}^{\mathrm{S}}_{2\,\mathrm{add}}$ is split into two parts: one over $[0,1]$ and the other over $[1, \Lambda/\Lambda_0]$.
The first integral is finite and computed exactly, while the second is treated by subtracting its logarithmic asymptotic at $p \to \infty$. 
After this, it is sufficient to calculate the remaining integral already with an infinite upper limit.
The final result at $\epsilon = 0$, combining the finite part over 
which with the sum of the integral over unit interval $[0,1]$, the integral with subtraction and the logarithmic asymptotics, is
\begin{align}\label{eq:calculated_counterterms}
    &\mathcal{J}^{\mathrm{S}}_{\mathrm{add}} = \rho \nu_0\big[i \varepsilon_{i j l}k_l \big]h_0 \left(g_0 \Lambda_0^{- 2\epsilon}\right) \mathfrak{F}_{\mathrm{add}}, \notag \\
    &\mathfrak{F}_{\mathrm{add}} = \frac{\ln{\left(\frac{\Lambda }{\Lambda _0}\right)}}{6 \pi ^2 \left(u_0 + 1\right)^2} - \frac{\left(u_0 - 3 u_0\ln \left(u_0\right) + 3\right)}{36 \pi ^2 u_0 \left(u_0 + 1\right)^2}.
\end{align}
Substituting into \eqref{eq:calculated_counterterms} the one-loop $h_0$ with $c_1^{[\Lambda]}$ at $\epsilon = 0$ from \eqref{eq:one_loop_curl-term} shows that this expression exactly doubles $\mathcal{J}^{\mathrm{S}[\Lambda]}_{2}$ in Eq. \eqref{eq:lsubdivergence}.
Thus, we arrive at an expression for the counterterm $\delta h$ in the form 
\begin{align}\label{eq:delta_h_after_shift}
    \delta h = -2\mathcal{J}^{\mathrm{S}[\Lambda]}_{2}.
\end{align}

\subsubsection{\label{sec:epsilon-renormalization}\pdfmath{\epsilon}-renormalization}

Once the magnetic field has been shifted and the $\Lambda$-renormalization performed, one obtains an expression for the bare spontaneous field, $B_0 = B_0 (h_0(\Lambda), g_0, \nu_0, u_0(\rho), \epsilon)$, as discussed in the previous Section.
Afterwards, one can similarly write $B = B (h, g, \nu, u, \rho, \mu, \epsilon)$ for the renormalized spontaneous field in terms of the corresponding renormalized parameters.
Alternatively, a slightly different route can be taken by performing the $\Lambda$-renormalization in the unshifted theory followed by $\epsilon$-renormalization (see Sec.~\ref{sec:renormalization_without_shift}), and only then shifting the magnetic field.
In both approaches, $B$ emerges as a UV-finite parameter renormalized by constant $Z_b$ as predicted by Eq. \eqref{eq:B_field_renormalization}.
Therefore, the resulting two-loop expressions for $B$ in either approach coincide up to a finite renormalization transformation.
The desired difference will be given exactly by the non-logarithmic term in \eqref{eq:calculated_counterterms}.

Putting this all together yields a renormalized theory with the action
\begin{align}\label{eq:MHD_action_after_shift_renormalized}
\mathcal{S}_{\boldsymbol{B} R} =&~\mathcal{S}_R + Z_3\boldsymbol{v'}\cdot(\boldsymbol{B} \cdot \boldsymbol{\nabla})\boldsymbol{b} + \boldsymbol{b'}\cdot(\boldsymbol{B} \cdot \boldsymbol{\nabla}) \boldsymbol{v} \notag \\
&- \rho \nu {h} Z_4 \boldsymbol{b'} \cdot \left(\boldsymbol{\nabla} \times \boldsymbol{b}\right),
\end{align}
where $\mathcal{S}_R$ is given in Eq.\eqref{eq:renormalized_MHD_action_without_shift}, and the two-loop value of $Z_4$ is provided by Eq.\eqref{eq:two-loop_Z4}.

The consistent $\epsilon$-renormalization in this shifted-field formalism proceeds just as described at the end of Sec. \ref{sec:renormalization_without_shift}, except that now one encounters an additional class of $B$-type diagrams.
Practically, this means that in every occurrence of bare parameters, we replace $h_0 + \delta h \to h Z_h$ while other model parameters and fields undergo the usual multiplicative transformations \eqref{eq:h0_epsilon_renormalization}, \eqref{eq:multiplicative_renorm}.
We then subtract the $\epsilon$-poles from the newly introduced ($B$-type) diagrams and take $\Lambda \to \infty$ in all remaining convergent integrals.
The net non-trivial contributions turn out to be precisely those extracted from Eq.~\eqref{eq:two-loop_B0_coeffisient}.
Symbolically, this can be written as 
\begin{align}\label{eq:C_2^B_general_expression}
    c_2^{[B]}(u, \epsilon) \coloneqq \lim \limits_{\Lambda \to \infty} R'_{\epsilon} \left[C_2^{[B]}\left(\nu\Lambda/B, u,\epsilon \right)\right],
\end{align}
where the coefficient $C_2^{[B]}\left(\nu\Lambda/B, u,\epsilon \right)$ is formally derived from Eq.~\eqref{eq:two-loop_B0_coeffisient} upon replacing all bare parameters by their renormalized counterparts; here $R'_{\epsilon}\left[\bullet \right]$ indicates the subtraction of $\epsilon$-divergences in the subgraphs, analogous to $R'_{\Lambda}$.

The procedure of determining the value of the renormalized field $\boldsymbol{B}$ that stabilizes the system proceeds analogously to the Dyson equation \eqref{eq:dyson_eq_for_Gammab'b_h0_to_propagator}, except that $\Lambda$-dependent terms are already absent (either by virtue of analytical regularization or by prior $\Lambda$-renormalization).
In particular, the sole logarithmically divergent contribution (from diagram $D_{\mathrm{ct}1}^{(2)}$ proportional to $h$) is canceled by a pole in $Z_4$.
Thus one recovers an expression of the same form as \eqref{eq:Lambda_renormalized_Gammab'b_after_shift}.
By requiring it to be zero, we find the corresponding two-loop $B$.
With the procedure for calculating $\mathcal{C}_2^{[B]}(u)$ in hand, we have all the necessary ingredients for an iterative solution of this equation.
\begin{widetext}
More explicitly,
\begin{align}\label{eq:two_loop_equation_determining_B_at_the_level_of_epsilon_renormalization}
    0 = h + gZ_g \mu^{2\epsilon} \left(\frac{B Z_b}{\nu Z_{\nu}}\right)^{1 - 2 \epsilon} \frac{(u Z_u)^{-1/2 + \epsilon}}{8\pi (1 + u Z_u)(2 - \epsilon)\cos{(\pi \epsilon)}} + \mathcal{J}_{2}^{[\epsilon]} + g^2 \left(\frac{B}{\nu}\right) c_2^{[B]}(u, 0) + h g \mathcal{C}_{\mathrm{add}}^{[h]}(u) + \mathcal{O}(\epsilon^3),
\end{align}
where the second term must be expanded to the second order in $\epsilon$ (note that $g = \mathcal{O}(\epsilon)$) using one-loop RG constants provided below, the term $\mathcal{J}_{2}^{[\epsilon]}$ includes the relevant $\epsilon$-poles subtracted via $R'_{\epsilon}$, and the two-loop coefficient $c_2^{[B]}(u, 0)$ is given by \eqref{eq:C_2^B_general_expression}.
An additional term, $\mathcal{C}_{\mathrm{add}}^{[h]}(u)$ is clarified below. 
Summarily, $\mathcal{J}_{2}^{[\epsilon]}$ is given by
\begin{eqnarray}\label{eq:total_pole_contribution_from_two_loop_diagrams}
\mathcal{J}_{2}^{[\epsilon]} &= B g^2\left[- \frac{ \mathcal{C}(u)\left(1 - 2 \epsilon \ln{
   \left(\frac{B}{\mu\nu}\right)}\right)}{3840 \pi^3 u^{3/2} \left(u + 1\right)^3 \epsilon} + \frac{(36 + 19 u + 6 u^2 + 3 u^3) - 2 \mathcal{C}(u)\ln{ \left( u \right)}}{7680 \pi^3 u^{3/2} \left(u + 1\right)^3}\right] \hspace{-0.13cm},~ \mathcal{C}(u) \coloneqq 9 u^3 + 26 u^2 + 49 u + 12. \qquad
\end{eqnarray}
\end{widetext}

\begin{table}
\caption{\label{tab:numeric_values_of_Lambda-and_B_coefficients}
Numerical values of the coefficients $c_2^{[\Lambda]}(u, 0)$ and $c_2^{[B]}(u, 0)$ for different orders of $u$.}
\begin{ruledtabular}
\begin{tabular}{lll}
$u$ & \multicolumn{1}{c}{$c_2^{[B]}(u, 0)$} & \multicolumn{1}{c}{$c_2^{[\Lambda]}(u, 0)$\footnote{Since $c_2^{[\Lambda]}(u, 0)$ does not directly contribute to the definition of $B$, their values for different $u$ are given only for illustrative purposes.}} \\
\hline \rule{-0.11cm}{0.4cm}
$10^5$ & $-1.5 \times 10^{-10}$ & $3.37 \times 10^{-10}$ \\
$10^4$ & $-4.2\times10^{-10}$ & $3.37 \times 10^{-9}$ \\
$10^3$ & $-9.9 \times 10^{-9}$ & $3.37 \times 10^{-8}$ \\
$10^2$ & $-1.7 \times 10^{-7}$ & $3.35 \times 10^{-7}$ \\
$10$ & $\hspace{0.25cm}3.5 \times 10^{-6}$ & $3.44 \times 10^{-6}$ \\
$u_{\star}^{(1)}$ & $\hspace{0.25cm}1.2 \times 10^{-4}$ & $~\,2.3 \times 10^{-5}$ \\
$10^{-1}$ & $\hspace{0.25cm}2.6\times10^{-2}$ & $~\,3.8 \times 10^{-5}$ \\
$10^{-2}$ & $\hspace{0.25cm}1.1 \times 10^{0}$ & $~\,3.6 \times 10^{-5}$ \\
$10^{-3}$ & $\hspace{0.25cm}3.7\times10^{1}$ & $~\,3.6 \times 10^{-5}$ \\
$10^{-4}$ & $\hspace{0.25cm}1.1\times10^{3}$ & $~\,3.6 \times 10^{-5}$ \\
$10^{-5}$ & $\hspace{0.5cm}3\times 10^{4}$ & $~\,3.6 \times 10^{-5}$ \\
\end{tabular}
\end{ruledtabular}
\end{table}

Technically, the pole contribution $\mathcal{J}_{2}^{[\epsilon]}$ originates from diagrams $\mathcal{J}_2^{\mathrm{P}}$, $\mathcal{J}_2^{\mathrm{L}}$, and $\mathcal{J}_2^{\mathrm{S}}$.
For each of these, one analyzes the asymptotic behavior according to \eqref{eq:masterform}.
The integrals $\mathcal{J}_2^{\mathrm{P}}$ and $\mathcal{J}_2^{\mathrm{L}}$ exhibit singularities only as $q \to \infty$ with asymptotic behavior given by \eqref{eq:expansionUV}.
Consequently, one must subtract and then add back the corresponding asymptotics in the $\mathcal{I}_2^{\mathrm{L}}$, $\mathcal{I}_4^{\mathrm{L}}$ parts.
Note that for $\mathcal{J}_2^{\mathrm{L}}$ only the second term appears in their decomposition (in the spirit of \eqref{eq:two-loop_integral_after_rescaling_by_B0}), since all superficially divergent parts are absent in our approach.
In contrast, the integrals $\mathcal{J}_2^{\mathrm{S}}$ generally follow the asymptotic behavior described by \eqref{eq:expansionUVLambda}. 
More precisely, as noted in the previous Section, the problematic behavior arises from the functions $\mathcal{F}_{\epsilon}^{(2)}$ and $\mathcal{F}_{\epsilon}^{(3)}$, which exhibit asymptotics of the type given in  \eqref{eq:expansionUV} as $q \to \infty$.
This corresponds to the second term in \eqref{eq:expansionUVLambda}, while the first term is responsible for the $\Lambda$-divergences. 
Therefore, we treat these functions in the same way as the integrals $\mathcal{J}_2^{\mathrm{P}}$ and $\mathcal{J}_2^{\mathrm{L}}$.

As expected, the contribution $\mathcal{J}_{2}^{[\epsilon]}$ is exactly canceled by the $\epsilon$-poles arising the expansion of one-loop diagrams (the second term in the right hand side of Eq. \eqref{eq:two_loop_equation_determining_B_at_the_level_of_epsilon_renormalization}), serving as a powerful tool for checking our calculations. 
For convenience, we present here all the values of the one-loop renormalization constants, easily obtained from the data in the Table \ref{tab:2loop_MHD_Ren_consts} and formulas \eqref{eq:RG_functions} and \eqref{eq:RG_func_through_first_poles}: 
\begin{align}\label{eq:one-loop_RG_functions}
&Z_{\nu} = 1 - \frac{g}{40 \pi^2 \epsilon}, \quad Z_u = 1 + \frac{(3u^2 + 3u -10)g}{120 \pi^2 u (1 + u) \epsilon}, \notag \\
&Z_g = 1 + \frac{3 g}{40 \pi^2 \epsilon}, \quad Z_{b} = 1 + \frac{g}{120 \pi^2 u \epsilon}.
\end{align}
Let us also present here the two-loop value of $Z_h$, obtained combining two-loop $Z_{\nu}$ from Table \ref{tab:2loop_MHD_Ren_consts} with $Z_4$ from 
\eqref{eq:two-loop_Z4}:
\begin{eqnarray}\label{eq:two-loop_Zh}
     Z_h \coloneqq Z_4 Z_{\nu}^{-1} = 1 + \frac{(3u^2 + 6 u + 13)g}{120 \pi^2 (1 + u)^2 \epsilon} - \frac{z_{21}^{(1)}(u) g^2}{\epsilon}.
     \qquad
\end{eqnarray}

The contribution $\mathcal{C}_{\mathrm{add}}^{[h]}(u)$ represents the UV-finite part of the sum of the correction diagrams shown in Fig. \ref{fig:diagrams_with_divergent_subgraph_and_their_counterterms} when the magnetic field is shifted after $\epsilon$-renormalization in the unshifted theory. 
Its explicit form is given by
\begin{align}\label{eq:UV-finite_h-correction_from_correction_diagrams}
    \mathcal{C}_{\mathrm{add}}^{[h]}(u) \coloneqq \frac{\left(u + 3 - 3 u\ln \left(u\right)\right)}{36 \pi ^2 u \left(u + 1\right)^2}.
\end{align}
In contrast, if the shift is performed before $\epsilon$-renormalization, as discussed in the previous Section, then $\mathcal{C}_{\mathrm{add}}^{[h]}(u) = 0$.
This is equivalent to choosing $\delta h$ in the form given in \eqref{eq:delta_h_after_shift} rather than \eqref{eq:deltah_counterterm}.

Numerical evaluations of the two-loop coefficients ${c}_2^{[B]}(u, 0)$ were carried out using Wolfram Mathematica \cite{Mathematica} in conjunction with an adaptive Monte-Carlo method (the implementation of the VEGAS algorithm \cite{Lepage1978}).
Using such approach, we maintain only one significant digit of accuracy for our final results, which reflects both the complexity of the integration procedure and the fact that a precise determination of a coefficient that multiplies an arbitrary parameter $|h|$ is not essential.
Let us stress that it is apparently impossible to achieve greater accuracy for fivefold integrals contributing to $c_2^{[B]}(u, 0)$ using currently available numerical procedures.
Notably, the accessible range of $u$ values for our numerical procedure is comparable to that currently achieved by alternative approaches such as DNS.
A summary of the numerical results from \eqref{eq:two_loop_equation_determining_B_at_the_level_of_epsilon_renormalization} is presented in Table \ref{tab:numeric_values_of_Lambda-and_B_coefficients}.
As seen in Table \ref{tab:numeric_values_of_Lambda-and_B_coefficients}, when $u \ll 1$, the two-loop coefficient $c_2^{[B]}(u, 0)$ appears to diverge as $u^{-3/2}$, similar to its one-loop counterpart $c_1^{[B]}(u,0) \sim u^{-1/2}$.
The corresponding coefficient $c_2^{[\Lambda]}(u, 0)$ in this regime also duplicates the behavior of its one-loop analog, reaching a constant value.

Finally, after canceling all $\epsilon$-poles in Eq. \eqref{eq:two_loop_equation_determining_B_at_the_level_of_epsilon_renormalization}, one arrives at a UV-finite relationship between $B$ and $h$ in purely renormalized quantities.
Explicitly,
\begin{align}\label{eq:two-loop_B}
    &B = \frac{16 \pi \sqrt{u} (1 + u)\nu |h|}{g}\biggl[1 - 16 \pi \sqrt{u} (1 + u) c_2^{[B]}(u,0) g \notag \\
    &- \left(\frac{1}{2} - 2 \ln{\left(\frac{16 \pi (1 + u)|h|}{\mu g} \right)} \right)\epsilon + g \mathcal{C}_{\mathrm{add}}^{[h]}(u)\biggr],
\end{align}
where the presence of $\mathcal{C}_{\mathrm{add}}^{[h]}(u)$ depends on statement of the problem (one may say, from choosing a subtraction scheme).
The parameter $h \propto g$ suppose to have the same sign as $h_0$ (negative).
At one-loop, it is known that $h_0$ is negative; our two-loop analysis for physically relevant $u$-values indicates that this remains the case (the values of $c_2^{[\Lambda]}(u, 0)$ are small (see the Table \ref{tab:numeric_values_of_Lambda-and_B_coefficients}) and should not change the sign of $h_0$ as a whole).

Let us also emphasize that the renormalization performed shows that $B$ is successfully calculated for $m = 0$ and can be considered independent of $m$ within the framework of the model used.
This circumstance agrees with the first Kolmogorov hypothesis described in Sec. \ref{sec:chiral_MHD}.
Recall that $B$ is, in its meaning, a one-point Green function.

\section{\label{sec:RG_in_dynamo_regime}Renormalization group in the dynamo regime}

After removing the curl terms, the parameter $B$ in the renormalized theory is fixed by formula \eqref{eq:two-loop_B}.
For the Green functions of the renormalized model \eqref{eq:MHD_action_after_shift_renormalized} obtained in this way, the usual RG equations of type \eqref{eq:RG_equation_for_W} hold -- with an additional term related to the anomalous dimension of the extra dimensional parameter that emerges in dynamo regime.
Note that \eqref{eq:two-loop_B} is a constraint between two IR-significant parameters $B$ and $h$, which makes only one of them independent and therefore included in the RG operator $\overline{\mathrm{D}}_{\mu}$.
One may choose this additional parameter to be either $B$ or $h$.
If $h$ is the chosen one, then the corresponding two-loop anomalous dimension $\gamma_h$ is easily obtained from the corresponding $Z_h$ via the formula \eqref{eq:RG_func_through_first_poles} yielding
\begin{align}\label{eq:two-loop_gammah}
    \gamma_h = -\frac{(13 + 6 u + 3 u^2)g}{60 \pi^2 (1 + u)^2} + 2 z_{21}^{(1)}(u)g^2.
\end{align}

\subsection{\label{sec:B_through_RG}Green functions in the dynamo regime}

We investigate the magnetic dynamo in the turbulent regime, i.e., the regime of critical scaling from the RG perspective, treating $B$ as the independent parameter in the model \eqref{eq:MHD_action_after_shift_renormalized}.
In this formulation, the RG equation \eqref{eq:RG_equation_for_W} for the generating functional of the renormalized coupled Green functions $\mathcal{W}_R$ acquires an additive term $-\gamma_B \mathrm{D}_B$, where $\gamma_B = \gamma_b$ as follows from Eq. \eqref{eq:B_field_renormalization}.
The corresponding RG equation for an $n$-point renormalized connected Green function $\mathcal{W}_{Rn} \coloneqq F_R$ (with the total field count $n_{\Phi} \coloneqq \sum_{\varphi} n_{\varphi}$) takes the form
\begin{eqnarray}\label{eq:RG_equation_for_Wn_with_h_after_shift}
    \left[\mathrm{D}_{\mu} + \sum \limits_{g, u}\beta_{\alpha} \partial_{\alpha} - \sum \limits_{\nu, B} \gamma_{\sigma} \mathrm{D}_{\sigma} + \sum \limits_{\boldsymbol{b}, \boldsymbol{b'}} \gamma_{\varphi} n_{\varphi}\right] F_{R} = 0. \qquad
\end{eqnarray}
Solutions of \eqref{eq:RG_equation_for_Wn_with_h_after_shift} are conveniently studied in terms of dimensionless momentum and frequency variables.
Suppose
\begin{align}\label{eq:arbitrary_renormalized_GF_after_shift}
    F_R = k^{d_F} \nu^{d_F^{\omega}} R\left(k/\mu, \omega/\nu k^2, B/\nu k, g, u, \ldots \right),
\end{align}
where $R$ is a scaling function of dimensionless arguments, $d_F$ and $d_F^{\omega}$ are the canonical dimensions of $F_R$, and the ellipsis indicates possible additional dimensionless ratios $\omega/\omega_i$, $k/k_i$, etc.
In what follows, we employ $s \coloneqq k/\mu$ as a RG-trajectory parameter except where otherwise stated.

Since the renormalized theory satisfies the constraint \eqref{eq:two-loop_B}, the same condition must be satisfied for $R$.
Inserting \eqref{eq:arbitrary_renormalized_GF_after_shift} into the RG equation \eqref{eq:RG_equation_for_Wn_with_h_after_shift} yields
\begin{align}\label{eq:eq:arbitrary_GF_after_shift_through_invariant_variables}
    &F_{\mathrm{RG}} = k^{d_F} \mybar{0.8}{1pt}{\nu}^{d_F^{\omega}} \mybar{1}{1pt}{R}\left(1, \omega/ \mybar{0.8}{1pt}{\nu} k^2, \mybar{0.8}{1pt}{B}, \mybar{0.8}{1pt}{g},  \mybar{0.8}{1pt}{u}, \ldots \right) \notag \\
    & \times \exp{\left(\sum \limits_{\boldsymbol{b}, \boldsymbol{b'}} n_{\varphi}\int \limits_1^s \frac{\mbox{d}s'}{s'}\gamma_{\varphi}\Big(\mybar{0.8}{1pt}{g}(s', g), \mybar{0.8}{1pt}{u}(s', g, u) \Big)\right)}.
\end{align}
in which $\mybar{0.8}{1pt}{g}(s', g)$ and $ \mybar{0.8}{1pt}{u}(s', g, u)$ are the invariant variables, traditionally defined as 
\begin{align}\label{eq:MHD_invariant_charges}
\begin{split}
    &\mathrm{D}_s \mybar{0.8}{1pt}{g} = \beta_g(\mybar{0.8}{1pt}{g}), \qquad\qquad \mathrm{D}_s \mybar{0.8}{1pt}{u} = \beta_u(\mybar{0.8}{1pt}{g}, \mybar{0.8}{1pt}{u}), \\
    &\mybar{0.8}{1pt}{g}|_{s = 1} = g,  \qquad\qquad\quad~ \mybar{0.8}{1pt}{u}|_{s = 1} = u.
\end{split}
\end{align}
The remaining first integrals \cite{Vasilev_RG} of Eq. \eqref{eq:RG_equation_for_Wn_with_h_after_shift} do not explicitly depend on $s$ and satisfy
\begin{align}\label{eq:invariant_h_and_nu}
\begin{split}
    &\quad~\mathrm{D}_s \mybar{0.8}{1pt}{\nu} = -\mybar{0.8}{1pt}{\nu} \gamma_{\nu}(\mybar{0.8}{1pt}{g}), \quad \quad~~~ \mathrm{D}_s \mybar{0.8}{1pt}{B} = -\mybar{0.8}{1pt}{B} \gamma_{b}(\mybar{0.8}{1pt}{g}, \mybar{0.8}{1pt}{u}), \\
    &\quad~\mybar{0.8}{1pt}{\nu}|_{s = 1} = \nu,  \quad\qquad\qquad~~ \mybar{0.8}{1pt}{B}|_{s = 1} = B/\nu k.
\end{split}
\end{align}
For the invariant viscosity $\mybar{0.8}{1pt}{\nu}$ in the first of Eqs. \eqref{eq:invariant_h_and_nu}, there is a well-known \textit{exact} solution
\begin{align}\label{eq:exact_RG_solution_for_invariant_nu}
    \mybar{0.8}{1pt}{\nu} = \nu \exp{\left(-\int \limits_g^{\mybar{0.8}{1pt}{g}} \mbox{d} x\,\frac{\gamma_{\nu}(x)}{\beta_g(x)} \right)} = \left(\frac{g \nu^3}{\mybar{0.8}{1pt}{g} s^{2 \epsilon}} \right)^{1/3},
\end{align}
while for $\mybar{0.8}{1pt}{B}$, the corresponding solution may be formally written as
\begin{eqnarray}\label{eq:RG_solution_for_invariant_h}
    \mybar{0.8}{1pt}{B} = \left(\frac{B}{{\nu} k}\right) \exp{\left(-\int \limits_1^s \frac{\mbox{d}s'}{s'} \gamma_b\big(\mybar{0.8}{1pt}{g}(s', g), \mybar{0.8}{1pt}{u}(s', g, u)\big) \right)}.\quad~~
\end{eqnarray}
The scaling function $\mybar{1}{1pt}{R}$ in \eqref{eq:eq:arbitrary_GF_after_shift_through_invariant_variables} is then fixed by requiring that it match the corresponding perturbative expansion of $R$ from \eqref{eq:arbitrary_renormalized_GF_after_shift} at $s = 1$. where all invariant variables are equal to corresponding renormalized ones.

An IR-attractive (kinetic) fixed point described by Eqs. \eqref{eq:two-loop_kinetic_fixed_point} and \eqref{eq:two-loop_kinetic_fixed_point_vaslues} guarantees IR scaling for asymptotically small $\omega$, $k$ at fixed $g_0$, $\nu_0$, and $B_0$ (or $g$, $\nu$, $B$ and $\mu$ in renormalized variables).
On the other hand, according to the Kolmogorov's phenomenology of developed turbulence, one typically studies scaling at fixed  $\nu$ and $W$.
Thus, to obtain final results, one should express $g_0$ in terms of $W$.

As $s \to 0$, the invariant charges $\mybar{0.8}{1pt}{g}$ and $\mybar{0.8}{1pt}{u}$ approach the kinetic fixed point $(g_{\star}, u_{\star})$, which two-loop coordinates are given by Eqs. \eqref{eq:two-loop_kinetic_fixed_point} and \eqref{eq:two-loop_kinetic_fixed_point_vaslues}.
From \eqref{eq:exact_RG_solution_for_invariant_nu} and \eqref{eq:RG_solution_for_invariant_h}, it follows that the remaining invariant variables $\mybar{0.8}{1pt}{\nu}$ and $\mybar{0.8}{1pt}{B}$ exhibit nontrivial scaling:
\begin{align}
\label{eq:IR_scaling_of_invariant_nu_and_h}
    &\mybar{0.8}{1pt}{\nu}_{\star} \sim (g \nu^3/g_{\star})^{1/3}s^{-\gamma_{1\star}}, \quad \mybar{0.8}{1pt}{B}_{\star} \sim (B/\nu k) s^{-\gamma_{b \star}} {C}_{\star}.
\end{align}
where $\gamma_{b \star} = \gamma_{3\star}/2$ with $\gamma_{3\star}$ from \eqref{eq:two-lopp_gamma3}, and 
\begin{eqnarray}\label{eq:invariant_h_scaling_const}
    C_{\star} = \exp{\left(\int \limits_0^1 \frac{\mbox{d}s'}{s'} \Big(\gamma_b\big(\mybar{0.8}{1pt}{g}(s', g), \mybar{0.8}{1pt}{u}(s', g, u)\big) - \gamma_{b\star} \Big)\right)}. \quad~~
\end{eqnarray}

Perturbative expansions for $R$ yield a series in the dimensionless parameter $s^{-2\epsilon} g$, which tends to infinity as $s \to 0$.
The solution of RG equation reorganizes these terms into the improved variables $\mybar{0.8}{1pt}{g}(s)$ and $\mybar{0.8}{1pt}{u} (s)$, which approach the finite limit $(\mybar{0.8}{1pt}{g}(s), \mybar{0.8}{1pt}{u} (s)) \to (g_{\star}, u_{\star})$ as $s \to 0$, providing a partial resummation of the naive perturbation series.
Note, however, that the limit cannot be taken literally: in developed turbulence, the inertial range is bounded below by $m$ (see Eq. \eqref{eq:pump_function}), thus the smallest $s$ is of the order of $l_{\mathrm{mac}}^{-1}/\mu \propto (\mathrm{Re})^{-3/4}$.
Consequently, the maximum value of the expansion parameter $s^{-2\epsilon} g$ is $ \propto (\mathrm{Re})^{3\epsilon/2}$.

\subsection{\label{sec:spectrum}Energy spectrum}

The energy spectrum of turbulent pulsations is characterized by the critical dimensions of the corresponding fields.
In the dynamo regime, the full set of critical dimensions can be computed by standard arguments (see, e.g., \cite{Vasilev_RG})
Specifically, combining the RG equation \eqref{eq:RG_equation_for_Wn_with_h_after_shift} for an arbitrary $n$-point function $\mathcal{W}_{Rn}$ in coordinate space (where the dimension of $\mathcal{W}_{Rn}$ is the sum of the field dimensions) with the appropriate scale-invariant equations allows one to eliminate $\mu$ and $\nu$ fixed in the studied asymptotic.
Then, under the additional assumption that the scaling function $R$ remains finite\footnote{Note that in this case, $R$ indeed remains finite, and the ``problem of zeros of $R$-functions`` mentioned in Sec. \ref{sec:summary_of_previous_work} is absent.
This is due to the fact that in dynamo regime the model has a propagator $\Delta^{bb}$ (see Eq. \eqref{eq:propagator_matrix_after_shift}), the parameter $B$ in which is initially considered a critical dimensional quantity.
This distinguishes it from the charge $g_2$ from Eq. \eqref{eq:MHD_with_magnetic_noise}, which becomes critically dimensional only in the special case $g_{2\star} = 0$.}, using canonical dimensions from Table \ref{tab:MHD_canonical_dimensions} for $d = 3$ one obtains the MHD critical dimensions in the dynamo regime in the form
\begin{eqnarray}\label{eq:critical_dimensions_in_MHD}
\begin{split}
    &\Delta_{\omega} = - \Delta_t = 2 - \gamma_{1\star}, \quad \Delta_{\boldsymbol{b}} = \Delta_{\boldsymbol{v}} + \gamma_{3\star}/2, \qquad\\ 
    &\Delta_{\boldsymbol{v}} + \Delta_{\boldsymbol{v'}} = \Delta_{\boldsymbol{b}} + \Delta_{\boldsymbol{b'}} = 3, \quad \Delta_{\boldsymbol{v}} = 1 - \gamma_{1\star}.
\end{split}
\end{eqnarray}
Hence, the key result of the RG analysis for MHD in the dynamo regime is that, in contrast to the velocity field $\boldsymbol{v}$, the magnetic fields $\boldsymbol{B}$ and $\boldsymbol{b}$ acquire critical dimensions at the physical value $\epsilon = 2$ that differ from the Kolmogorov value $\Delta_{\boldsymbol{v}} = \Delta_{\boldsymbol{b}} = -1/3$ valid in the kinematic MHD regime (see Eqs. \eqref{eq:b_critical_dimension} and \eqref{eq:b'_critical_dimension} as well as the discussion in the end of page \hyperpage{11}).
This difference arises through $\gamma_{b\star}$, comes from the nontrivial renormalization of $\boldsymbol{b}$.
This fact is usually not taken into account in phenomenological generalizations of Kolmogorov theory to fully developed MHD turbulence.

Because $\Delta_{\boldsymbol{v}}$ and $\Delta_{\boldsymbol{b}}$ differ by $\gamma_{3\star}/2$, the equipartition between velocity and magnetic energies (present in the kinematic approximation) is \textit{violated} in the dynamo regime. 
The corresponding three-dimensional energy spectra coincide with static correlators $G^{v}(\boldsymbol{k})$ and $G^{b}(\boldsymbol{k})$ defined in Eq. \eqref{eq:static_magnetic_and_kinetic_correlators}.
In coordinate representation the critical dimensions of $G^{v}(\boldsymbol{k})$ and $G^{b}(\boldsymbol{k})$ are $2 \Delta_v = -2/3$ and $2 \Delta_b = -2/3 + \gamma_{3\star}$.
Therefore, in the Fourier representation, these become $2 \Delta_v = -11/3$ and $2 \Delta_b = -11/3 + \gamma_{3\star}$.
Although kinematic MHD would yield pure power-law scaling in the absence of an IR cutoff, once the new dimensional parameter $B$ appears, the RG analysis indicates that
\begin{align}
\label{eq:kinetic_and_magnetic_spectrums_in_dynamo_regime}
    &G^{v}(\boldsymbol{k}) = k^{-11/3} W^{2/3}R_v(\boldsymbol{\theta}, \rho), \\
    &G^{b}(\boldsymbol{k}) = k^{-11/3} W^{2/3} (k/\mu)^{\gamma_{3\star}} R_b(\boldsymbol{\theta}, \rho)
\end{align}
where $R_v$ and $R_b$ some scaling functions that depend on the dimensionless parameter $\boldsymbol{\theta} \coloneqq \boldsymbol{B}(k/W)^{1/3}(\mu/k)^{\gamma_{3}/2}$ and the helicity parameter $\rho$.
Let us emphasize once again that the Kolmogorov character of the velocity spectrum in our model is associated with the exact dimension $\Delta_{\boldsymbol{v}}$, which is a consequence of the exact relationship between the renormalization constants $Z_g = Z_{\nu}^{-3}$ (see Eq. \eqref{eq:Ren_consts_for_parameters}).
Also note that, as shown in Sec.~\ref{sec:two-loop_B_field_calculations}, $B$ does not require an IR cutoff $m$, nor do the other relevant model's parameters; hence, effectively, in our model the large-scale magnetic field is simply the homogeneous field. 

\section{\label{sec:conclusions}
Concluding remarks and speculations
}

Advances in modern statistical field theory applied to fully developed turbulence, together with recent rapid progress in computer algebra methods, have enabled us to make substantial progress in studying, probably, one of the most complex models known in statistical field theory: helical stochastic magnetohydrodynamics.
After nearly 40 years, we have succeeded in extending the one-loop results \cite{Adzhemyan1987} for the dynamo regime in the studied model to a two-loop order.

From a field-theoretical perspective, the model under consideration is, in many respects, a unique example.
The divergent loop corrections give rise to a mass-like contribution -- the so-called curl term.
Unlike in conventional field theory, where dynamical mass generation is typically desirable, here such a ``massive'' theory turns out to be unstable.
Therefore, the only way to make the model meaningful is to acknowledge this instability and then stabilize it by some means.
It is also worth noting that, unlike in the theory of critical phenomena -- where a model’s dynamics is often described by a phenomenological Langevin equation -- the statistical properties we study are governed by the genuine MHD equations, whose validity is well-established.

Performed two-loop ultraviolet renormalization has uncovered two possible mechanisms of stabilization. 
Each mechanism is associated with a distinct regime of the model, referred to as the kinematic approximation and the turbulent dynamo regime.
In the first regime, one effectively ``hard-wires'' a bare ``mass'' into the theory so that it precisely cancels the contributions arising from loop corrections.
In the second, it is assumed that alongside the dynamical generation of ``mass'', there is a spontaneous breaking of rotational symmetry, also induced by loop corrections.
This spontaneous symmetry breaking drives the system from the old vacuum state $\langle \boldsymbol{b}\rangle = 0$ toward a new, ``lower-energy'' vacuum with $\langle \boldsymbol{b}\rangle = \boldsymbol{B}$.
Technically, the amplitude of $\boldsymbol{B}$ is chosen to cancel the instability associated with the massive terms in the action.
Its direction remains arbitrary, as expected for spontaneous symmetry breaking.

In spirit, this latter approach resembles the well-known Coleman-Weinberg mechanism in field theory, though its practical implementation here is entirely different.
In the usual scenario of spontaneous symmetry breaking induced by radiative (loop) corrections, the mass of the field emerges from logarithmic quantum corrections to the effective potential (effective action, free energy, or first Legendre transform), determined by dimensionless coupling constants under fixed renormalization conditions.
In our model, however, the conventional effective-action approach does not fix the mean field and does not provides a complete picture of all the model’s vacua.
Instead, we proceed ``by contradiction'': on physical grounds, we assume the generation of a mean field and then demonstrate that the system can be stabilized by an appropriate choice of this field.
In fact, we assume that the observed mean field is the true ``ground state'' of the system, because the condition of masslessness (i.e., absence of the curl term) imposes a strong constraint that prevents the system from existing stably below the usual ``phase transition point'', as commonly encountered in critical phenomena.
In this sense, our situation is closer to that found in quantum mechanics and quantum field theory.

It should also be noted that the emergence of $\boldsymbol{B}$ introduces additional anisotropic structures aligned with its direction.
In particular, ``exotic'', ``bizarre'', and  ``drift'' contributions linear in the external momentum appear in the small-scale equations.
Under ideal MHD conditions (zero viscosity and resistivity), these would generate corrections to Alfv\'en waves that grow polynomially in time.
In field-theoretical language, spontaneous breaking of the continuous symmetry group ($\mathbf{SO}(3) \to \mathbf{SO}(2)$) entails Goldstone modes, which in our case manifest as polynomial corrections to Alfv\'en waves.
As in the Goldstone model, these modes are polarized perpendicularly to $\boldsymbol{B}$
and tend to restore the isotropy that was broken.
This subtle effect -- discovered first in \cite{Adzhemyan1987} -- demonstrates that understanding the requirement of ``insignificance of viscosity'' in the inertial interval, stemming from Kolmogorov’s phenomenology, \textit{literally as the absence of viscosity in the formulas} is incorrect.

In addition to its fundamental field-theoretic aspects, the model involves highly complex computational techniques that may interest researchers in pure computational mathematics.
In this work, we provide as comprehensive a description as possible of the method we developed for computing loop diagrams within the dynamo regime.
In principle, this framework can serve as a foundation for further exploration of the regime, including studies of anisotropic, stratified, or rotating systems, as well as higher-order loop calculations -- which, given their enormous complexity, are likely to remain topics for the distant future.

Turning to the physical context, the large-scale turbulent dynamo studied here -- which produces large-scale magnetic fields -- is an indisputable physical reality, critically important for describing the nonlinear dynamics of astrophysical and geophysical systems, as well as for various technological applications.
Moreover, our formalism is ideally suited to describing the dynamo directly in the regime of fully developed turbulence.

It is worth noting that our analysis of the two-loop stabilizing mean field was performed without an infrared cutoff (i.e. at $m = 0$).
Consequently, in our model, the large-scale field naturally becomes homogeneous.
In real settings, however, such a homogeneous field often assumes a dipole structure.
Our approach employs statistical averaging over an ensemble of turbulent states, whereas in practice one typically averages over space or time (e.g., over a sphere or over one rotation period).

From a spectral perspective, the appearance of a spontaneous field $\boldsymbol{B}$ yields a three-dimensional anisotropic spectrum.
This spectrum is characterized not only by a power-law dependence on momentum but also by an additional scaling function. 
By virtue of the identity $Z_g = Z_{\nu}^{-3}$, the inertial-range velocity sector maintains the Kolmogorov slope $-11/3$.
By contrast, in the dynamo regime, the magnetic sector acquires a slope steeper than $-11/3$, specifically $-11/3 + \gamma_{3 \star}$, reflecting a fundamental departure from energy equipartition.
Because the one-loop renormalization constants are independent of $\rho$, a one-loop analysis alone cannot fully illuminate the role of helicity in the model.
At two-loop order, however, this limitation is resolved; notably, the helicity-dependent two-loop contribution to $\Delta_{\boldsymbol{b}}$ exceeds its nonhelical counterpart by a factor of $10$ (see Eq. \eqref{eq:two-lopp_gamma3}).

Unfortunately, a loop analysis does not answer all possible questions.
Our results suggest that helicity, rather than magnetic noise, is the fundamental cause of the dynamo.
On the other hand, the one-loop value of the spontaneous magnetic field $\boldsymbol{B}$ is independent of $\rho$, and the two-loop helical contribution is $100$ times smaller than the nonhelical one.
As can be seen from Eq. \eqref{eq:two-loop_B}, the $\rho$-dependence under the $\epsilon$-expansion only emerges through the expansion of the relevant one-loop term.
Hence, in general, even an infinitesimal helicity generates a finite spontaneous field. 
One might then ask \cite{Adzhemyan1987} whether a dynamo could arise in a normal fluid with $\rho = 0$?
Using only the local stability criterion for a stochastic dynamical system -- i.e., the tendency of its response function to vanish at large times -- one cannot conclusively answer.
In a normal fluid, the state $\langle \boldsymbol{b}\rangle = 0$ is stable against small perturbations (since curl terms are forbidden by symmetry at $\rho = 0$), just as a helical fluid with $\langle \boldsymbol{b}\rangle = \boldsymbol{B}$ is also stable.
There is no, up to our knowledge, a known global criterion that can distinguish which of these two locally stable states is the genuine one, leaving this question open, as it was four decades ago.

There is, however, an important unresolved issue that we hope to address in the near future.
It is straightforward to see that the parameter $h_0$ -- which triggers the instability and fixes the amplitude of the spontaneous field -- is directly related to the transport coefficient $\alpha_0$ from Eq.~\eqref{eq:gradient_expansion_for_EMF}, i.e., to the so-called $\alpha$-effect.
This $\alpha$-effect, in turn, is linked to the turbulent electromotive force $\mathcal{E}$, determined by the correlator $\langle \boldsymbol{v} \otimes \boldsymbol{b}\rangle$ arose in the dynamo regime through the second relation in Eq.~\eqref{eq:mean-field_dynamo}.
Using the two-loop estimates for the spontaneous magnetic field derived in this paper, we intend to perform a detailed renormalization group study of the $\alpha$-effect in helical turbulent MHD in the near future.

\begin{acknowledgments}
The authors are indebted to M. Nalimov for useful discussions.
The work was supported by VEGA Grant \textnumero 1/0297/25 of the Ministry of Education, Science, Research and Sport of the Slovak Republic, and project VVGS‐2023‐2567 of the internal
grant scheme of Faculty of Science, \v{S}af\'arik University in Ko\v{s}ice.
\end{acknowledgments}

\appendix

\begin{widetext}
\section{\label{app:two-loop_Gammab'b_expressions}Two-loop \pdfmath{\Lambda}-divergent parts of \texorpdfstring{$\Sigma^{b'b}$}{} diagrams}

The explicit form of the integrands $F_l \coloneqq F_l(p, q, z)$ for $l = 1,\ldots,8$ in Eqs. \eqref{eq:two_loop_bb'_linear-asymp} and \eqref{eq:representation_for_diagrams_comes_from_homogeneity} is as follows:
\begin{eqnarray*}
    &F_1 = -\frac{q^2 \left(z^2-1\right) (p-q) \left(p^2-q z (2 p+q)\right) \left(2 u_0 \left(2
   p^2+p q z+q^2\right)+p^2 u_0^2+7 p^2+6 p q z+6 q^2\right)}{192 \pi ^4
   \left(u_0+1\right){}^2 \left(p^2-2 p q z+q^2\right) \left(p^2+p q z+q^2\right)
   \left(p^2 u_0+p^2+2 p q z+2 q^2\right)}, \\
   &F_2 = -\frac{q^2 \left(z^2-1\right) (p+q)}{48 \pi ^4 \left(u_0+1\right){}^2 \left(u_0
   \left(p^2+2 p q z+q^2\right)+p^2+q^2\right)}, \qquad
   F_3 = -\frac{q^2 \left(z^2-1\right) (p-q) \left(p^2+q z (2 p+q)\right)}{192 \pi ^4
   \left(u_0+1\right) \left(p^2-p q z+q^2\right) \left(p^2+2 p q z+q^2\right)}, \\
   &F_4 = \frac{p^2 q^2 \left(z^2-1\right) (p-q)^2 (p+q) \left(u_0 \left(p^2+2 p q z+q^2\right)+2
   \left(p^2+p q z+q^2\right)\right)}{96 \pi ^4 \left(u_0+1\right) \left(p^2+p q
   z+q^2\right) \left(p^2+2 p q z+q^2\right)^2 \left(u_0 \left(p^2+2 p q
   z+q^2\right)+p^2+q^2\right)}, \\
   &F_5 = \frac{p^2 q \left(z^2-1\right) (p-q) (p z-q) \left(p^2+p (q-q z)+q^2\right)}{96 \pi ^4 \left(u_0+1\right){}^2 \left(p^2-2 p q z+q^2\right)^2} \\
   &\times \frac{\left(6
   p^4-10 p^3 q z+q^2 u_0^2 \left(p^2-2 p q z+q^2\right)+2 u_0 \left(p^2-2 p q z+3
   q^2\right) \left(p^2-p q z+q^2\right)+p^2 q^2 \left(4 z^2+11\right)-8 p q^3 z+5
   q^4\right)}{
   \left(p^2-p q z+q^2\right) \left(2 p (p-q z)+q^2 u_0+q^2\right) \left(u_0
   \left(p^2-2 p q z+q^2\right)+p^2+q^2\right)},\\
   &F_6 = \frac{p q z \left(z^2-1\right) (p+q)}{48 \pi ^4 \left(u_0+1\right){}^2 \left(u_0
   \left(p^2+2 p q z+q^2\right)+p^2+q^2\right)}, \\
   &F_7 = \frac{p q z \left(z^2-1\right) (p-q) \left(p^2+p (q-q z)+q^2\right) \left(2 p (p-q
   z)+q^2 u_0+3 q^2\right)}{96 \pi ^4 \left(u_0+1\right){}^2 \left(p^2-2 p q
   z+q^2\right) \left(p^2-p q z+q^2\right) \left(2 p (p-q z)+q^2 u_0+q^2\right)}, \\
   &F_8 = -\frac{p q^2 \left(z^2-1\right) (p-q) (p+q z) \left(p^2+p q (z+1)+q^2\right)}{96 \pi ^4
   \left(u_0+1\right) \left(p^2+p q z+q^2\right) \left(p^2+2 p q z+q^2\right) \left(u_0
   \left(p^2+2 p q z+q^2\right)+p^2+q^2\right)}.
\end{eqnarray*}

\section{\label{app:scalarization_of_typical_integral}Example of tensor reduction of a typical integral}

For brevity and clarity, we label vector indices with Greek letters and omit the explicit dependence on the integrand’s arguments, $F \coloneqq F\left(p,q, (\boldsymbol{p} \cdot \boldsymbol{q}), (\boldsymbol{p} \cdot \boldsymbol{B}_0), (\boldsymbol{q} \cdot \boldsymbol{B}_0)\right)$.
A typical two-loop integral from our calculations is of the form \eqref{eq:tensor_integral_general_form} with $\boldsymbol{k}_1 = \ldots = \boldsymbol{k}_s = \boldsymbol{0}$ and $n = m = 2$: 
\begin{align}\label{eq:typical_tensor_integral_with_four_indices}
    J_{\nu \mu \alpha \beta}^{(2,2)} = \int \mbox{d}^d p \,\, p_{\nu} p_{\mu} \left[\int \mbox{d}^d q \,\, q_{\alpha} q_{\beta} F \right] = \int \mbox{d}^d p \,\, p_{\nu} p_{\mu} J_{\alpha \beta}^{(0,2)}, \qquad
    J_{\alpha \beta}^{(0,2)} \coloneqq \int \mbox{d}^d q \,\, q_{\alpha} q_{\beta} F.
\end{align}
Here, for two-loop integrals $J_{\alpha_1\cdot \alpha_m \beta_1 \cdot \beta_n}^{(m,n)}$: the superscripts $m$ and $n$ indicate the tensor rank with respect to the momenta $\boldsymbol{p}$ and $\boldsymbol{q}$, respectively.
Without loss of generality, we also assume that in two-loop integrals the integration over $\boldsymbol{q}$ is always carried out first, so that the vector $\boldsymbol{p}$ then effectively serves as an external momentum.

In accordance with steps (a) and (b) of the scalarization procedure outlined in Sec. \ref{sec:scalarization}, the integral $J_{\alpha \beta}$ can be written as
\begin{eqnarray}\label{eq:typical_tensor_integral_with_two_indices}
    J_{\alpha \beta}^{(0,2)} = \delta_{\alpha \beta}f_0 + \hat{B}_{0\alpha}\hat{B}_{0\beta}f_1 + p_{\alpha}p_{\beta}f_2 + (p_{\alpha}\hat{B}_{0\beta} + p_{\beta}\hat{B}_{0\alpha})f_3, \qquad f_i \coloneqq \int \mbox{d}^d q \,\, C_i^{(0,2)}(\boldsymbol{p}, \boldsymbol{q})F. \quad i = 0,1,2,3.
\end{eqnarray}
Here, the coefficients $C_i^{(0,2)}(\boldsymbol{p}, \boldsymbol{q})$, $i = 0,1,2,3$ are obtained from the associated system of linear equations, which is derived (under the $\boldsymbol{q}$-integration) by successively contracting both sides of \eqref{eq:typical_tensor_integral_with_two_indices} with $\delta_{\alpha\beta}$, $\hat{B}_{0\alpha}\hat{B}_{0\beta}$, $p_{\alpha}p_{\beta}$ and $p_{\alpha}\hat{B}_{0\beta}$.
Introducing the angular variables $z \coloneqq (\boldsymbol{p} \cdot \boldsymbol{q})/pq$, $z_p \coloneqq (\boldsymbol{p} \cdot \boldsymbol{B}_0)/p B_0$, and $z_q \coloneqq (\boldsymbol{q} \cdot \boldsymbol{B}_0)/q B_0$ for the coefficients $C_i^{(0,2)}(\boldsymbol{p}, \boldsymbol{q})$, we obtain:
\begin{eqnarray}\label{eq:scalarization_coefficients_with_subscript2}
    &C_0^{(0,2)} = \frac{q^2 \left(-2 z z_p z_q+z_p^2+z_q^2+z^2-1\right)}{(d-2) \left(z_p^2-1\right)}, \quad C_1^{(0,2)} = \frac{q^2 \left(-2 (d-1) z z_p z_q+\left((d-2) z^2+1\right) z_p^2+(d-1) z_q^2+z^2-1\right)}{(d-2)
   \left(z_p^2-1\right)^2}, \notag \\
   &C_2^{(0,2)} = \frac{q^2 \left(-2 (d-1) z z_p z_q+z_p^2 \left((d-2) z_q^2+1\right)+(d-1) z^2+z_q^2-1\right)}{(d-2) p^2
   \left(z_p^2-1\right)^2}, \\
   &C_3^{(0,2)} = \frac{q^2 \left(-\left((d-1) z_p z_q^2\right)+z z_q \left(d z_p^2+d-2\right)+z_p \left(-\left((d-1)
   z^2\right)-z_p^2+1\right)\right)}{(d-2) p \left(z_p^2-1\right)^2}. \notag
\end{eqnarray}
We formally provide these scalarization coefficients for an arbitrary dimension $d > 2$.
However, for the purposes of Sec. \ref{sec:one_loop_approximation_after_shift}, these expressions are relevant primarily for $d = 3$.

Substituting \eqref{eq:typical_tensor_integral_with_two_indices} into $J_{\nu \mu \alpha \beta}^{(2,2)}$ yields a representation of the form
\begin{eqnarray}\label{eq:intermediate_representation_for_four-index_tensor_integral}
    J_{\nu \mu \alpha \beta}^{(2,2)} = \delta_{\alpha \beta} \int \mbox{d}^d p \,\, p_{\nu} p_{\mu} f_1 + \hat{B}_{0\alpha}\hat{B}_{0\beta} \int \mbox{d}^d p \,\, p_{\nu} p_{\mu} f_2 + \int \mbox{d}^d p \,\, p_{\nu} p_{\mu} p_{\alpha} p_{\beta} f_3 + \int \mbox{d}^d p \,\, p_{\nu} p_{\mu} (p_{\alpha}\hat{B}_{0\beta} + p_{\beta} \hat{B}_{0\alpha})f_4.
\end{eqnarray}
From \eqref{eq:intermediate_representation_for_four-index_tensor_integral} it follows that, for the final result, one needs the expansion coefficients of the following three prototypical integrals of the scalar function $f \coloneqq f(\boldsymbol{p}, (\boldsymbol{p} \cdot \boldsymbol{B}_0))$:
\begin{eqnarray}
    &&J_{\mu \nu}^{(2,0)} \coloneqq \int \mbox{d}^d p \,\, p_{\nu} p_{\mu} f = \delta_{\mu \nu} \int \mbox{d}^d p \,\, C_0^{(2,0)}(\boldsymbol{p}) f + \hat{B}_{0\mu}\hat{B}_{0\nu} \int \mbox{d}^d p \,\, C_1^{(2,0)}(\boldsymbol{p}) f, \label{eq:typical_tensor_integral_with_two_indices_without_external_momentum}\\
    &&J_{\mu \nu \alpha \beta}^{(4,0)} \coloneqq \int \mbox{d}^d p \,\, p_{\nu} p_{\mu} p_{\alpha} p_{\beta} f = \left(\delta_{\mu \nu}\delta_{\alpha \beta} + \delta_{ \alpha\nu} \delta_{\mu \beta} + \delta_{\nu \beta}\delta_{\alpha \mu}\right) \int \mbox{d}^d p \,\,C_0^{(4,0)}(\boldsymbol{p}) f + \hat{B}_{0 \nu}\hat{B}_{0 \mu}\hat{B}_{0\alpha} \hat{B}_{0\beta} \int \mbox{d}^d p \,\,C_2^{(4,0)}(\boldsymbol{p}) f \notag \\
    &&+ \left(\delta_{\nu \mu}\hat{B}_{0\alpha}\hat{B}_{0\beta} + \delta_{\alpha \beta}\hat{B}_{0\mu}\hat{B}_{0\nu} + \delta_{\nu \alpha}\hat{B}_{0\mu}\hat{B}_{0\beta} +  \delta_{\mu \beta}\hat{B}_{0\nu}\hat{B}_{0\alpha} + \delta_{\nu \beta}\hat{B}_{0\mu}\hat{B}_{0\alpha} + \delta_{\mu \alpha}\hat{B}_{0\nu}\hat{B}_{0\beta} \right)\int \mbox{d}^d p \,\,C_1^{(4,0)}(\boldsymbol{p}) f, \label{eq:typical_tensor_integral_with_four_indices_without_external_momentum} \\
    &&J_{\mu \nu \alpha}^{(3,0)} \coloneqq \int \mbox{d}^d p \,\, p_{\nu} p_{\mu} p_{\alpha} f = \left(\delta_{\nu \mu}\hat{B}_{0 \alpha} + \delta_{\nu \alpha} \hat{B}_{0 \mu} + \delta_{\alpha \mu} \hat{B}_{0 \nu}\right)\int \mbox{d}^d p \,\,C_0^{(3,0)}(\boldsymbol{p}) f + \hat{B}_{0\nu}\hat{B}_{0\mu}\hat{B}_{0\alpha} \int \mbox{d}^d p \,\,C_1^{(3,0)}(\boldsymbol{p}) f \label{eq:typical_tensor_integral_with_three_indices_without_external_momentum}
\end{eqnarray}
By solving the associated systems of equations in the same manner as was done for $J_{\alpha \beta}^{2,0}$ in \eqref{eq:typical_tensor_integral_with_two_indices}, we find the remaining three sets of scalar coefficients:
\begin{align}\label{eq:scalarization_coefficients_with_subscripts1_3_4}
\begin{split}
    &C_0^{(2,0)} = \frac{p^2(1 - z_p^2)}{d - 1}, \quad C_1^{(2,0)} = \frac{p^2(z_p^2 d - 1)}{(d-1)}, \quad
    C_0^{(4,0)} = \frac{p^3 z_p \left(1-z_p^2\right)}{(d-1)}, \quad
    C_1^{(4,0)} = \frac{p^3 z_p \left((d+2) z_p^2-3\right)}{(d-1)}, \\
    &C_0^{(3,0)} = \frac{p^4 \left(1-z_p^2\right)^2}{d^2-1}, \quad 
    C_1^{(3,0)} = \frac{p^4 \left(1-z_p^2\right) \left((d+2) z_p^2-1\right)}{\left(d^2-1\right)}, \quad 
    C_2^{(3,0)}= \frac{p^4 \left((d+2) z_p^2 \left((d+4) z_p^2-6\right)+3\right)}{ \left(d^2-1\right)}.
\end{split}
\end{align}
Finally, substituting Eqs. \eqref{eq:typical_tensor_integral_with_two_indices_without_external_momentum}--\eqref{eq:typical_tensor_integral_with_three_indices_without_external_momentum} into \eqref{eq:intermediate_representation_for_four-index_tensor_integral} yields the following decomposition of the tensor integral $J_{\nu \mu \alpha \beta}$ in terms of the metric and the external magnetic field $\boldsymbol{B}_0$:
\begin{align}\label{eq:final_decomposition_for_four_index_tensor_integral}
    J_{\nu \mu \alpha \beta}^{(2,2)} &= \int \mbox{d}^d p\,\mbox{d}^d q\,\,F \Bigg[
    \delta_{\alpha \beta}\delta_{\mu \nu}\left( C_0^{(2,0)} C_0^{(0,2)}  + C_0^{(3,0)} C_2^{(0,2)} \right)
    + \left(\delta_{\nu \alpha}\delta_{\mu \beta} + \delta_{\nu \beta}\delta_{\mu \alpha}\right)C_0^{(3,0)} C_2^{(0,2)} \notag \\
    &+ \delta_{\alpha \beta}\hat{B}_{0 \nu}\hat{B}_{0 \mu} \left(C_1^{(2,0)} C_0^{(0,2)} + C_1^{(3,0)} C_2^{(0,2)} \right)
    + \delta_{\nu \mu}\hat{B}_{0\alpha}\hat{B}_{0\beta} \left(C_0^{(2,0)} C_1^{(0,2)} + C_1^{(3,0)} C_2^{(0,2)} + 2C_0^{(4,0)} C_3^{(0,2)} \right) \notag\\
    &+ \left(\delta_{\nu \alpha} \hat{B}_{0\mu}\hat{B}_{0\beta} + \delta_{\mu \beta}\hat{B}_{0\nu}\hat{B}_{0\alpha} + \delta_{\nu \beta}\hat{B}_{0\mu}\hat{B}_{0\alpha} + \delta_{\mu \alpha}\hat{B}_{0\nu} \hat{B}_{0\beta}\right) \left(C_1^{(3,0)} C_2^{(0,2)} + C_0^{(4,0)} C_3^{(0,2)} \right) \notag\\
    &+ \hat{B}_{0\alpha} \hat{B}_{0\beta} \hat{B}_{0\nu}\hat{B}_{0\mu} \left(C_1^{(2,0)} C_1^{(0,2)} + C_2^{(3,0)} C_2^{(0,2)} + 2 C_1^{(4,0)} C_3^{(0,2)} \right)
    \Bigg].
\end{align}

The sequential construction described above leads to a well-known approach: for a rank-$r$ tensor integral, one can directly write the most general linear combination of all rank-appropriate symmetric tensors (with respect to a single momentum) formed from the metric and the external vector parameters of the integral.
In that case, the required scalar coefficients are derived in one step (rather than incrementally, as done above), thereby significantly reducing the computational effort.
Naturally, this procedure is consistent, and the final scalar coefficients are independent of whether they are computed incrementally or from a single, fully symmetric expansion.
To illustrate, let us consider a simple example with $J_{\nu \beta}^{(1,1)}$.
According to the above procedure, one obtains for $J_{\nu \beta}^{(1,1)}$ the decomposition
\begin{align}\label{eq:step-by-step_decomposition_for_typical_integral_J(1,1)}
    J_{\nu \beta}^{(1,1)} &= \int \mbox{d}^d p \,p_{\nu}\left[p_{\beta}\int \mbox{d}^d q\, C_0^{(0,1)}F + \hat{B}_{0\beta}\int \mbox{d}^d q \,C_1^{(0,1)}F\right] \notag \\
    &= \int \mbox{d}^d p \,\mbox{d}^d q\,\,F \left[\delta_{\nu \beta} C_0^{(2,0)} C_0^{(0,1)} + \hat{B}_{0\nu}\hat{B}_{0\beta} \left(C_1^{(2,0)} C_0^{(0,1)} + C_0^{(1,0)}C_1^{(0,1)}\right)\right].  
\end{align}
Conversely, from general arguments, one obtains
\begin{align}\label{eq:general_decomposition_for_typical_integral_J(1,1)}
    &J_{\nu \beta}^{(1,1)} \coloneqq \int \mbox{d}^d p\, \mbox{d}^d q\,\,p_{\nu}q_{\beta} F = \delta_{\nu \beta} \int \mbox{d}^d p \,\mbox{d}^d q\,\,C_0^{(1,1)}F + \hat{B}_{0\nu}\hat{B}_{0\beta} \int \mbox{d}^d p \,\mbox{d}^d q\,\,C_1^{(1,1)}F.
\end{align}
For consistency, the following identities must hold:
\begin{align}\label{eq:scalarization_coefficient_identities_for_J(1,1)}
    C_0^{(1,1)} \equiv C_0^{(2,0)} C_0^{(0,1)}, \qquad \qquad C_1^{(1,1)} \equiv \left(C_1^{(2,0)} C_0^{(0,1)} + C_0^{(1,0)}C_1^{(0,1)}\right),
\end{align}
which can be readily confirmed by computing all the coefficients explicitly.

Let us also make an important remark regarding scalar products.
In diagrammatic tensor structures, one encounters not only monomials $p_{\alpha_1}\cdots p_{\alpha_m} q_{\beta_1}\cdots q_{\beta_n}$ with free indices, but also scalar products of loop momenta, $(\boldsymbol{p}\cdot\boldsymbol{q}) = p_{\mu}q_{\mu}$.
If one wishes, such scalar products can be subsumed into monomials $p_{\alpha_1}\cdots p_{\alpha_m} q_{\beta_1}\cdots q_{\beta_n}$ in a general manner, with the convolution over the relevant indices performed at the conclusion.
Although this raises the rank of the tensor integrals and thereby increases computational complexity, the end result remains unchanged from the approach in which all scalar products are incorporated into the definition of the scalar function $F$.
For instance, for the integral $J_{\nu \beta}^{(1,1)}$ with the function $(\boldsymbol{p}\cdot\boldsymbol{q})F$, the expansion is of the form \eqref{eq:general_decomposition_for_typical_integral_J(1,1)}.
One can also obtain the same expansion by contracting a fourth-rank tensor $J_{\nu \mu \alpha \beta}^{(2,2)}$ from \eqref{eq:final_decomposition_for_four_index_tensor_integral} with $\delta_{\mu \alpha}$.
In particular, this produces the identity
\begin{align}\label{eq:eq:scalarization_coefficient_identities_for_J(1,1)_through_J(2,2)}
   (\boldsymbol{p}\cdot\boldsymbol{q}) C_0^{(1,1)} \equiv \left(C_0^{(2,0)} C_0^{(0,2)}  + C_0^{(3,0)} C_2^{(0,2)}\right) + (1 + d)\left(C_0^{(3,0)} C_2^{(0,2)}\right) + \left(C_1^{(3,0)} C_2^{(0,2)} + C_0^{(4,0)} C_3^{(0,2)}\right) 
\end{align}
as well as an analogous relation for $C_1^{(1,1)}$.
Verifying these identities is straightforward.
The analysis of tensor structures at $B_0 = 0$ can also be carried out using the technique described above.
In fact, at $B_0 = 0$, our diagrams can be regarded as vacuum diagrams in a $\varphi^{4}$-type theory.
For these, our final result for tensor reduction coincides with that presented in \cite{Davydychev1991}.

\end{widetext}

\section{\label{app:tensor_structure_of_specific_diagram}Tensor structure of a specific diagram}

Here, we present an example of the scalarization of the linear asymptotics $\boldsymbol{k} \rightarrow 0$ (the curl term) for the diagram $D_{\mathrm{L}2}^{(2)}$, depicted in Fig. \ref{fig:two-loop_surface_divergent_diagrams_before_shift}.
The tensor structure of any diagram is determined by the product of the tensor operators $\mathbb{P}_{i j}$ and $\mathbb{R}_{ij}$, which represent the diagram lines \eqref{eq:propagator_matrix_after_shift}, along with the associated vertex factors $\mathbb{U}_{i j l}$, $\mathbb{W}_{i j l}$, and $\mathbb{V}_{i j l}$ from \eqref{eq:interaction_vertices}.
Upon fully contracting all tensor indices, the term in $D_{\mathrm{L}2}^{(2)}$ linear in the incoming momentum $\boldsymbol{k}$ takes the form
\begin{widetext}
\begin{eqnarray}
\label{eq:curl_asymptotic_of_problematic_diagram_tensor_structure}
    \mathbb{T}_{ij} = 2i \rho p_j \Bigg[\varepsilon_{i \nu \alpha} \Bigg(p_{\nu}q_{\alpha}\left(\frac{(\boldsymbol{k} \cdot \boldsymbol{q})}{q} - \frac{(\boldsymbol{k} \cdot \boldsymbol{p})(\boldsymbol{q} \cdot \boldsymbol{p})}{q p^2}\right) - k_{\nu}p_{\alpha}\left(p - \frac{(\boldsymbol{q}\cdot \boldsymbol{p})^2}{p q^2}\right)\Bigg) + \varepsilon_{\mu \nu \alpha}k_{\mu}q_{\alpha}p_{\nu}\left(p_i \frac{(\boldsymbol{p}\cdot\boldsymbol{q})}{q p^2} -\frac{q_i}{q} \right)\Bigg].\quad
\end{eqnarray}
\end{widetext}

Therefore, the scalarization of the linear-in-$\boldsymbol{k}$ asymptotics of $D_{\mathrm{L}2}^{(2)}$ reduces to scalarizing the three prototypical integrals in the formula:
\begin{align}\label{eq:expression_of_problematic_diagram_linear_asymptotic_in_terms_of_typical_tensor_integrals}
    &D_{\mathrm{L}2\, i j}^{(2)} = i\rho \varepsilon_{i \nu \alpha}k_{\mu} J_{\nu j \alpha \mu}^{(2,2)} + i\rho \varepsilon_{i \nu \alpha}k_{\mu} J_{\mu \nu j \alpha}^{(3,1)} + i \rho \varepsilon_{\mu \nu \alpha}k_{\mu} J_{\nu j\alpha i}^{(2,2)} \notag \\
    &+ i \rho \varepsilon_{\mu \nu \alpha}k_{\mu} J_{\nu i j\alpha}^{(3,1)} + i \rho \varepsilon_{i \mu \alpha}k_{\mu} J_{\alpha j}^{(2,0)},
\end{align}
where the notation for the tensor integrals follows Appendix \ref{app:scalarization_of_typical_integral}.
The pertinent scalar functions for the integrals in \eqref{eq:expression_of_problematic_diagram_linear_asymptotic_in_terms_of_typical_tensor_integrals} arise by multiplying all the scalar factors (including scalar products of loop momenta) in \eqref{eq:curl_asymptotic_of_problematic_diagram_tensor_structure} by some general function $F\left(p,q, (\boldsymbol{p} \cdot \boldsymbol{q}), (\boldsymbol{p} \cdot \boldsymbol{B}_0), (\boldsymbol{q} \cdot \boldsymbol{B}_0)\right)$.
This function is obtained by integrating over frequencies the product of the scalar parts of the diagram’s propagators.
For integrals of the type $J_{\mu \nu j \alpha}^{(3,1)}$, one can build their decomposition in terms of the metric and the external field $\boldsymbol{B}_0$ following the same strategy outlined in Appendix \ref{app:scalarization_of_typical_integral}.
We omit it here because these integrals (as well as any integrals with an odd number of $\boldsymbol{p}$ or $\boldsymbol{q}$ factors) do not contribute to the curl terms, i.e. to structures of the type $i \rho \varepsilon_{i j \mu} k_{\mu}$ (they do, however, affect other linear-in- $\boldsymbol{k}$ structures, as discussed below).

By substituting the decomposed tensor integrals (in terms of the metric and the external field $\boldsymbol{B}_0$ into \eqref{eq:expression_of_problematic_diagram_linear_asymptotic_in_terms_of_typical_tensor_integrals}, we obtain
\begin{align}\label{eq:final_form_of_problematic_diagram_linear_asymptotic}
    &D_{\mathrm{L}2\, i j}^{(2)} = i \rho \varepsilon_{i j \mu}k_{\mu} J_1 + i \rho \left[\boldsymbol{k} \times  \boldsymbol{\hat{B}}_0\right]_i \hat{B}_{0 j} J_2 \notag \\
    &+ i \rho \left[\boldsymbol{k}\times\boldsymbol{\hat{B}}_0 \right]_j \hat{B}_{0 i} J_3 + i \rho (\boldsymbol{k} \cdot \boldsymbol{\hat{B}}_0)\varepsilon_{i j \mu}\hat{B}_{0 \mu}J_4,
\end{align}
where $J_s$ are certain scalar integrals. 

\begin{widetext}
\section{\label{app:Alfven_waves_with_exotic_corrections}Alfv\'en waves with exotic corrections}

Here we present a complete one-loop solution of the system \eqref{eq:linearized_equations_of_motion_in_matrix_form}, containing power-law corrections to the exponentially decaying Alfv\'en waves.
\begin{align}
    b_2(t) = \frac{\mathrm{e}^{-\frac{1}{2} k^2 \nu _0 t \left(u_0 + 1\right)}}{2 a \gamma }&\left((k^2 \nu _0 |u_0 - 1| (i c_3 k^2 \nu _0 |u_0 - 1| - 2 \gamma  c_1) + i a^2 c_3)\sin \left(\frac{a
    t}{2}\right) + 2 a \gamma  c_1 \cos
    \left(\frac{a t}{2}\right)\right), \label{eq:b2_component_of_Alfven_wave_with_exotic_terms}\\
    b_3(t) = \frac{\mathrm{e}^{-\frac{1}{2} k^2 \nu _0 t \left(u_0 + 1\right)}}{2 a^2 \gamma}&\Bigg[ \sin \left(\frac{a t}{2}\right) \Big(a k^4 \nu _0^2 (u_0 - 1)^2 (i c_4 - c_3 \lambda  t
   (a^2 - 4 \gamma ^2)) - 2 a k^2 \nu _0 |u_0 - 1| (a^2 c_3 \lambda +\gamma  c_2 - 8 i
   \gamma^3 c_1 \lambda  t ) \notag\\
   &+ a (4 \gamma ^2 \lambda  \left(a^2 c_3 t - 4 i \gamma  c_1\right) + i a^2
   c_4) - a c_3 \lambda  k^8 \nu_0^4 t (u_0 - 1)^4 - 2 a \lambda  k^6 \nu_0^3 |u_0 - 1|^3
   (2 i \gamma  c_1 t + c_3)\Big) \notag\\
   &+ \cos \left(\frac{a
   t}{2}\right) \Big(-c_3 \lambda  k^6 \nu _0^3 t |u_0 - 1|^3 (a^2 - 4 \gamma ^2) + 4 a^2 \gamma
   ^2 c_3 \lambda  k^2 \nu_0 t |u_0 - 1| + 2 \gamma  (a^2 c_2 - 16 i \gamma ^4 c_1 \lambda  t) \notag\\
   &- c_3 \lambda  k^{10} \nu _0^5 t |u_0-1|^5-4 i \gamma  c_1 \lambda  k^8 \nu _0^4 t
   (u_0 - 1)^4 + 24 i \gamma ^3 c_1 \lambda  k^4 \nu_0^2 t \left(u_0 - 1\right)^2\Big) \Bigg] \label{eq:b3_component_of_Alfven_wave_with_exotic_terms}\\
    v_2(t) = \frac{\mathrm{e}^{-\frac{1}{2} k^2 \nu _0 t \left(u_0+1\right)}}{a} &\left( (c_3 k^2 \nu _0 |u_0 - 1| + 2 i \gamma  c_1 )\sin \left(\frac{a
   t}{2}\right) + a c_3 \cos \left(\frac{a t}{2}\right)\right), \label{eq:v2_component_of_Alfven_wave_with_exotic_terms}\\
    v_3(t) = \frac{\mathrm{e}^{-\frac{1}{2} k^2 \nu _0 t \left(u_0 + 1\right)}}{a^2}& \Bigg[i \sin \left(\frac{a t}{2}\right) \Big(2 a (a^2 c_3
   \lambda + \gamma  c_2 - 4 i \gamma ^3 c_1 \lambda  t ) + 2 a \lambda  k^4 \nu_0^2 (u_0 - 1)^2 (i
   \gamma  c_1 t + c_3) \notag \\
   &+i a k^2 \nu_0 |u_0 - 1| (4 \gamma  c_1 \lambda - c_4) \Big) + i \cos
   \left(\frac{a t}{2}\right) \Big(c_3 \lambda  k^4 \nu _0^2 t
   \left(u_0 - 1\right)^2 (a^2 - 4 \gamma ^2) - a^2 (4 \gamma ^2 c_3 \lambda  t + i c_4) \notag \\
   &- 8 i \gamma^3 c_1 \lambda  k^2 \nu_0 t |u_0 - 1| + \lambda  k^6 \nu_0^3 t |u_0-1|^3 (2 i \gamma 
   c_1 + c_3 k^2 \nu_0 |u_0 - 1|)\Big) \Bigg].\label{eq:v3_component_of_Alfven_wave_with_exotic_terms}
\end{align}  
In Eqs. \eqref{eq:b2_component_of_Alfven_wave_with_exotic_terms}-\eqref{eq:v3_component_of_Alfven_wave_with_exotic_terms} we have used the notation from Eq. \eqref{eq:matrix_A0} along with $a \coloneqq D(\boldsymbol{k})$ from \eqref{eq:minus_discriminant} and arbitrary amplitude vector $\boldsymbol{c}$ from \eqref{eq:perturbation_theory_solution_for_linearized_system}.
Note that, from two independent solutions with $\omega_{1,2}(\boldsymbol{k)}$ (see Eq. \eqref{eq:omega12(k)}), we have selected the one for which the wave propagates in the direction $\boldsymbol{k}$, i.e. $\omega_{1}(\boldsymbol{k)}$.
\end{widetext}

\section{\label{app:supplementary_info_about_two-loop_Gammab'b_diagrams} Two-loop contributions in \texorpdfstring{$\Sigma^{b'b}$}{} corresponding to the curl terms in presence of \texorpdfstring{$B_0$}{B\_0}}

\begin{figure}[t]
\begin{overpic}[percent,grid=false,tics=5,width=1\linewidth]{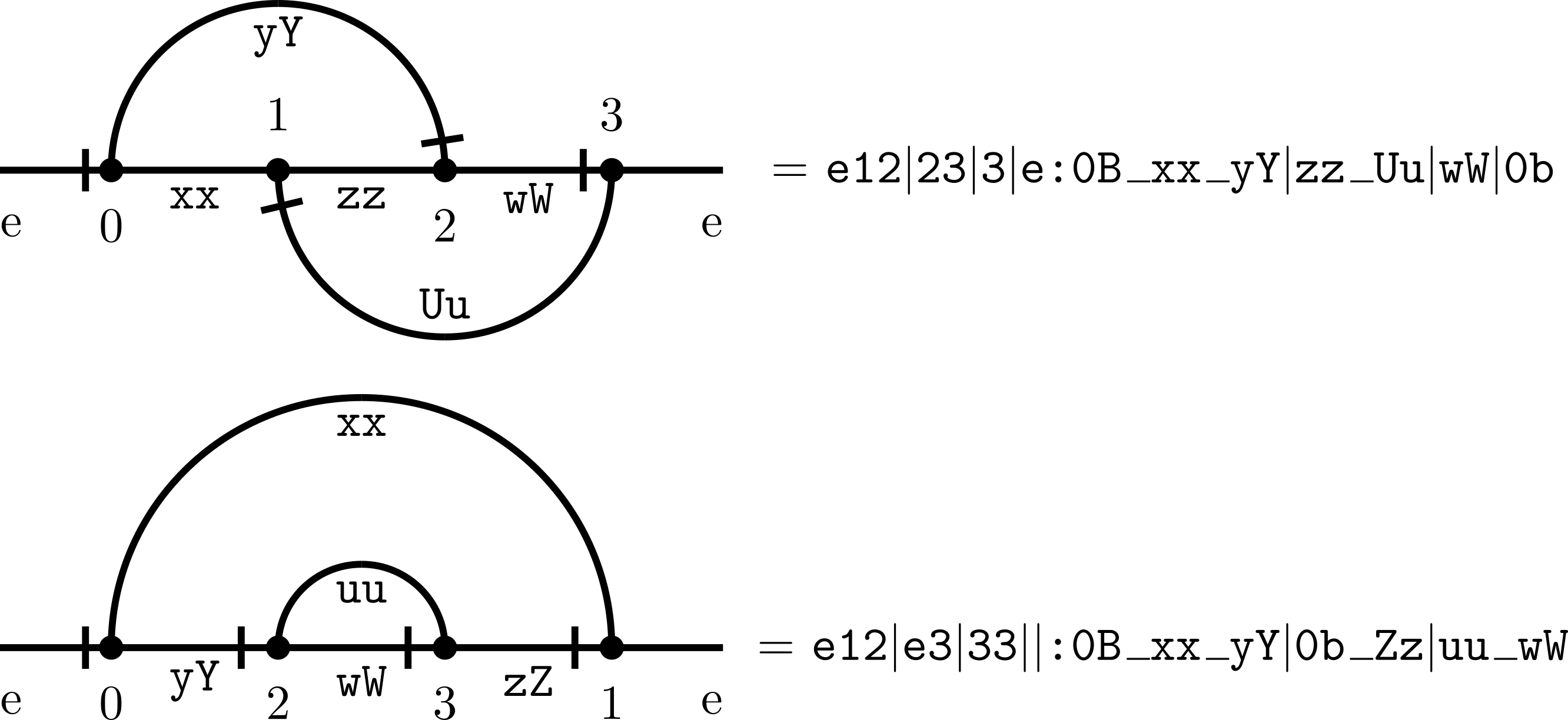}
\end{overpic}
\caption{Two topologies of two-loop $\Sigma^{b'b}$ graphs and corresponding Nickel indices illustrating the ordering of vertices and lines according to the labeling.
The endpoints of lines corresponding to auxiliary fields ($\boldsymbol{v}'$ or $\boldsymbol{b}'$) are represented by uppercase Latin letters, while those corresponding to physical fields ($\boldsymbol{v}$ or $\boldsymbol{b}$) are represented by lowercase letters.
For the lines defined in Eq. \eqref{eq:propagator_matrix_after_shift}, the possible combinations include: \texttt{vv}, \texttt{vV}, \texttt{Vv}, \texttt{bb}, \texttt{bB}, \texttt{Bb}, \texttt{vb}, \texttt{bv}, \texttt{Vb}, \texttt{bV}, \texttt{vB}, and \texttt{Bv}.
}
\label{fig:Nickel_index_illustration}
\end{figure}

Each of the $488$ two-loop self-energy diagrams (graphs) contributing to $\Sigma^{b'b}$ is uniquely specified by its Nickel index defined in \cite{NickelIndex} (see also \cite{Batkovich2014_GraphState, Kompaniets2017}).
This index serves as an intuitive notation for an adjacency list with respect to a particular minimal labeling of the vertices.
For reader's convenience, Fig. \ref{fig:Nickel_index_illustration} illustrates this notation for both topological types of self-energy diagrams in the two-loop approximation.

This notation is particularly useful for obtaining explicit expressions for two-loop integrals within our diagrammatic framework. Once these expressions are derived, the linear term in the external momentum expansion is isolated for each integral, frequency variables are integrated out, and tensor reductions are performed. The resulting expressions take the form given in Eq. \eqref{eq:two-loop_integral_after_rescaling_by_B0}, representing the integrands corresponding to the two-loop curl contributions to $\Sigma^{b'b}$.
Detailed theoretical descriptions of the frequency integration and scalarization procedures are provided in Sections \ref{sec:frequency_integration} and \ref{sec:scalarization}, respectively.
In practice, these computations were carried out using two independent methods: one implemented in Python and another in Wolfram Mathematica.
The agreement between these two approaches provided a robust intermediate validation of our results, which are presented in the next Section as the most practical form for further use.

\subsection{\label{app:description_of_ancillary_files}Description of ancillary files}

This article is supplemented by a comprehensive dataset in the form of human-readable text files provided in Wolfram Mathematica \cite{Mathematica} (\verb|.wl|) format.
Specifically, the Supplemental Material includes:
\begin{itemize}
    \item[(i)] Files \verb|ExportSunSet.wl| and \verb|ExportTriangles.wl|, which contain all necessary information about the integrals for each topology: the former represents topology I1 (sunset diagrams) and the latter topology I2 (double-scoop diagrams).
    \item[(ii)] Files \verb|I1-F.wl|, \verb|I1-L.wl|, \verb|I1-P.wl|, \verb|I1-IR.wl|, \verb|I1-S.wl|, 
    \verb|I2-F.wl|, \verb|I2-L.wl|, and \verb|I2-P.wl| containing the Nickel indices for the subtypes of diagrams as described in Sec. \ref{sec:Two-loop_diagrams_classification}. 
    \item[(iii)] The file \verb|IRCancellation.nb|, containing the IR asymptotic of each diagram of type I1-IR, written without the factor $(p q)^{-2\epsilon}$ for clarity.
    In addition, this file contains a proof that the total sum of all these asymptotics is zero.
    \item[(iv)] The file \verb|Rules.nb| contains various patterns necessary for transforming the integrands in \verb|ExportSunSet.wl| and \verb|ExportTriangles.wl| from compact notation to a form suitable for numerical or other types of manipulations.
\end{itemize}


Each file from (i) can be loaded into a list-type variable (e.g., \verb|DiagramInfo|) using the \verb|Get| command.
Each entry in \verb|DiagramInfo| is of the form
\begin{verbatim}
DiagramInfo[[i]] = [NI(D),Integrand(D),Sym(D)]
\end{verbatim}
indexed by the integer \verb|i| corresponding to diagram $D$ in the file.
This triplet contains the diagram's Nickel index \verb|NI(D)|, its integrand for the curl term \verb|Integrand(D)| after frequency integration and tensor contractions and symmetry factor \verb|Sym(D)|.
All integrands are given without multiplication by the Jacobian 
\begin{align}\label{eq:Jacobian}
    \mathrm{J} = 4\pi p^{2} q^{2} \sin{\theta_p} \sin{\theta_q}
\end{align}
arising from the measure $\mbox{d} \Omega$ given 
in Eq. \eqref{eq:angular_measure}.
Note also that we consider ``leg-fixed'' diagrams, so the symmetry factor is given by $\mathrm{Sym}(D) = 1/\mathrm{Aut}(D)$, where the automorphisms are allowed to permute only the internal diagram lines. 

Each of the files (ii) contain the Nickel indices of diagrams of the corresponding subtype.
All relevant information corresponding to these indices can be retrieved from \verb|ExportSunSet.wl| or \verb|ExportTriangles.wl| using the \verb|Cases| command.
For example, suppose we have loaded the Nickel indices from \verb|I1-F.wl| into a variable \verb|x| and the complete information from \verb|ExportSunSet.wl| into a variable \verb|y|.
Then the following command will filter out the required diagrams:
\begin{verbatim}
z=Flatten/@(Cases[y,{N_,D_,S_}/;MemberQ[x,N]]);
\end{verbatim} 
Here, the pattern \verb|{N_, D_, S_}| corresponds to the Nickel index, integrand, and symmetry factor, respectively, and the resulting variable \verb|z| will be an array of triplets.

The integrand \verb|N| is written in compact notation.
To obtain an expression suitable for numerical integration, one must use the rules described in the file \verb|Rules.nb|.
More specifically, let \verb|Int1| be the one particular integrand.
To obtain its expression in the full form, the following command have to be written:
\begin{verbatim}
Int2 = Int1/.InverseRules /. InverseRules2 /. 
NumericRules11m /. NumericRules2
\end{verbatim}
Additionally, to study expression analytically, the variable $u$ can be kept as a symbol using the following sequence of rules:
\begin{verbatim}
Int2 = Int1/.InverseRules/.InverseRules2
/.NumericRules11m/.NumericRules12
\end{verbatim}
\nocite{*}
\bibliography{bibliography}
\end{document}